\documentclass[superscriptaddress,groupedaddress,nofootnoteinbib,11pt]{article}
\pdfoutput=1 
\usepackage{jheppub}

\usepackage{amsmath, amsfonts, amsthm, amssymb, graphicx, color, hyperref, slashed}
\usepackage{subfigure}
\usepackage{pstool}
\usepackage{graphicx}
\usepackage{float} 
\usepackage[utf8]{inputenc}
\usepackage[normalem]{ulem}

\allowdisplaybreaks[3]

\def \bal#1\eal  {\begin{align} #1 \end{align}}
\def\({\left(}
\def\){\right)}
\def\[{\left[}
\def\]{\right]}
\def\<{\left\langle}
\def\>{\right\rangle}
\def\d{\mathrm{d}}

\newcommand{\eref}[1]{Eq.~(\ref{#1})}
\newcommand{\f}[2]{\frac{#1}{#2}}
\newcommand{\bim} {\begin{itemize}[noitemsep]}   
   
\newcommand{\eim}{\end{itemize}}
\newcommand{\be} {\begin{equation}} 
\newcommand{\ee} {\end{equation}}
\newcommand{\bc}{\begin{center}}   
\newcommand{\ec}{\end{center}}

\newcommand{\nn} {\nonumber\\}

\newcommand{\ie}{{\it i.e.,}~}

\newcommand{\mc} {\mathcal}

\newcommand{\ai}{{\alpha}}
\newcommand{\bi}{{\beta}}

\newcommand{\li}{{\lambda}}
\newcommand{\ti}{{\tau}}
\newcommand{\ei}{{\eta}}

\newcommand{\epi}{\epsilon}

\title{Generalized positivity bounds on chiral perturbation theory}
\author[a]{Yu-Jia Wang,}
\author[b,c]{Feng-Kun Guo,}
\author[d,c]{Cen Zhang}
\author[a]{and Shuang-Yong Zhou}

\affiliation[a]{Interdisciplinary Center for Theoretical Study, University of Science and Technology of China, Hefei, Anhui 230026, China and Peng Huanwu Center for Fundamental Theory, Hefei, Anhui 230026, China}
\affiliation[b]{CAS Key Laboratory of Theoretical Physics, Institute of Theoretical Physics, Chinese Academy of Sciences, Beijing 100190, China}
\affiliation[c]{School of Physical Sciences, University of Chinese Academy of Sciences, Beijing 100049, China}
\affiliation[d]{Institute of High Energy Physics, Chinese Academy of Sciences, Beijing 100049, China}

\emailAdd{wyjiavri@mail.ustc.edu.cn}
\emailAdd{fkguo@itp.ac.cn} 
\emailAdd{cenzhang@ihep.ac.cn}
\emailAdd{zhoushy@ustc.edu.cn}

\preprint{\small USTC-ICTS/PCFT-20-12}

\date{\today}

\abstract{Recently, a new set of positivity bounds with $t$ derivatives have been discovered. We explore the generic features of these generalized positivity bounds with loop amplitudes and apply these bounds to constrain the parameters in chiral perturbation theory up to the next-to-next-to-leading order. We show that the generalized positivity bounds give rise to stronger constraints on the $\bar l_i$ constants, compared to the existing axiomatic bounds. The parameter space of the $b_i$ constants is constrained by the generalized positivity bounds to be a convex region that is enclosed for many sections of the total space. We also show that the improved version of these positivity bounds can further enhance the constraints on the parameters. The often used Pad\'e unitarization method however does not improve the analyticity of the amplitudes in the chiral perturbation theory at low energies.}

\begin{document}
\maketitle
\flushbottom

\section{Introduction}

Chiral perturbation theory (ChPT) is one of the oldest and most widely studied effective field theories (EFTs)~\cite{Weinberg:1966kf,Li:1971vr,Gasser:1983yg,Gasser:1984gg,Pich:1995bw,Scherer:2002tk,Bernard:2006gx}. It is the low energy description of quantum chromodynamics (QCD), which is perturbative at high energies but strongly coupled at low energies, giving rise to the vast richness of hadron physics. The essential feature of the theory is that a chiral symmetry group, a product of two groups of the same structure, is spontaneously (and often mildly explicitly) broken down to the diagonal subgroup, which generates a nonlinear realization of the chiral symmetry. The structure of ChPT thus is largely determined by the nonlinearly realized symmetry~\cite{Weinberg:1966kf,Coleman:1969sm,Callan:1969sn}. Because of the universal structure, ChPT is also relevant for other physical scenarios as long the underlying symmetry breaking pattern is the same. Indeed,  the SU(2) ChPT was also widely used as a low energy EFT for the electroweak symmetry breaking before the discovery of the light Higgs boson (see, {\it e.g.},~\cite{Dobado:1990jy, Distler:2006if}). 

Although the construction of higher dimensional operators of an EFT is determined by the symmetry of the system, there are often many of them that are relevant for a given problem. Each of these operators is accompanied by a Wilson coefficient or low energy constant (LEC). The number of LECs proliferates very quickly with the dimension of the operators, so if the LECs are allowed to take any values, the parameter space of the EFT is very large. However, the LECs are not allowed to take arbitrary values if one is to assume the ultraviolet (UV) completion of the EFT satisfies axiomatic principles of quantum field theory or the scattering amplitude such as Lorentz invariance, unitarity, locality, crossing symmetry and analyticity. In particular, there are the so called positivity bounds which the LECs must satisfy~\cite{Pham:1985cr, Pennington:1994kc, Ananthanarayan:1994hf, Comellas:1995hq, Dita:1998mh, Adams:2006sv}. The simplest positivity bound (see, {\it e.g.,}~\cite{Adams:2006sv}) states that the second $s$ derivative ($s,t,u$ being the conventional Mandelstam variables) of the pole subtracted scattering amplitude has to be positive. At low energies, the scattering amplitude can be well described by the LECs. As a result, the LECs are constrained by positivity bounds. This bound utilizes the optical theorem that is valid in the forward scattering limit, but the extension away from the forward limit is also possible~\cite{Dita:1998mh, Manohar:2008tc, Mateu:2008gv, Nicolis:2009qm, Bellazzini:2016xrt}. Recently, an infinite set of generalized positivity bounds have been discovered~\cite{deRham:2017avq}, which makes use of the simple fact that arbitrary numbers of $t$ derivatives on the imaginary part of the amplitude is positive at and away from the forward limit, though quite a few technical points need to be resolved for the generalization of these bounds for particles with spin~\cite{deRham:2017zjm}. We will briefly review the derivation of the generalized positivity bounds in Section~\ref{sec:improbounds} for the case of multiple scalars. Recently, there has been a renewed interest in applying positivity bounds, the forward limit bound and its generalizations, in various EFTs from particle physics and cosmology~\cite{deRham:2018qqo, deRham:2017imi, Zhang:2018shp, Bi:2019phv, Remmen:2019cyz, Bellazzini:2018paj, Baumann:2015nta, Bellazzini:2015cra, Cheung:2016yqr,  Cheung:2016wjt, Bellazzini:2017fep, Bonifacio:2016wcb, Hinterbichler:2017qyt, Bonifacio:2017nnt, Bellazzini:2017bkb,  Bonifacio:2018vzv, Bellazzini:2019xts, Melville:2019wyy, deRham:2019ctd, Alberte:2019xfh, Alberte:2019zhd, Ye:2019oxx, Herrero-Valea:2019hde}.

Chiral perturbation theory often serves as an exemplary EFT where features of EFTs are shown with or referred to, and as a test ground where new ideas are applied up on. In this paper, we shall apply the newly discovered generalized positivity bounds to ChPT for $\pi\pi$ scatterings. The purpose is twofold. Positivity bounds have been previously used to constrain the LECs in ChPT~\cite{Pennington:1994kc, Ananthanarayan:1994hf, Dita:1998mh, Distler:2006if, Manohar:2008tc, Mateu:2008gv} (see also Refs.~\cite{Sanz-Cillero:2013ipa,Du:2016tgp} for applications to ChPT with matter fields). In terms of the generalized positivity bounds of Refs.~\cite{deRham:2017avq, deRham:2017zjm}, the earlier bounds are roughly the lowest order bound with only two $s$ derivatives in an infinite set of bounds with the new ingredients of arbitrary numbers of $t$ (and $s$) derivatives. Another new ingredient that one may add into the generalized bounds is the use of the improved bounds, where the low energy part of the dispersion integral is subtracted to improve the strength of the bounds. With these new ingredients, it is of interest to see how the generalized positivity bounds can improve the previous bounds on the LECs, and we will see that, indeed, the improvements are obvious, as will be shown in a number of ways. Secondly, we also use ChPT to uncover some salient properties of generalized positivity bounds for an EFT amplitude computed to higher loops. This is interesting despite previous applications of the generalized bounds in other models, as most previous applications of generalized positivity bounds are for tree level amplitudes in cosmology, where high derivative interactions can provide high momentum powers  in the amplitude already at tree level. Particularly, we will explore the properties of the improved version of the generalized bounds with an explicit high-loop amplitude.

The rest of the paper is organized as follows: We start by introducing ChPT and the amplitudes for the $\pi\pi$ scattering processes in Section~\ref{sec:chpt}, with some long formulas of the amplitudes presented in Appendix \ref{appendix: loop fun}. In Section~\ref{sec:Pbounds}, we briefly review how to derive the generalized positivity bounds, the $Y$ bounds and the improved version. In Section~\ref{sec:Y bounds cons}, we first explore the structure of the infinite set of the $Y$ bounds, charting out the strengths of the different bounds and preparing for the later applications, and then use the strongest bounds to constrain the $\bar l_i$ and $b_i$ constants, which are the LECs in the next-to-leading order (NLO) chiral Lagrangian and combinations of the NLO and the next-to-next-to-leading (NNLO) LECs, respectively; part of the results (the 3D sections) are deferred to Appendix \ref{sec:3D}. In Section~\ref{sec:impYb}, we discuss the structure of the improved $Y$ bounds and use them to extract the energy scale where ChPT breaks down and, more importantly, to enhance the constraints on the $\bar l_i$ and $b_i$ constants; we also show that the Pad{\'e} unitarized amplitude has worse analytical properties than the original amplitude. We conclude in Section~\ref{sec:concl}.

\section{Chiral perturbation theory}
\label{sec:chpt}

Chiral perturbation theory is a low energy EFT of a field theory with (approximate) chiral symmetry group ${\rm G}_L\times {\rm G}_R$ that is spontaneously broken to the diagonal vector subgroup ${\rm G}_V$. The chiral symmetry is often approximate because it may be explicitly broken. For example, in QCD, it is explicitly broken by the quark mass terms (which in turn come from spontaneous breaking of other symmetries), so the degrees of the freedom associated with the breaking are pseudo-Goldstone pseudoscalars, called pions, and in this paper we only consider pion scatterings. One of the simplest chiral EFTs is where the chiral symmetry ${\rm SU}(2)_{L} \times {\rm SU}(2)_{R}$ is spontaneously broken to ${\rm SU}(2)_{V}$, which contains 3 pseudoscalars $\pi^a$. This model is the prototype of modern EFTs and has been extensively used to describe the low-energy dynamics of QCD involving the lightest $u$ and $d$ quarks~\cite{Weinberg:1966kf,Li:1971vr,Gasser:1983yg,Pich:1995bw,Scherer:2002tk,Bernard:2006gx}. It was also used to parametrize possible symmetry breaking patterns of electroweak interactions in the limit of a heavy Higgs~\cite{Appelquist:1980vg,Longhitano:1980tm,Longhitano:1980iz,Herrero:1993nc}; now, with the discovery of a light Higgs, it has also been extended to include the light Higgs explicitly~\cite{Buchalla:2012qq,Buchalla:2013rka,Alonso:2012px,Guo:2015isa,Buchalla:2017jlu,Alonso:2017tdy,Jenkins:2017dyc} (for a review see Ref.~\cite{Pich:2018ltt}).  

The quark mass terms explicitly break the chiral symmetry, but one may add the St\"uckelberg or spurion fields to introduce the same symmetry breaking pattern into the effective Lagrangian, and then the chiral Lagrangian can be systematically obtained by the coset construction order by order in positive powers of the momenta and the quark masses. 
In the standard ChPT power counting, each insertion of the light quark mass matrix is counted as order $\mathcal{O}(p^2)$ with $p$ the typical small momentum. For ChPT without any matter field, the leading order is $\mathcal{O}(p^2)$.
Using the sigma model parametrization, the basic building blocks for pion scatterings in the isospin symmetry limit are given by
\be
U= \sqrt{1-\frac{\pi^a \pi^a}{F^2} }\mathbf{1}   + i\frac{  \pi^{a} \tau^{a}}{F} ~~~~ {\rm and}~~~~ \chi= M^2 \mathbf{1} ,
\ee
where $F$ and $M$, being positive constants, are the pion decay constant in the chiral limit and the leading order pion mass, respectively, $\ti^a$ is the Pauli matrix, $\mathbf{1}$ is the $2\times2$ identity matrix and the summation for the repeated index $a$ from 1 to 3 is implied. Taking the square root of matrix $U$: $u = \sqrt{U}$, we can construct the following Hermitian matrices
\be
u_{\mu}=i u^{\dagger} \partial_{\mu} U u^{\dagger} = u_{\mu}^\dagger  ,~~~{\rm and} ~~~~ \chi_{+}=u^{\dagger} \chi u^{\dagger}+u \chi^{\dagger} u =  \chi_{+}^\dagger .
\ee
Then the chiral Lagrangian needed to calculate the pion scatterings up to $\mc{O}(p^6)$, the next-to-next-to leading order, is~\cite{Gasser:1983yg,Bijnens:1997vq}
\bal
\mathcal{L}_{\mathrm{ChPT}}&=\mathcal{L}_{2}+  \mathcal{L}_{4} + \mathcal{L}_{6}
\\
&=\frac{F^{2}}{4}\left\langle u_{\mu} u^{\mu}+\chi_{+}\right\rangle + \frac{l_1}{4}\left\langle u^{\mu} u_{\mu}\right\rangle^{2} + \frac{l_2}{4}\left\langle u_{\mu} u_{\nu}\right\rangle\left\langle u^{\mu} u^{\nu}\right\rangle + \frac{l_3}{16}\left\langle\chi_{+}\right\rangle^{2} +\sum_{i} c_i Y_i ,
\eal
where $\<~ \>$ denotes taking the trace of the matrix in the flavor space, $l_i$ and $c_i$ are the LECs in the $\mc{O}(p^4)$ and $\mc{O}(p^6)$ Lagrangians, respectively, and the $Y_i$ operators can be found in Table 2 of Ref.~\cite{Bijnens:1999sh}. The amplitude of $\pi\pi$ scatterings have been calculated up to two loops~\cite{Bijnens:1995yn, Bijnens:1997vq}, which for the $\pi^a\pi^b\to\pi^c\pi^d$ scattering is given by 
\begin{equation}
\label{amplitude}
T_{a b \rightarrow c d}(s,t,u)= A(s, t, u) \delta^{a b} \delta^{c d}+A(t, s, u) \delta^{a c} \delta^{b d} +A(u, t, s) \delta^{a d} \delta^{b c} ,
\end{equation}
with
\begin{align}
 A(s, t, u)&= x_{2}[s-1] +x_{2}^{2}\left[b_{1}+b_{2} s+b_{3} s^{2}+b_{4}(t-u)^{2}\right] 
\nn
&~~~ +x_{2}^{2}\left[F^{(1)}(s)+G^{(1)}(s, t)+G^{(1)}(s, u)\right] +x_{2}^{3}\left[b_{5} s^{3}+b_{6} s(t-u)^{2}\right] 
\nn
&~~~ +x_{2}^{3}\left[F^{(2)}(s)+G^{(2)}(s, t)+G^{(2)}(s, u)\right] +O\left(x_{2}^{4}\right) ,
\end{align}
where $x_{2}={M_{\pi}^2}/{F_{\pi}^{2}}$ is the power counting parameter, with $M_{\pi}$ the pion mass and $F_\pi$ the pion decay constant, and $s,t,u$ are the dimensionless Mandelstam variables 
\be
\label{studef}
s=\f{\left(p_{a}+p_{b}\right)^{2}}{M_{\pi}^2},~ t=\f{\left(p_{a}-p_{c}\right)^{2}}{M_{\pi}^2},~ u=\f{\left(p_{a}-p_{d}\right)^{2}}{M_{\pi}^2}  .
\ee
$F^{(i)}(s)$ and $G^{(i)}(s, t)$ are loop functions defined in Appendix~\ref{appendix: loop fun}, and the $b_{i}$ constants are functions of  the renormalised LECs $l_{i}^{r}(\mu)$ and $r_{i}^{r}(\mu)$ in  $\mathcal{L}_{4}$ and $\mathcal{L}_{6}$, respectively (see Appendix~\ref{appendix: loop fun}). Using the positivity bounds, we can impose bounds on the $b_i$ constants. For the case of QCD, we take $M_{\pi}=139.57~{\rm MeV}$ and  $F_{\pi}=92.28(9)$~MeV~\cite{Tanabashi:2018oca}. Note that in the above expressions we apparently used $F_\pi$ or $x_2$ to order the EFT expansion. However, we would like to emphasize that the real perturbative expansion parameter is actually ${M_{\pi}^2}/\Lambda^2$ with  $\Lambda=4\pi F_\pi$~\cite{Manohar:1983md}.

We will consider a generic elastic scattering process $\left|\pi^{ \alpha}\right\rangle +  \left|\pi^{ \beta}\right\rangle \rightarrow  \left|\pi^{ \alpha}\right\rangle+  \left|\pi^{ \beta}\right\rangle$ with $\left|\pi^{ \alpha}\right\rangle$ and $\left|\pi^{ \beta}\right\rangle$ states being superpositions of the isospin states
\bal
\left|\pi^{ \alpha}\right\rangle&=\alpha_{a}\left|\pi^{ a}\right\rangle=\alpha_{1}\left|\pi^{ 1}\right\rangle+\alpha_{2}\left|\pi^{ 2}\right\rangle+ \alpha_{3}\left|\pi^{ 3}\right\rangle , \\
\left|\pi^{ \beta}\right\rangle&=\beta_{a}\left|\pi^{a}\right\rangle=\beta_{1}\left|\pi^{ 1}\right\rangle+\beta_{2}\left|\pi^{ 2}\right\rangle+ \beta_{3}\left|\pi^{ 3}\right\rangle ,
\eal
where $\alpha_{a}$ and $\beta_{a}$ are arbitrary complex constants satisfying the normalized conditions: $|\alpha_{a}|^2=1$ and  $|\beta_{a}|^2=1$. In this case, the scattering amplitude is given by
\begin{equation}
T_{\alpha \beta \rightarrow\alpha \beta}(s,t,u)=\eta_1 A(s, t, u) +A(t, s, u)  +\eta_2 A(u, t, s) ,
\label{general amplitude}
\end{equation}
with $\eta_1=|\alpha_{a}\beta_{a}|^2$ and $\eta_2=|\alpha_{a}\beta_{a}^*|^2$. By the Cauchy–Schwarz inequality, we have 
\be
0\leq \ei_1 \leq 1~~{\rm and}~~ 0\leq \eta_2 \leq 1  .
\ee 
As we shall see below, positivity bounds can be derived for a general elastic scattering $\left|\pi^{ \alpha}\right\rangle +  \left|\pi^{ \beta}\right\rangle \rightarrow  \left|\pi^{ \alpha}\right\rangle+  \left|\pi^{ \beta}\right\rangle$ with general $\eta_1$ and $\eta_2$.

\section{Generalized positivity bounds}
\label{sec:Pbounds}

The properties of the UV theory of an EFT such as unitarity, locality, analyticity and crossing symmetry can be used to derive some conditions that constrain the LECs of the EFT. A forward limit positivity bound was derived in Ref.~\cite{Adams:2006sv}, and this has been generalized away from the forward limit in ChPT~\cite{Manohar:2008tc}. In this section, we will briefly summarize the generalized positivity bounds proposed in Refs.~\cite{deRham:2017avq,deRham:2017zjm}, which includes the one in Ref.~\cite{Manohar:2008tc} as a special case.

\subsection{The $Y$ bounds}

We shall use the shorthand notation $T(s, t)=T_{\alpha \beta \rightarrow\alpha \beta}(s,t,u)= \ai_a \bi_b  \ai_c^* \bi_d^*  T_{ab \rightarrow cd}(s,t,u)$. By Cauchy's integral theorem, crossing symmetry and the Froissart-Martin bound \cite{Froissart:1961ux, Martin:1962rt, Jin:1964zza} 
\begin{equation}
  \lim _{s \rightarrow \infty}|T(s, t)|<C s^{1+\varepsilon(t)}, ~ \varepsilon(t)<1, ~ 0 \leq t<4, ~ C={\rm\;const}, 
\end{equation}
we can derive a fixed-$t$ dispersion relation for the amplitude $T(s,t)$ 
\bal
B(s,t)  &\equiv T(s, t)-\frac{\lambda}{1-s}-\frac{\lambda}{1-t}-\frac{\lambda}{1-u}
\\
\label{Bvt}
&= a(t)+\int_{4 }^{\infty} \frac{\mathrm{d} \mu}{\pi\left(\mu-\mu_{p}\right)^{2}}\left[\frac{\left(s-\mu_{p}\right)^{2}}{\mu-s}  \operatorname{Im} T(\mu, t)+\frac{\left(u-\mu_{p}\right)^{2}}{\mu-u}  \operatorname{Im} \tilde T(\mu, t)\right]   ,
\eal
where we have used the dimensionless Mandelstam variables defined in \eref{studef}, $\li$ is a constant, the subtraction point is chosen to be $\mu_p=-t/2$, and the $s\leftrightarrow u$ crossed amplitude is $\tilde T(s, t)= \ai_a \bi^*_b  \ai_c^* \bi_d  T_{ab \rightarrow cd}(s,t,u)$. Expanding the amplitude in terms of partial waves,
\be
T(s, t)=16 \pi \sqrt{\frac{s}{s-4 }} \sum_{\ell=0}^{\infty}(2 \ell+1) P_{\ell}\(1+\frac{2 t}{s-4 }\) t_{\ell}(s),
\ee
the partial wave unitarity requires $\left| t_{\ell}(s)\right|^{2} \leq \operatorname{Im} t_{\ell}(s)$ for $s \geq 4$. Utilizing the positivity properties of the Legendre polynomials $\d^n P_\ell(x=1)/\d x^n\geq 0$ for $n\geq 0$ and Martin's extension of analyticity~\cite{Martin:1965jj} we can obtain~\cite{deRham:2017avq}
\be
\label{ImA0}
\frac{\mathrm{d}^{n}}{\mathrm{d} t^{n}} \operatorname{Im} T(s, t)>0, \quad s > 4 , \quad 0 \leq t<4,\quad n\geq 0.
\ee
(See Appendix B of Ref.~\cite{deRham:2017imi} for the reason why this is a strict positivity (rather than semi-positivity) condition for a nontrivial scattering.) Since the $t$ variable is proportional to the cosine of the scattering angle, the inequality (\ref{ImA0}) essentially gives rise to constraints for all of the partial waves. The same relation holds for the $s\leftrightarrow u$ crossed amplitude $\tilde T(s, t)$. It is also convenient to use $v= s+ t/2-2$ instead of $s$, for which we have $v=0$ when $s=u$. Thanks to the two ingredients  (\ref{Bvt}) and (\ref{ImA0}), we see that if an even number of $v$ derivatives act on ${B}(s, t)$, we get a quantity that is positive definite. That is, if we define 
\be
\label{Bdef}
B^{(2N, M)}(t)=\left.\frac{1}{M !} \partial_{v}^{2N} \partial_{t}^{M} {B}(s, t)\right|_{v=0} ,
\ee
we have $B^{(2 N, 0)}(t)>0$ for $N\geq 1$.  $N$ needs to be greater than 0 because we need to eliminate the unknown subtraction function $a(t)$. Actually, when $M=0$, these inequalities also hold away from $v=0$, as explained in Ref.~\cite{Manohar:2008tc}. The same, however, is not true for the $t$ derivatives because the sign of $B^{(2N, M)}(t)$ alternates for different $M$, due to the $u$ channel part of the integrand of Eq.~(\ref{Bvt}). This can be overcome if we linearly combine different $B^{(2N, M)}(t)$ and use the ``relaxing'' inequality of a positive integration  $\int_{4}^{\infty} \d \mu\, (...)/({\mu}+{t} / 2-2) < \int_{4}^{\infty} \d \mu\, (...)/\mc{M}^2$, where $(...)$ denotes a positive quantity and $\mc{M}^2=2+\f{t}2$ is the minimum of $({\mu}+{t} / 2-2)$. With some algebra, we can get~\cite{deRham:2017avq,deRham:2017zjm} 
\begin{equation}
\label{PB}
Y^{(2 N, M)}(t) =
            \sum_{r=0}^{M / 2} c_{r} B^{(2(N+r), M-2 r)}+\frac{1}{\mathcal{M}^{2}} \sum_{\text {even } k=0}^{(M-1) / 2}(2(N+k)+1) \beta_{k} Y^{(2(N+k), M-2 k-1)}  > 0 ,
\end{equation}
where  $ N \geq 1$, $M \geq 0$, $0\leq t< 4$, $Y^{(2 N, 0)}(t)=B^{(2 N, 0)}(t)$ and $c_{k}$ and $\beta_{k}$ are given by
\begin{equation}
c_{k}=-\sum_{r=0}^{k-1} \frac{2^{2(r-k)} c_{r}}{(2(k-r)) !}, \quad c_{0}=1\quad \text { and } \quad \beta_{k}=(-1)^{k} \sum_{r=0}^{k} \frac{2^{2(r-k)-1}}{(2(k-r)+1) !} c_{r} .
\end{equation}
An intriguing fact is that $c_{k}$ and $\beta_{k}$ are simply the Taylor expansion coefficients of the ${\rm sech}(x/2)$ and $\tan(x/2)$ functions, respectively~\cite{deRham:2017imi}. The full amplitude satisfying analyticity, unitarity and crossing symmetry is used to derive the positivity bounds derived above. However, at low energies, a decent EFT amplitude must approximate the full amplitude perturbatively to a desired accuracy by power counting, so the positivity bounds can be obtained with the EFT amplitude within that accuracy, and the bounds then translate into inequalities on the LECs.

\subsection{The improved $Y$ bounds}
\label{sec:improbounds}

The dispersion relation (\ref{Bvt}) integrates the imaginary part of the amplitude from $4$ to $\infty$. However, within the EFT, we can actually compute ${\rm Im}A(s,t)$ at low energies below the cutoff, so we can subtract out the low energy part of the integral to get \cite{deRham:2017imi, Bellazzini:2016xrt, Bellazzini:2017fep, deRham:2017xox}
\begin{equation}
B_{\epsilon \Lambda}(t)=B(t)-\int_{4}^{(\epi\Lambda)^2} \frac{\mathrm{d} \mu}{\pi\left(\mu-\mu_{p}\right)^{2}}\left[\frac{\left(s-\mu_{p}\right)^{2}}{\mu-s}  \operatorname{Im} T(\mu, t)+\frac{\left(u-\mu_{p}\right)^{2}}{\mu-u}  \operatorname{Im} \tilde T(\mu, t)\right],
\label{IPB}
\end{equation}
where we choose $\epi\ll  1/M_\pi$ ($\epi\Lambda$ being dimensionless) to stay well below the cutoff.  By doing so, we have assumed that the imaginary part of the amplitude can be determined with a desired accuracy below $\epi\Lambda$ within a given order of EFT -- note that unitarity is only perturbatively satisfied in an EFT constructed from a derivative expansion. Since the integrand is positive, we can use $B_{\epsilon \Lambda}^{(2 N, M)}(t)$ and go through the same steps and get
\begin{equation}
Y_{\epsilon \Lambda}^{(2 N, M)}(t)  >0 .
\end{equation}
where now $\mathcal{M}^{2} = \epsilon^{2} \Lambda^{2}+t / 2-2$. This is an improvement compared to the $Y$ bounds in the previous subsection, because essentially the improved bounds state that $Y^{(2N,M)}$ is actually greater than a positive number, as opposed to 0 as in the original $Y$ bounds. In addition, it raises the scale of $\mathcal{M}^{2}$ from $\mc{O}(1)$ to $\epsilon^{2} \Lambda^{2}\gg 1$, which in turn leads to enhancement of the higher order $Y$ bounds.

\section{$Y$ bounds on ChPT}
\label{sec:Y bounds cons}

As discussed in Section \ref{sec:Pbounds}, the generalized positivity bounds are a large set of new constraints with different choices of $\{\eta_1,\eta_2,t, N,M\}$. Up to two loops, the LECs in ChPT are bundled in the $b_i$ constants and the amplitude is linear in $b_i$, so for a given set of $\{\eta_1, \eta_2, t, N,M\}$ the positivity bound is a linear inhomogeneous inequality for $b_i$. To get the best bounds on the LECs in ChPT, we should survey all the different choices and solve a large, principally infinite, set of inequalities, and, as we will explain later, the final bounded region has to be convex. Before doing that, we shall first explore how the positivity bounds look like for different choices of the parameters mentioned above, so as to set up a guide to proceed more effectively.  Note that, for ChPT up to $\mc{O}(p^6)$, despite that the analytic polynomial terms, which contain the $b_i$ LECs, are at most to the third power of the Mandelstam variables, the amplitude nevertheless contains complicated logarithmic functions (see Appendix \ref{appendix: loop fun}) which give rise to the $Y$ bounds with large $N,M$. 

However, numerically evaluating the higher order bounds with those logarithmic functions is relatively slow, as imaginary numbers would appear and cancel later for the $s$ and $t$ values we are interested in. To alleviate this problem, we replace the logarithmic functions in the amplitude with the arc tangent functions using the identity ${\arctan}(z)=\frac{1}{2i}{\ln}(\frac{1+i\cdot z}{1-i \cdot z})$. Additionally, we shall Taylor expand the arc tangent functions at $t_0$ ($0<t_0<4$) to order $\mc{O}(\left(t-t_0\right)^{60})$, which further speeds up the numerical evaluations for large $N,M$.

\subsection{Structure of the bounds}
\label{sec:structure}

\begin{figure}[tb]
    \centering 
    \begin{subfigure}{}
      \includegraphics[width=.405\linewidth]{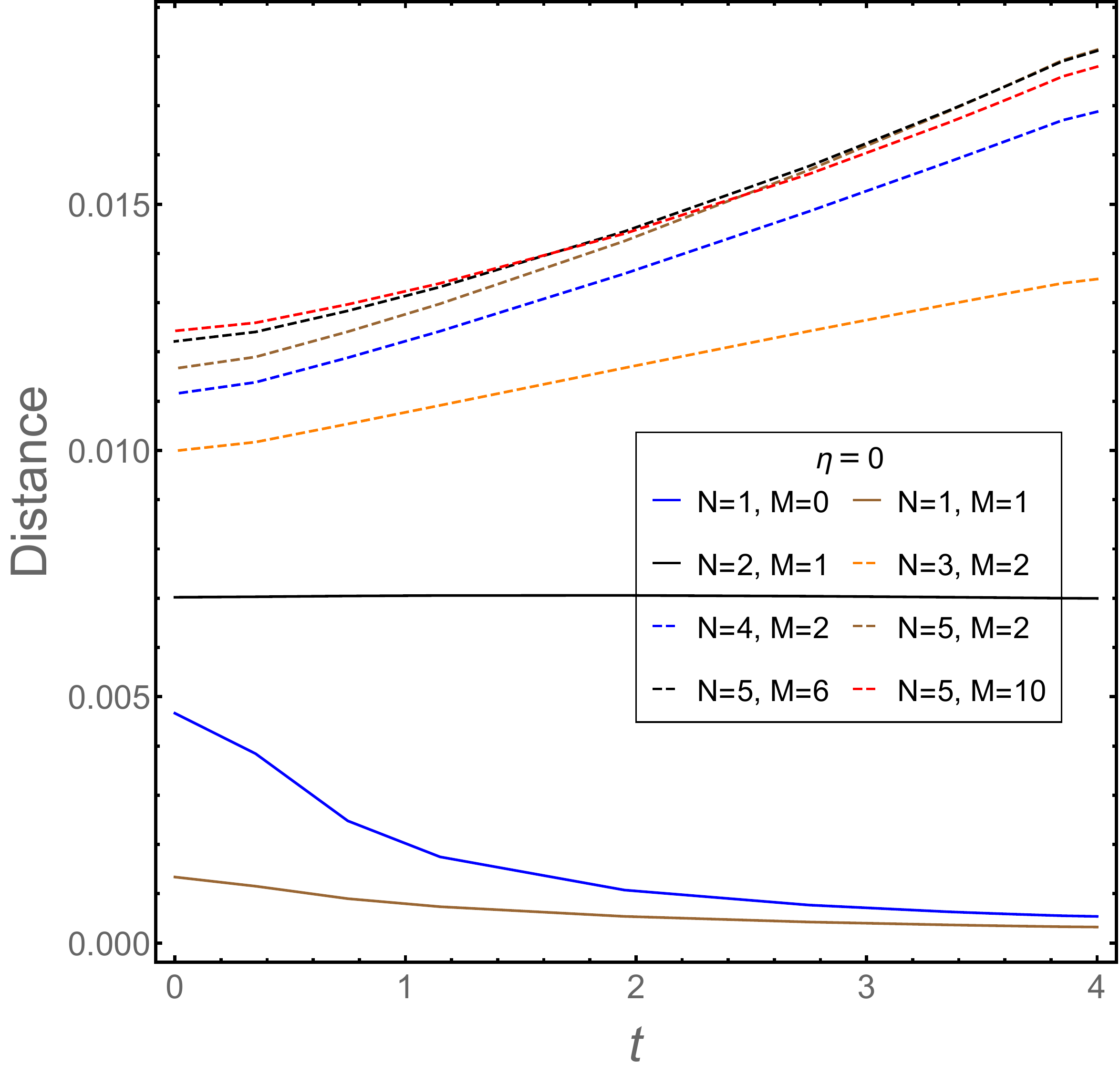}
    \end{subfigure}
     \begin{subfigure}{}
\includegraphics[width=.4\linewidth]{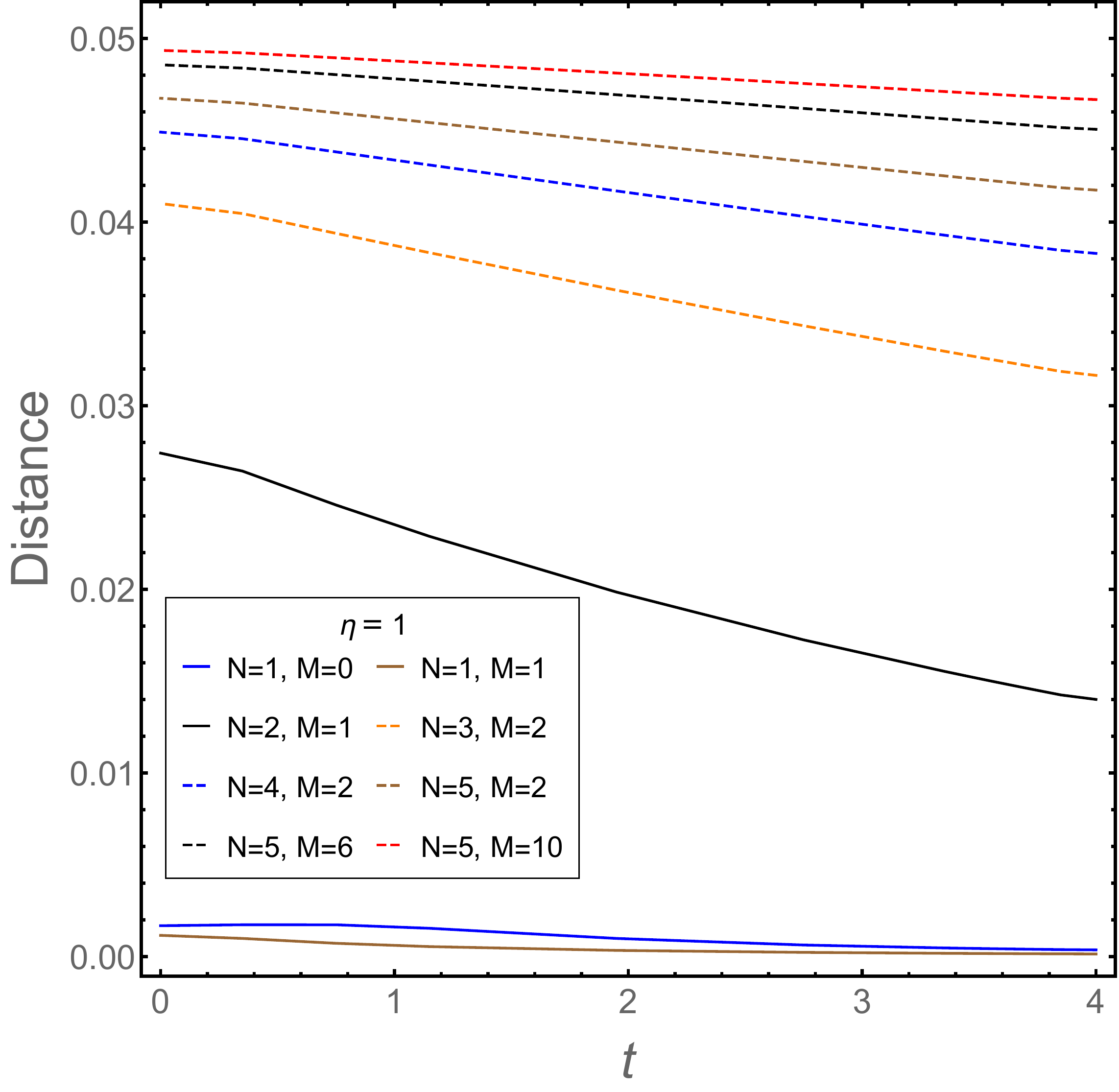}
    \end{subfigure}
\caption{Distances between the bound plane and a fiducial point of $b_i$ of in the $(b_1, b_2,b_3,b_4,b_5,b_6)$ space for the cases of $\eta=0,1$. The bound plane is depicted by $Y^{(2N,M)}(t)=a_0+\sum_{i=1}^6 a_i b_i=0$ for a given set of $\{\eta, t, N, M\}$. The fiducial point is taken to be the fitted values of $b_i$ in Eq.~\eqref{expri data1}. For most bounds, the distance varies monotonically with $t$, except for $\{\eta=0,N=2,M=1\}$ and $\{\eta=1, N=1,M=0\}$, but in all of these cases the shortest distances are always at $t=0$ or $4$.}
\label{fig:teta=01}
    \end{figure}

First of all, note that the $\eta_1$ and $\eta_2$ parameters appear linearly in the positivity bounds, so the strongest bounds can be obtained when $\eta_1$ and $\eta_2$ are evaluated at 0 and 1. Furthermore, in the $Y$ bounds,  $\eta_1$ and $\eta_2$ always appear in the combination of 
\be
\eta \equiv \f{\eta_1+\eta_2}{2} .
\ee
This is due to the fact that the $Y$ bounds by construction are evaluated at the $s\leftrightarrow u$ symmetric point, \ie at $v=s+\f{t}2-2=0$ (or $s=u$).
To see this at a technical level, we can recast the amplitude as
\begin{equation}
\label{eta=01}
T_{\alpha \beta \rightarrow\alpha \beta}(s,t,u)=\frac{\eta_1+\eta_2}{2} [A(s, t, u) +A(u, t, s)]+A(t, s, u)  +\frac{\eta_1-\eta_2}{2} [A(s, t, u) - A(u, t, s)]  .
\end{equation}
The $Y$ positivity bounds are just a linear combination of the $v$ and $t$ derivatives of the amplitude with $v$ evaluated at $0$.
Note that the last term in the amplitude $A(s, t, u) - A(u, t, s)$ is $s\leftrightarrow u$ antisymmetric, \ie $A(s, t, u) - A(u, t, s)$ changes its sign when $v$ goes to $-v$, so $A(s, t, u) - A(u, t, s)$ can be expanded in terms of odd powers of $v$. Thus, any even order $v$ derivatives of $A(s, t, u) - A(u, t, s)$ vanishes when evaluated at $v=0$. Since the $Y$ bounds only involve even order $v$ derivatives of the amplitude (see Eqs.~(\ref{Bdef}) and (\ref{PB})), this implies that $A(s, t, u) - A(u, t, s)$, thus in turn $(\eta_1-\eta_2)/2$, does not contribute to the $Y$ bounds, so the $Y$ bounds only depend on $\eta$. This can also be easily checked by explicit computations. Now, since the amplitude is linear on $\eta$, we only need to consider the extremum of $\eta$, \ie $\eta=0 ~{\rm or}~ 1$, which will give the strongest bounds for fixed $\{N,M,t\}$. That is, the strongest bounds are given by the choices of, for example, the $\pi^1\pi^1\to \pi^1\pi^1$ scattering ($\eta=1$) and the $\pi^1\pi^2\to \pi^1\pi^2$ scattering ($\eta=0$). This is probably not surprising considering the symmetry of the isospin space. 

\begin{figure}[tb]
\begin{center}
\includegraphics[width=.45\linewidth]{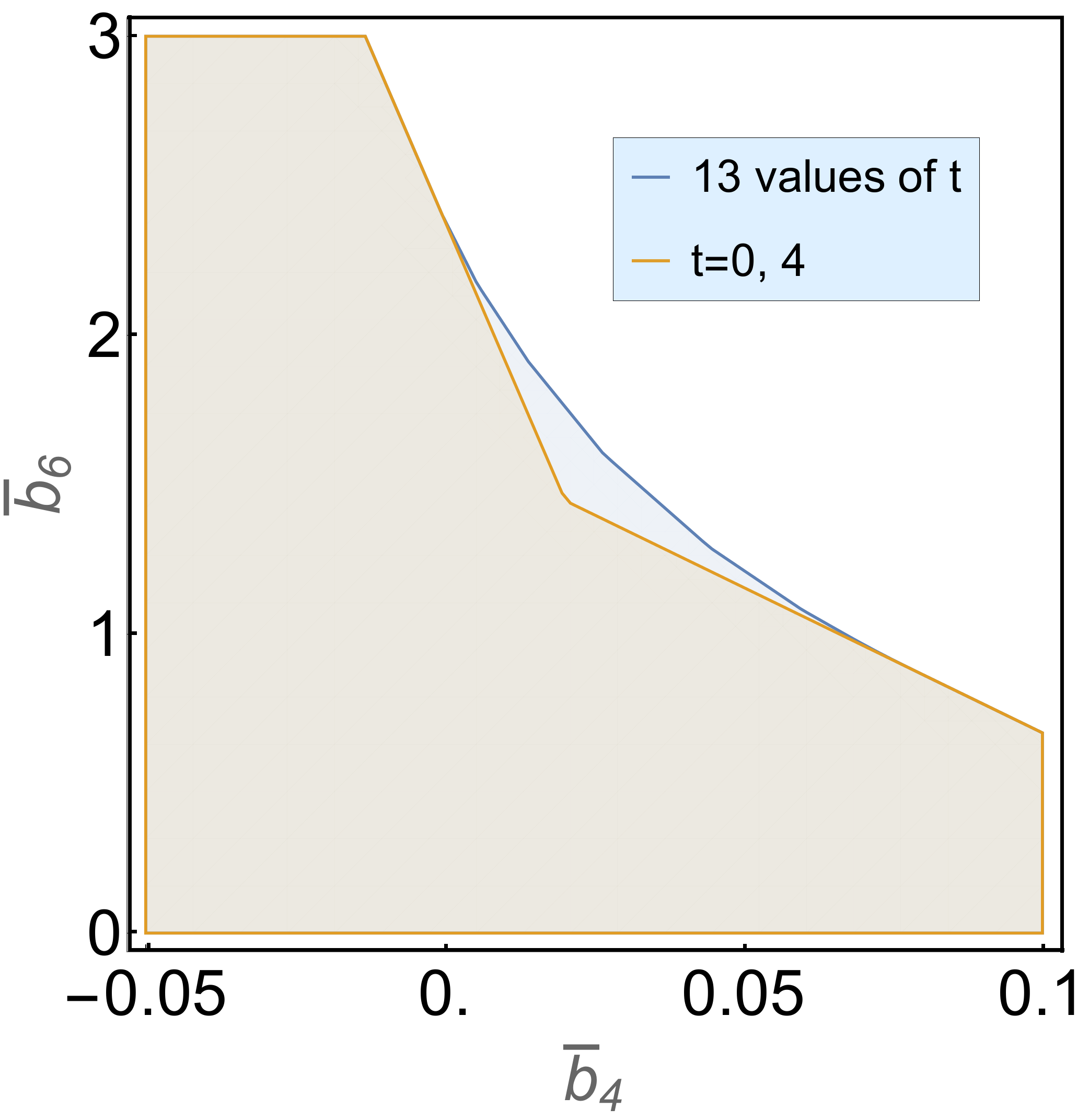}
\end{center}
\caption{ The positivity constraints on the parameter space of $\{b_4,b_6\}$ with other $b_i$ set to the central values of the fit \eqref{expri data1}. The yellow region is ruled out by the $Y$ bounds with $\eta=0,t=0,4$, and is smaller than the blue region that is ruled out by the $Y$ bounds with $\eta=0$ and $13$ values of $t$.  }
\label{fig:13t-vs-2t}
\end{figure}

For the other continuous parameter $t$, the situation is slightly subtler.  For a given set of $\{\eta, t, N, M\}$, any of the $Y$ bounds is of the form: $a_0+\sum_{i=1}^6 a_i b_i>0$, which after replacing $>$ with $=$ can be viewed as a plane in the 6D space of $(b_1, b_2,b_3,b_4,b_5,b_6)$.\footnote{The coefficients of $b_5$ and $b_6$ in the amplitude are third order polynomials of $v$, so only the $N= 1$ bounds contain the $b_5$ and $b_6$ constants.} Thus one may hope to devise a fiducial measure of the strength of the bound as the distance between the plane and a fiducial point. A reasonable fiducial point can be taken as the central values of the parameters determined in  Ref.~\cite{Colangelo:2001df} using Roy's dispersive equations with inputs from experiments (see \eref{expri data1}). In Figure~\ref{fig:teta=01}, we plot how the distances of various bounds vary with $t$. We see that for the bounds with $N<6$ and $M<11$ most of the bounds monotonically increase or decrease with $t$, the only two exceptions being $\{\eta=0,N=2,M=1\}$ and $\{\eta=1, N=1,M=0\}$. Even for these two exceptional cases, the bound plane at either $t=0$ or $4$ is the nearest to the fiducial point. However, this does not mean that we can simply take $t=0$ and $t=4$ when evaluating the bounds, because the $Y$ bounds are tilted in different directions, so bounds with greater distances from the fiducial point can also contribute to the strongest bounds -- the final convex region of the bounds, see Figure \ref{fig:13t-vs-2t} for an example. Therefore, we shall sample the whole range of $t$ where the $Y$ bounds are valid to get the strongest bounds. 

\begin{figure}[tb]
\begin{center}
\includegraphics[width=.5\linewidth]{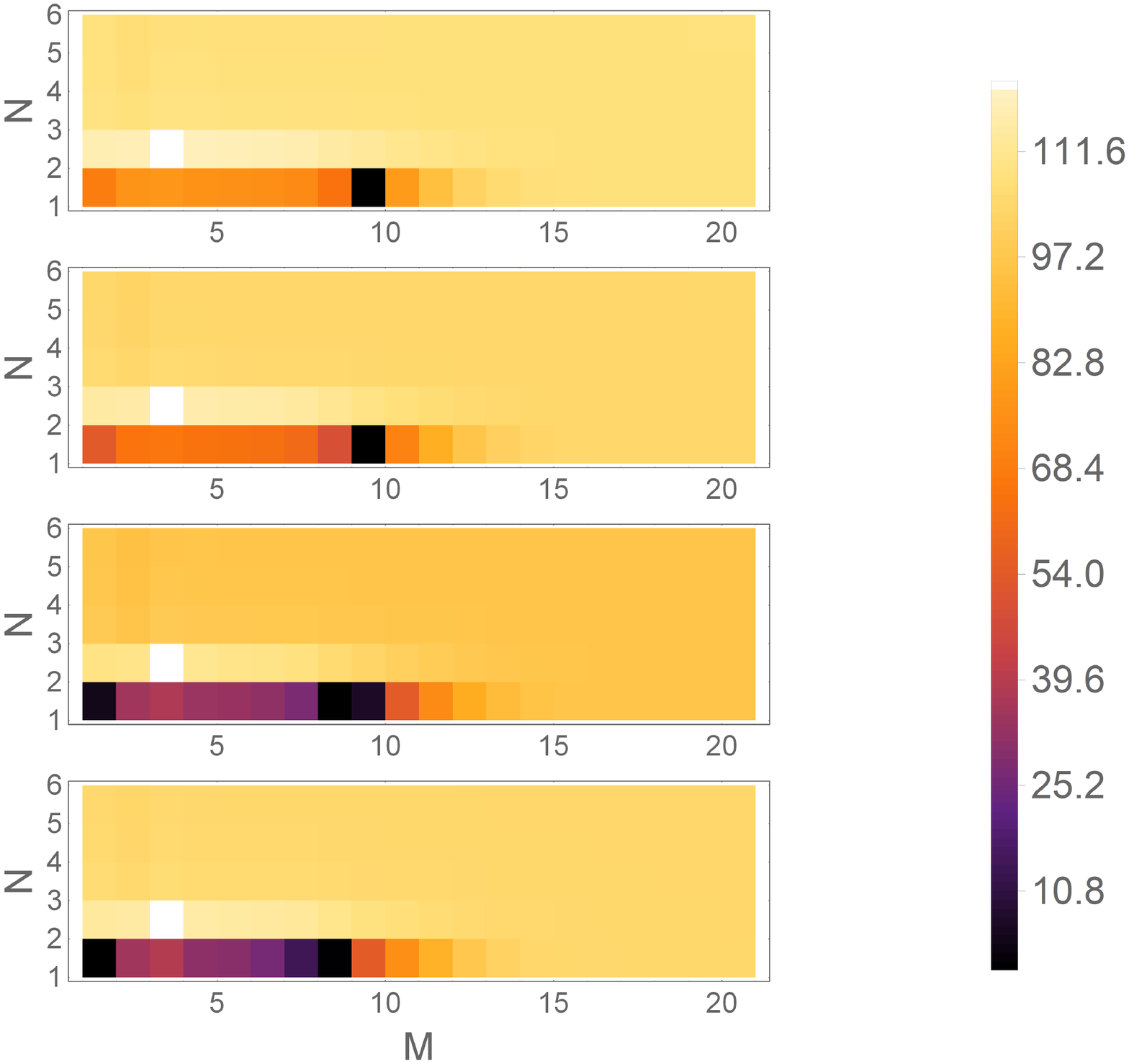}
\end{center}
\caption{Rescaled values of $Y^{(2N,M)}(t)$ for $\eta=0$ and different $\{N,M\}$ at $t=4,3.5,2,0.4$ (from top to bottom). The rescaled value of $Y^{(2N,M)}(t)$ is defined as $Y^{(2N,M)}(t)/a_0=1+\sum_{i=1}^6 a'_i b_i$. The strongest bounds are given by small $N$ and $M$.}
\label{fig:NMeta=0}
\end{figure}

\begin{figure}[tb]
\begin{center}
\includegraphics[width=.5\linewidth]{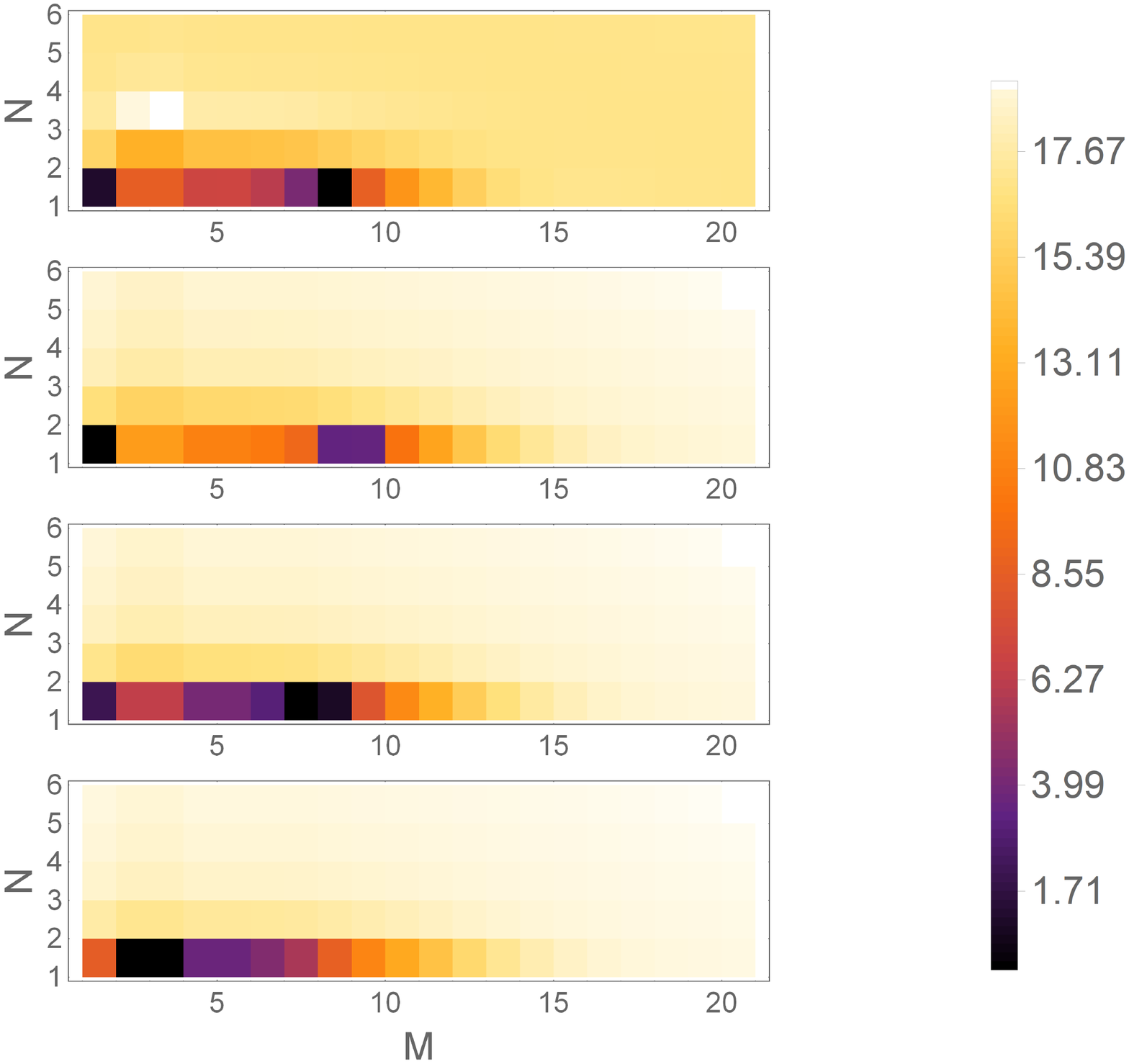}
\end{center}
\caption{Similar to Figure \ref{fig:NMeta=0} but for the case of $\eta=1$.}
\label{fig:NMeta=1}
\end{figure}

On the other hand, for fixed $t$ and $\eta$, the lowest few $N$ and $M$ give the strongest bounds. In Figures~\ref{fig:NMeta=0} and \ref{fig:NMeta=1}, we plot the strengths of the bounds for different $\{\eta,t,N,M\}$. The color value is the rescaled value of $Y^{(2N,M)}(t)$, that is, the value of $Y^{(2N,M)}(t)/a_0=1+\sum_{i=1}^6 a'_i b_i$ with $b_i$ substituted by the central values of Ref.~\cite{Colangelo:2001df}. So in the following, we will consider bounds with $\eta=0,1,N<6,M<11$ and with different values of $t$ sampling the allowed range.

\subsection{Bounds on $\bar l_1$ and $\bar l_2$}
\label{sec:compare}

The scale-independent LECs $\bar l_1$ and $\bar l_2$, which are related to $l^r_i$ by
\be
l_{1}^{r}=\frac{1}{96 \pi^{2}}\left(\bar{l}_{1}+\ln \frac{M_{\pi}^{2}}{\mu^{2}}\right) ,~~~l_{2}^{r}=\frac{1}{48 \pi^{2}}\left(\bar{l}_{2}+\ln \frac{M_{\pi}^{2}}{\mu^{2}}\right),
\ee  
have been constrained previously using field theoretical principles \cite{Pennington:1994kc,Ananthanarayan:1994hf,Dita:1998mh,Comellas:1995hq,Manohar:2008tc}. The strongest among them is given by Ref.~\cite{Manohar:2008tc}, which used essentially, in our notation, the $Y^{(2,0)}(t)$ positivity bound to constrain the two LECs. In this section, we will see how the bounds with $t$ or higher order $s$ derivatives can improve the constraints on $\bar l_1$ and $\bar l_2$.

To this end, we can truncate the amplitude to $\mc{O}(p^4)$. That means for $b_i$ we only keep the leading order LECs $16\pi^2 b_3 = \f13 \bar{l}_{1}+\f16\bar{l}_{2}-\frac{7}{12}$ and $16\pi^2 b_4=\f16\bar{l}_{2}-\frac{5}{36}$, while $b_1$ and $b_2$ do not enter the positivity bounds. Since the leading Weinberg tree amplitude has at most linear terms in Mandelstam variables, and thus do not contribute to the positivity bounds after two $s$ derivatives, the bounds obtained are independent of the pion mass and decay constant. Thus, the bounds on $\bar l_1$ and $\bar l_2$ at one loop are universal, not just for ChPT from QCD.

Up to $\mc{O}(p^4)$, the $\bar l_1$ and $\bar l_2$ coefficients only appear in the polynomial part of the amplitude, which is only up to quadratic order in Mandelstam's variables, so one may wonder whether the higher derivative $Y$ bounds can play any role here. As we see momentarily, they do provide further constraints. This is because the higher derivative bounds are built up on the lower derivative ones which contain the $\bar l_1$ and $\bar l_2$ coefficients, as one can see from \eref{PB}, and the loop logarithmic functions can contribute negatively to the positivity bounds. (By the same token, if we include the ${\cal O}(p^6)$ contributions we will not get new constraints on $\bar l_1$ and $\bar l_2$ alone --- the bounds will then contain LECs other than $\bar l_1$ and $\bar l_2$.)

Ref.~\cite{Manohar:2008tc} works with amplitudes for fixed total isospins. In this approach, the $s\leftrightarrow u$ crossing for a total isospin amplitude often generates terms with negative coefficients in front of the amplitudes, for which case one may not establish the positivity for the left hand cut in the fixed-$t$ dispersion relation. A linear combination of the total isospin amplitudes can overcome this problem, which allows Ref.~\cite{Manohar:2008tc} to apply the 2nd $s$ derivative bound  for the following processes: $\pi^{0} \pi^{0}\rightarrow \pi^{0} \pi^{0}$, $\pi^{+} \pi^{0}\rightarrow \pi^{+} \pi^{0}$, $\pi^{+} \pi^{+}\rightarrow \pi^{+} \pi^{+}$. Note that the  fields in the isospin basis are related to those in the Cartesian basis via $\pi^1 = (\pi^+ + \pi^-)/\sqrt{2}$, $\pi^2 = i(\pi^+ - \pi^-)/\sqrt{2}$ and $\pi^3=\pi^0$, while a conventional isospin basis can be chosen as $|1,+1\rangle=-\left|\pi^{+}\right\rangle$, $|1,-1\rangle=+\left|\pi^{-}\right\rangle$, $|1,0\rangle=\left|\pi^{0}\right\rangle$. The strongest bounds in that approach are given by
\begin{align}
\label{manohar cons}
\pi^{0} \pi^{0}\rightarrow \pi^{0} \pi^{0} :&~~\bar{l}_{1}+2 \bar{l}_{2} >  \frac{157}{40} = 3.9 \pm 0.4,  &\text{for}~\eta=1, t=4, s=0, N=1,M=0 ,
\\
\label{manohar cons2}
\pi^{+} \pi^{0}\rightarrow \pi^{+} \pi^{0}:&~~ \bar{l}_{2} > \frac{27}{20} = 1.4 \pm 0.4,  &\text{for}~\eta=0, t=4, s=0, N=1,M=0 ,
\\
\label{manohar cons3}
\pi^{+} \pi^{+}\rightarrow \pi^{+} \pi^{+}:& ~~ \bar{l}_{1}+3 \bar{l}_{2} > 5.6 \pm 0.4,  &\text{for}~\eta = \frac12, t=4, s=1.114, N=1,M=0.
\end{align}

Note that the third bound above corresponds to $\eta_1=0$ and $\eta_2=1$, \ie $\eta=1/2$ and $(\eta_1-\eta_2)/2=-1/2$, so this bound is not one of the $Y$ bounds. This marks a subtle difference between the Manohar-Mateu bounds and the $s$ derivative $Y$ bounds. Previously, below Eq.~(\ref{eta=01}), we have pointed out that only the parameter combination $\eta=(\eta_1+\eta_2)/2$ appears in the $Y$ bounds, and since $\eta$ appears linearly in the  $Y$ bounds, the strongest results can be obtained with either $\eta=0$ or $\eta=1$. This is valid because the $Y$ bounds by construction are evaluated at $v=s+\frac{t}2-2=0$, which is convenient for systematically obtaining all the higher order $t$ derivative bounds but does not include all possible valid values of $s$ and $t$ for the $s$ derivative bounds. Indeed, the third bound above is evaluated at $v\neq 0$ with $s=1.114$ and $ t=4$, and thus not covered by the $Y$ bounds. However, as we shall see shortly, the restriction of the $Y$ bounds being evaluated at $v=0$ is well compensated by the addition of the $t$ derivative bounds, and the $Y$ bounds ultimately provide stronger bounds. 

Following \cite{Manohar:2008tc}, we have also added the error estimates for the bounds Eqs.~(\ref{manohar cons}) to (\ref{manohar cons3}) from the $\mc{O}(p^6)$ contributions of the amplitude. There are three parts of the $\mc{O}(p^6)$ contributions: tree level contribution from the $\mc{O}(p^6)$ LECs, one loop contribution involving the $\mc{O}(p^4)$ LECs and two loop contribution from the leading chiral Lagrangian. Here we are bounding $\bar l_1$ and $\bar l_2$, while other ${\cal O}(p^4)$ LECs and the ${\cal O}(p^6)$ LECs are badly known. On the other hand, the two loop contribution only depends on $M_\pi$ and $F_\pi$, which we have better control over. Assuming naturalness, one may expect that the three contributions are around the same order, so a rough estimate of the errors is to multiply the two loop contribution by a factor of 3 and take the maximum of them as a common error estimate, which is the 0.4 quoted in Eqs.~(\ref{manohar cons}) to (\ref{manohar cons3}). 

\begin{figure}[tb]
\begin{center}
\includegraphics[width=.45\linewidth]{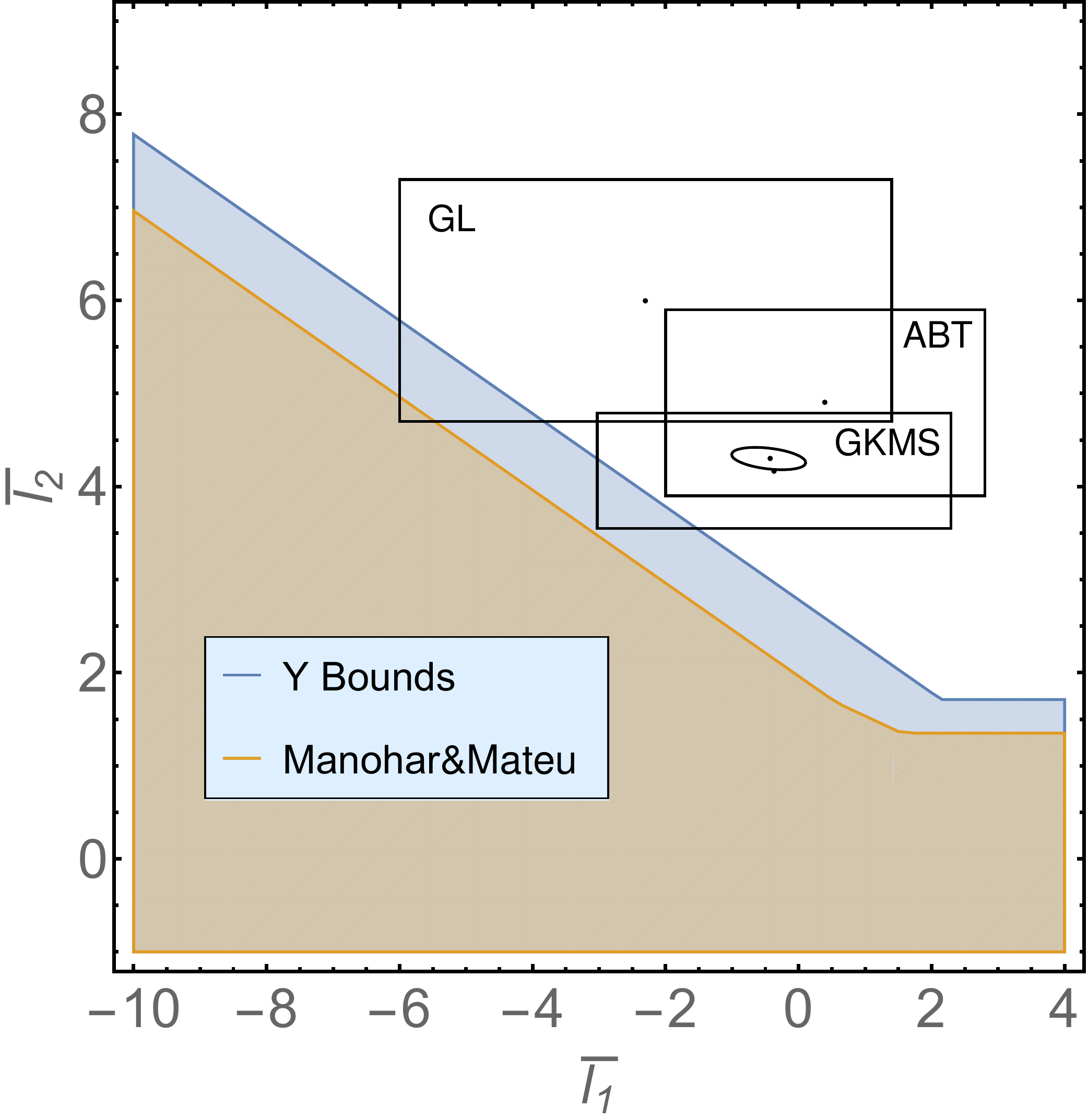}
\end{center}
\caption{Comparison of our positivity bounds on $\bar l_1$ and $\bar l_2$ with those of Manohar and Mateu~\cite{Manohar:2008tc} for the $\pi\pi$ scattering to one loop. See Eqs.~(\ref{manohar cons}) to (\ref{Yl2ndbound}) for the error estimates for these bounds. The rectangles GL, ABT, GKMS and the small ellipse inside it are the ranges of the  fitted values of $\bar l_1$ and $\bar l_2$ given in Refs.~\cite{Gasser:1983yg}, \cite{Girlanda:1997ed}, \cite{Amoros:2000mc} and \cite{Colangelo:2001df} respectively. }
\label{fig: 1-loop}
\end{figure}

In our approach, we consider a general elastic scattering $\pi^\ai\pi^\bi\to \pi^\ai\pi^\bi$: ${T}_{\alpha \beta \rightarrow\alpha \beta}(s,t,u)=\ei {A}(s, t, u)+{A}(t, s, u) +\ei {A}(u, t, s)$, with all possible $\ei$ ranging from 0 to 1, and also make use of bounds $Y^{(2N,M)}(t)>0$ with up to $N$-th $t$ derivatives and $2N$-th $s$ derivatives. As discussed in the previous Section \ref{sec:structure}, we only need to consider $\eta=0,1$. For fixed $\eta$, we find that all the $Y^{(2N,M)}(t)>0$ bounds with $N>1$ give rise to trivial results, as the coefficients of  $\bar{l}_{1}$ and $\bar{l}_{2}$ in the function $\tilde{A}(s,t,u)$ are polynomials of $v$ and $t$ with degrees less than $4$. On the other hand, all the $Y^{(2N,M)}(t)>0$ with $N=1$ but different $M$ can be cast as
\be
\ei  \bar{l}_{1}+(1+\ei) \bar{l}_{2} > \ei  f_{M}(t)+g_{M}(t),
\ee
where $f_{M}(t),g_{M}(t)$ are all monotonic increasing functions of $t$ within $0\leq t < 4$.  Thus, all the $N=1$ bounds are all parallel to each other and become the strongest at $t=4$. Numerically computing the different $N$ and $M$ bounds with these choices, we find that the strongest bounds are given by 
\bal
\pi^{0} \pi^{0}\rightarrow \pi^{0} \pi^{0} :&~~~~\bar{l}_1  +2  \bar{l}_2 > \frac{1559}{280} = 5.6 \pm 0.8,  &\text{for}~\eta=1, t=4, s=0, N=1,M=2 ,
\\
\label{Yl2ndbound}
\pi^{+} \pi^{0}\rightarrow \pi^{+} \pi^{0}:&~~~~ \bar{l}_2 > \frac{719}{420} = 1.7 \pm 0.8,  &\text{for}~\eta=0 ,t=4, s=0,N=1,M=2 .
\eal
These bounds are stronger than the bounds obtained by Manohar and Mateu~\cite{Manohar:2008tc} and others \cite{Pennington:1994kc,Ananthanarayan:1994hf,Distler:2006if}. Again, we have provided error estimates for these bounds following the method used in Ref.~\cite{Manohar:2008tc}, that is, taking the larger error of the two bounds from the two loop contributions and multiplying it by a factor of 3. The error estimates in these bounds are slightly greater than those of Eqs.~(\ref{manohar cons}) to (\ref{manohar cons3}), purely due to the technical steps of taking $t$ derivatives in the $Y$ bounds. Nevertheless, we would like to emphasize that those error estimates are quite rough for the contributions from the $\mc{O}(p^6)$ and $\mc{O}(p^4)$ LECs. In Figure \ref{fig: 1-loop}, we plot the improvement of our bounds against those of Manohar and Mateu~\cite{Manohar:2008tc} in Eq.~\eqref{manohar cons}, and also compare them with the fitted experimental values. We see that while the bounds by Manohar and Mateu~\cite{Manohar:2008tc} barely touch the one sigma regions of the empirical values, our bounds already eliminate some of those one sigma regions. 

\subsection{Bounds on the $b_{i}$ constants}
\label{sec:bi}

\begin{figure}[tbp]
    \centering 
    \begin{subfigure}{}
      \includegraphics[width=0.30\textwidth]{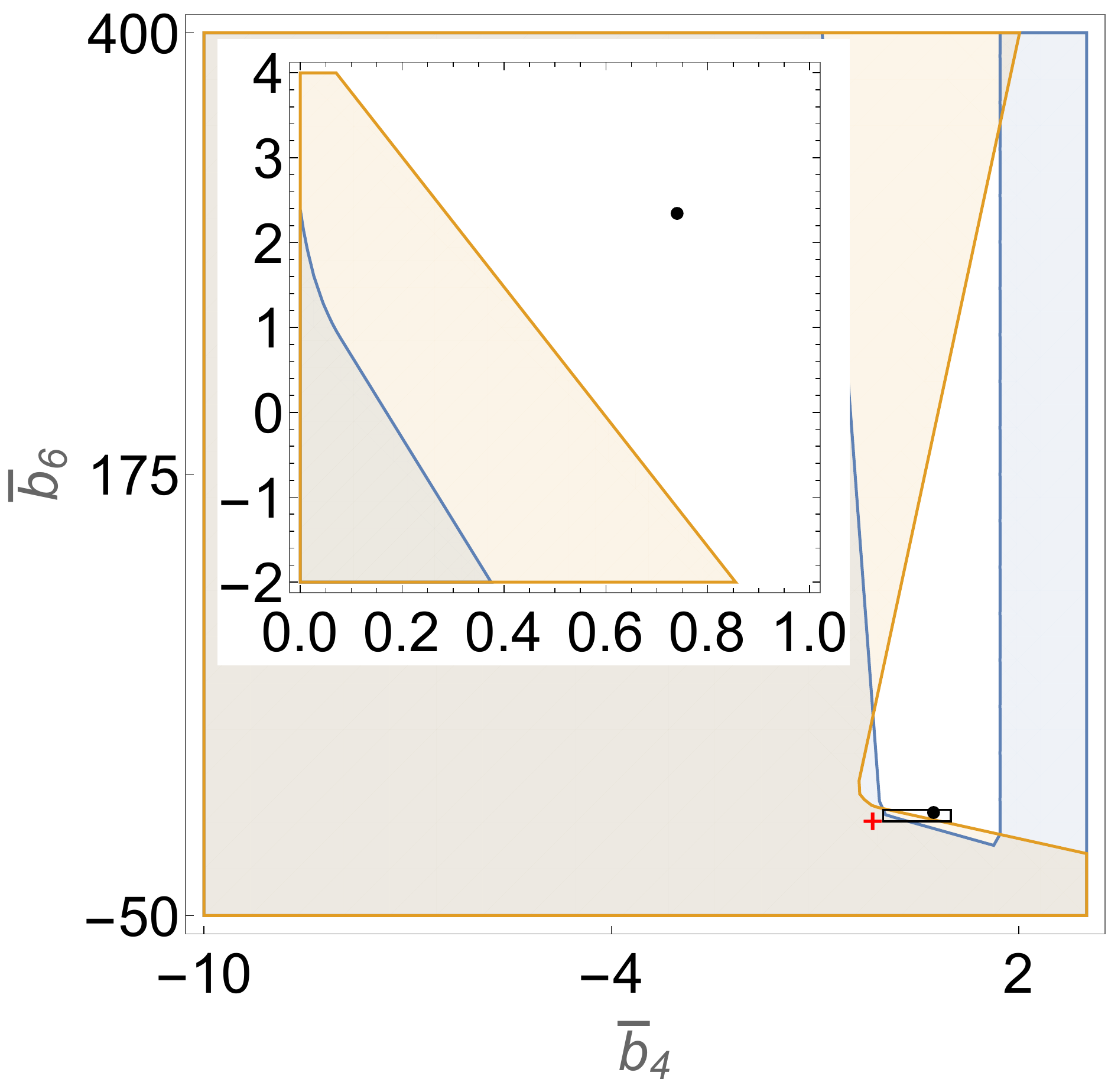}
    \end{subfigure}
     \begin{subfigure}{}
      \includegraphics[width=0.31\textwidth]{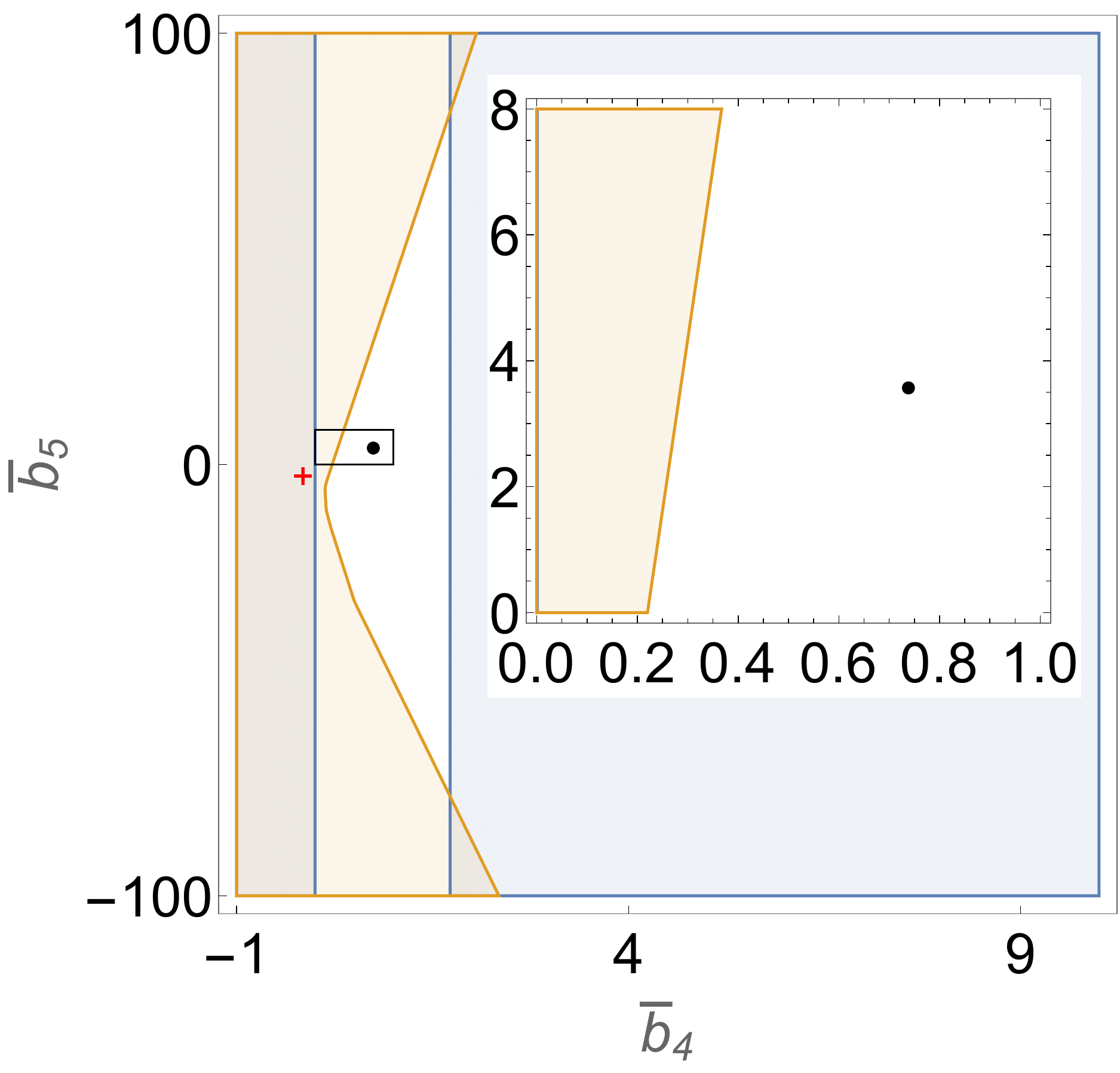}
    \end{subfigure}
    \begin{subfigure}{}
      \includegraphics[width=0.31\textwidth]{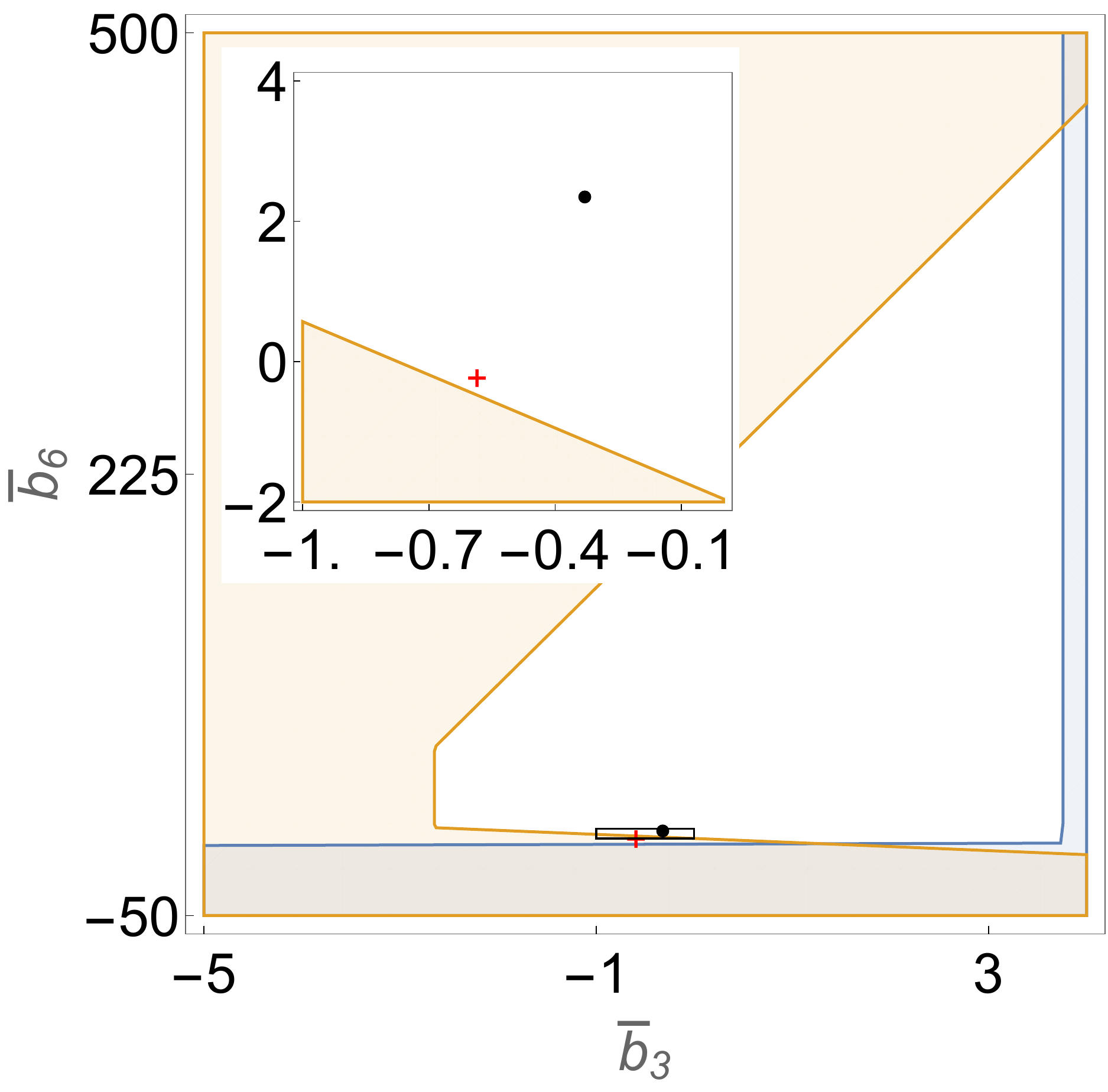}
    \end{subfigure}
    \\
    \centering 
    \begin{subfigure}{}
      \includegraphics[width=0.31\textwidth]{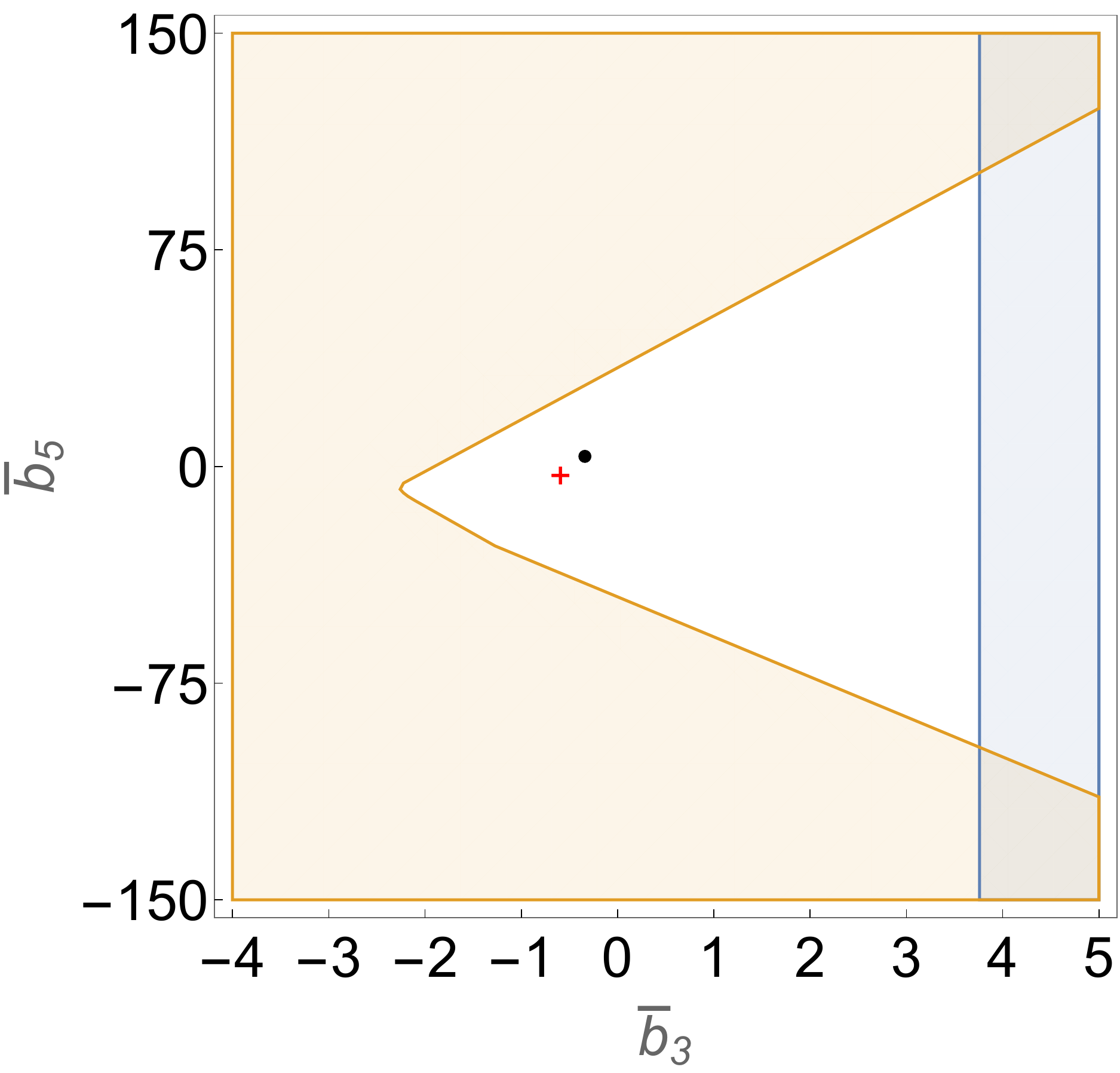}
    \end{subfigure}
     \begin{subfigure}{}
      \includegraphics[width=0.30\textwidth]{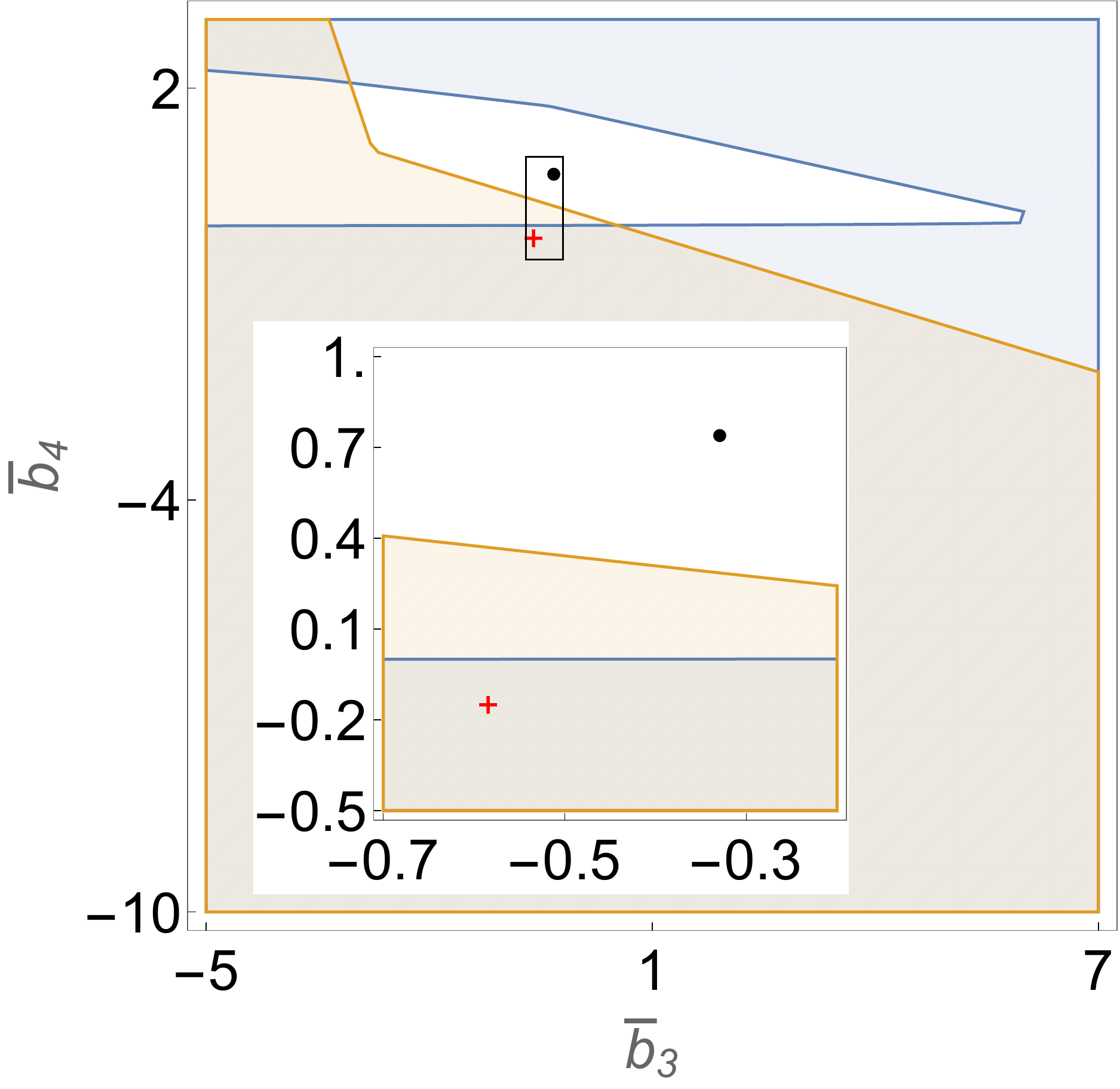}
    \end{subfigure}
    \begin{subfigure}{}
      \includegraphics[width=0.315\textwidth]{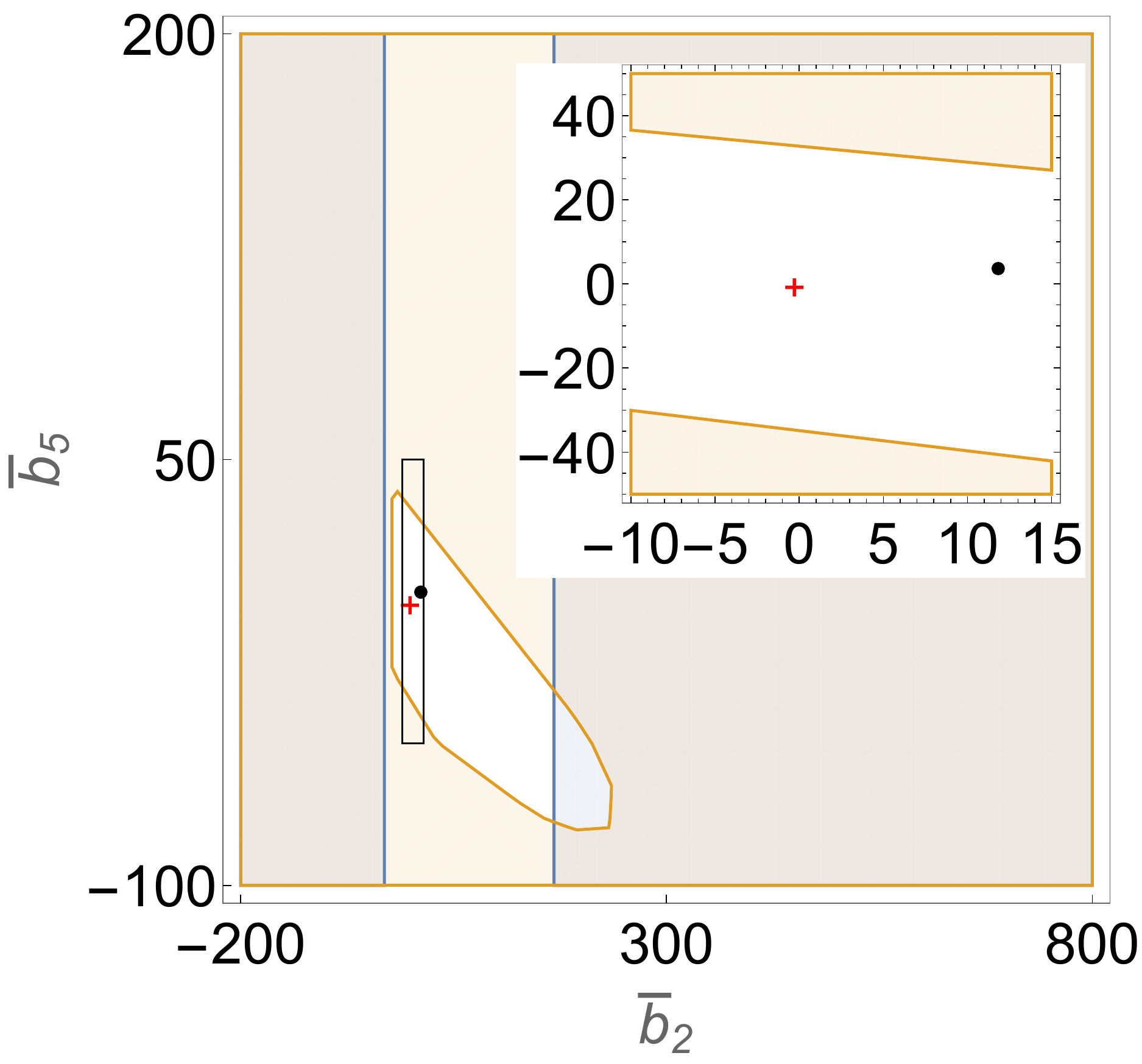}
    \end{subfigure}
    \\
    \centering 
    \begin{subfigure}{}
      \includegraphics[width=0.313\textwidth]{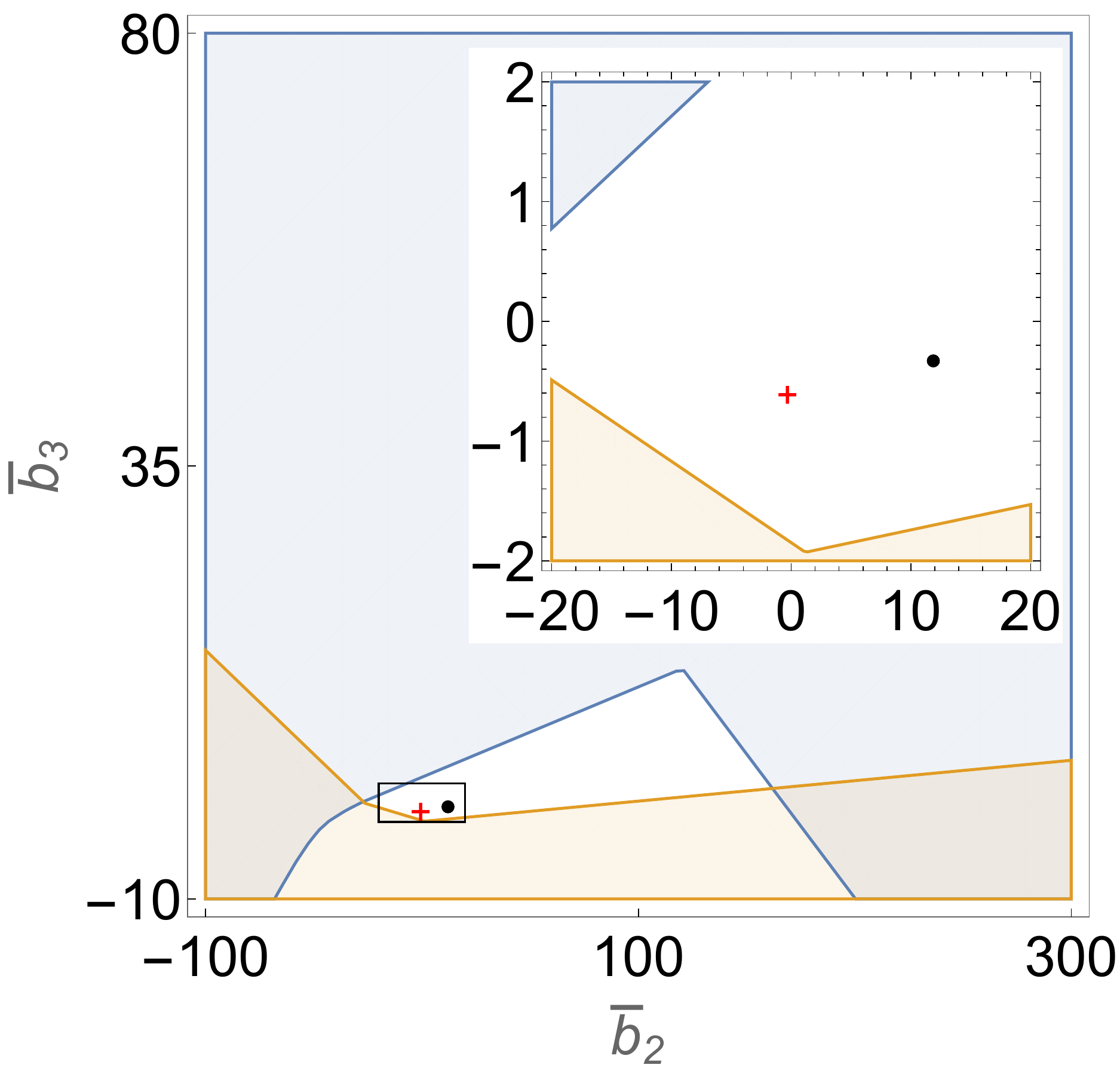}
    \end{subfigure}
     \begin{subfigure}{}
      \includegraphics[width=0.316\textwidth]{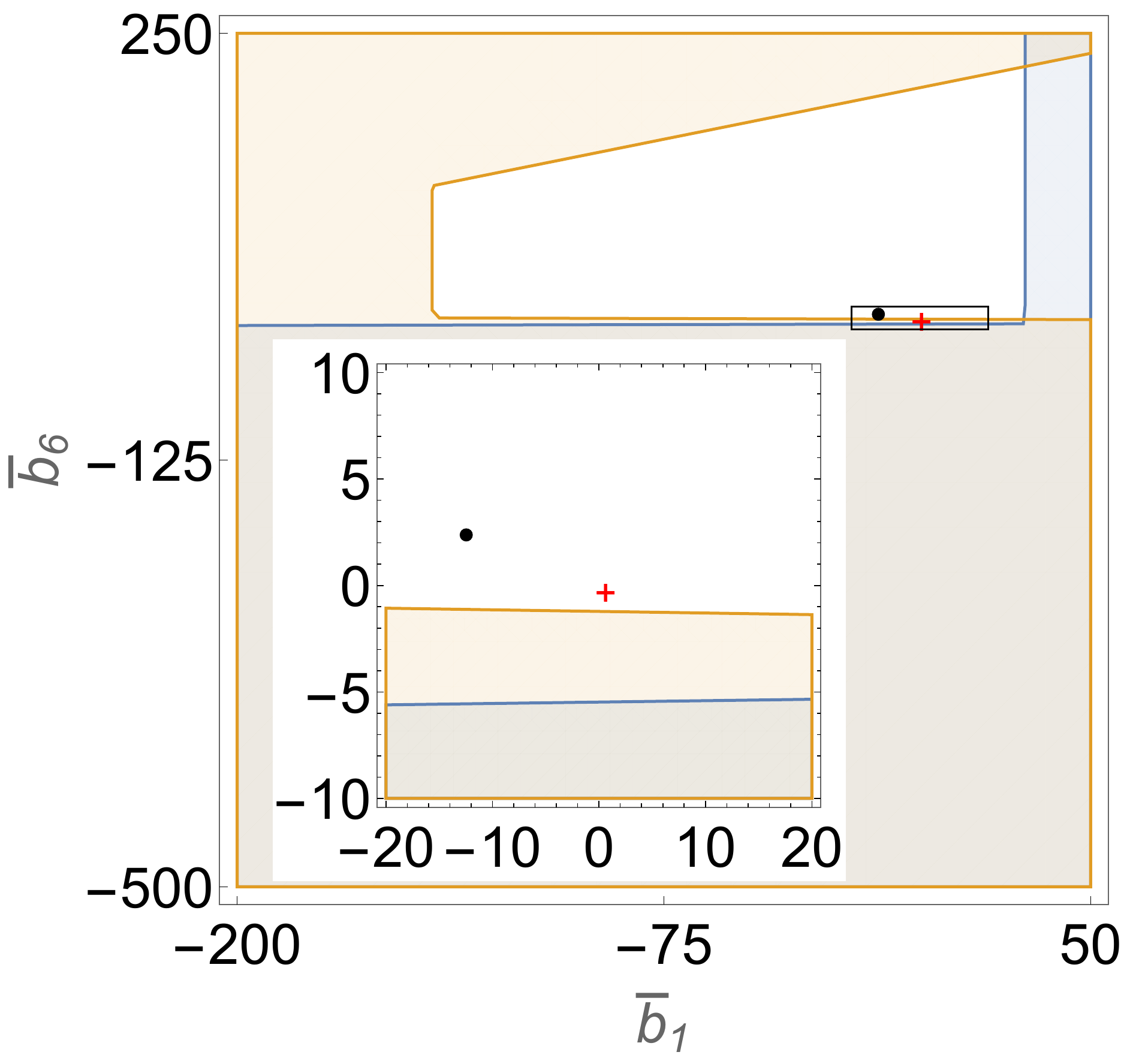}
    \end{subfigure}
    \begin{subfigure}{}
      \includegraphics[width=0.315\textwidth]{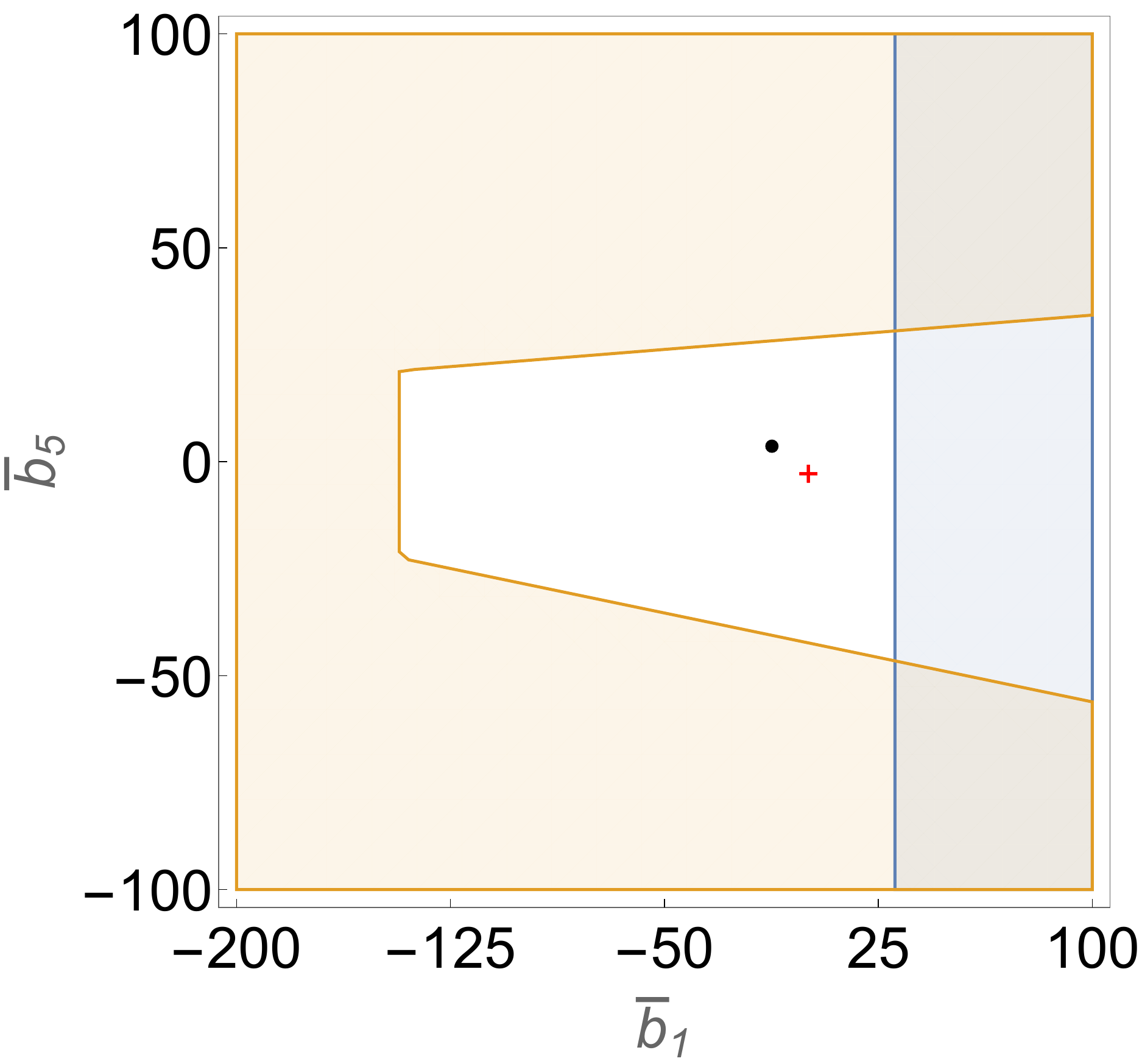}
    \end{subfigure}
    \caption{2D sections of the constrained $b_i$ space. The 2D sections are obtained by setting the other $4$ parameters to the central values of the fit \eqref{expri data1}. The yellow (blue) region is the region ruled out by the bounds with $\eta=1$ ($\eta=0$). The black point represents the central values of the fit \eqref{expri data1} with inputs from the experimental data and theoretical estimates, and the red cross represents the theoretical point computed from the Weinberg Lagrangian. To be continued in Fig~\ref{fig:2D2}.} 
    \label{fig:2D1}
\end{figure}

\begin{figure}[tbp]
    \centering 
    \begin{subfigure}{}
      \includegraphics[width=0.302\textwidth]{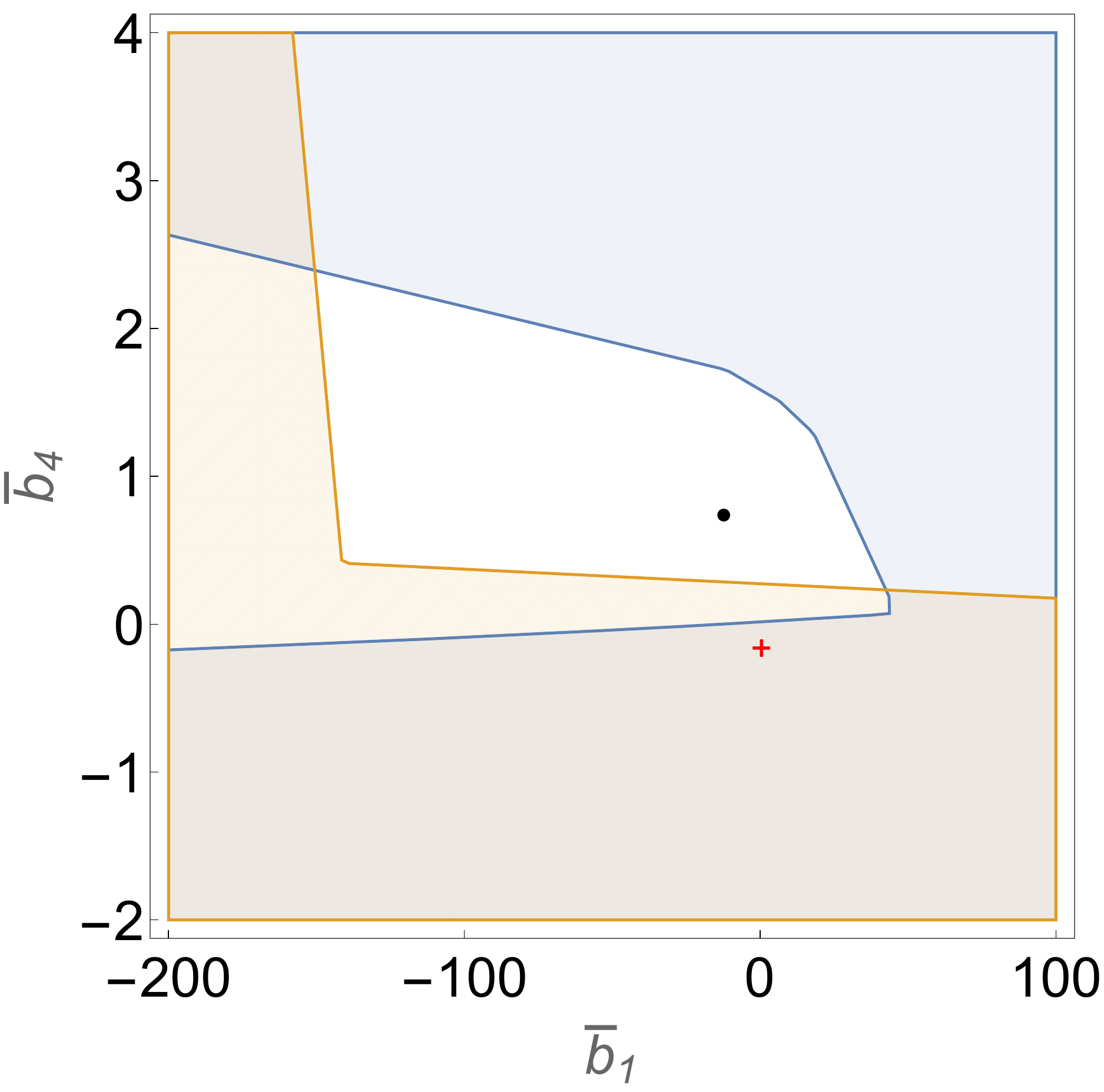}
    \end{subfigure}
     \begin{subfigure}{}
      \includegraphics[width=0.305\textwidth]{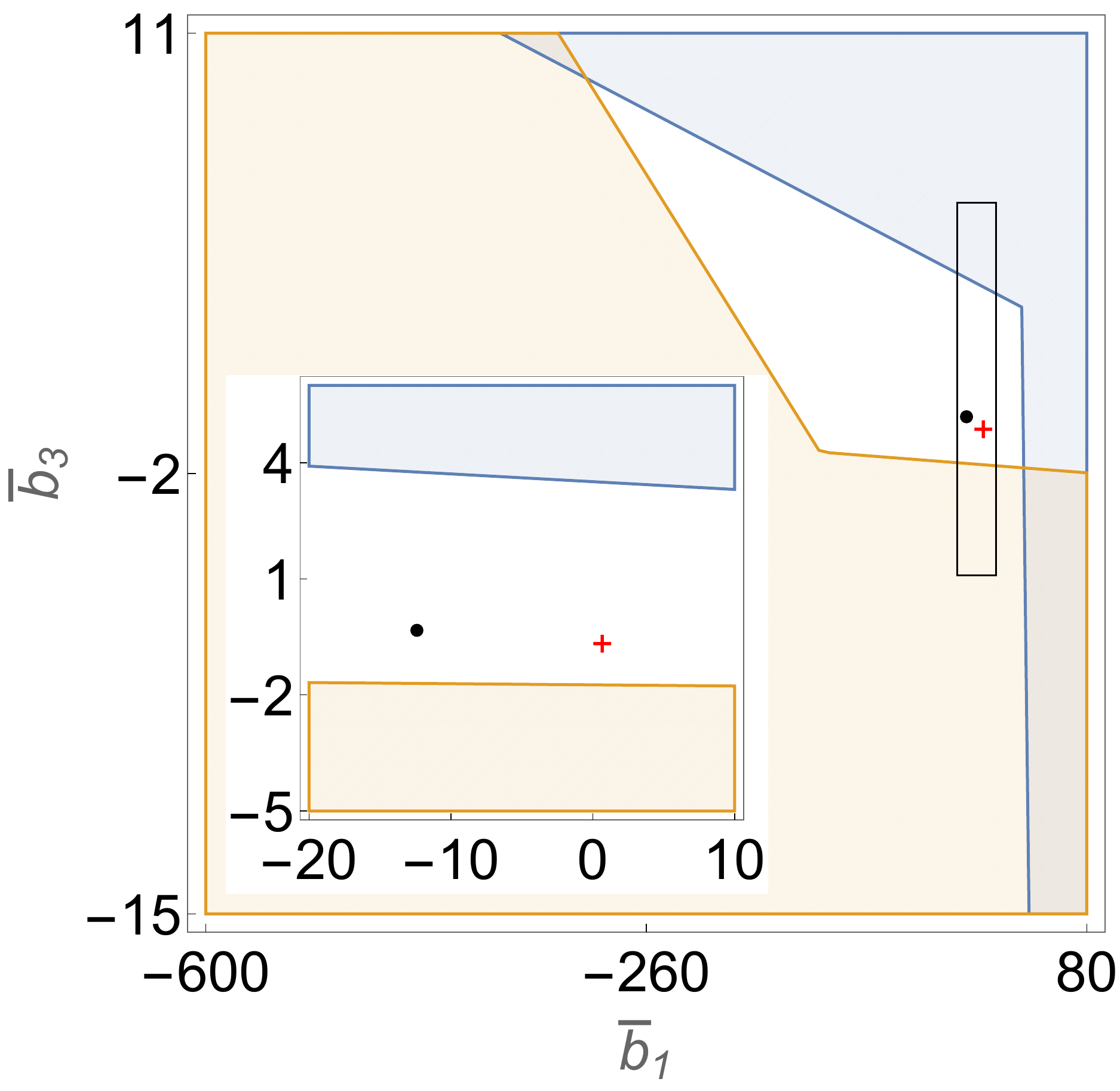}
    \end{subfigure}
    \begin{subfigure}{}
      \includegraphics[width=0.312\textwidth]{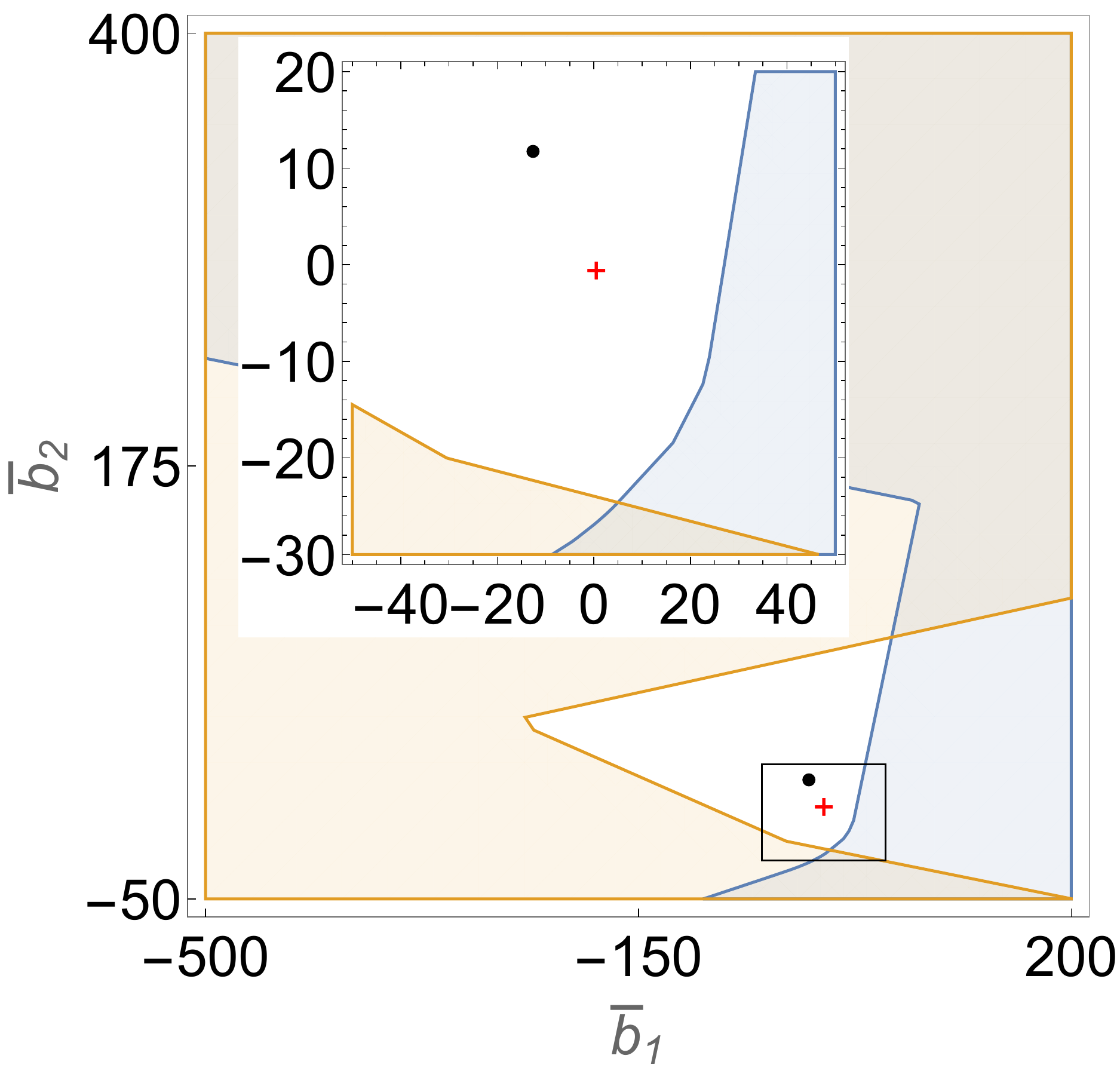}
    \end{subfigure}
    \\
      \centering 
    \begin{subfigure}{}
      \includegraphics[width=0.316\textwidth]{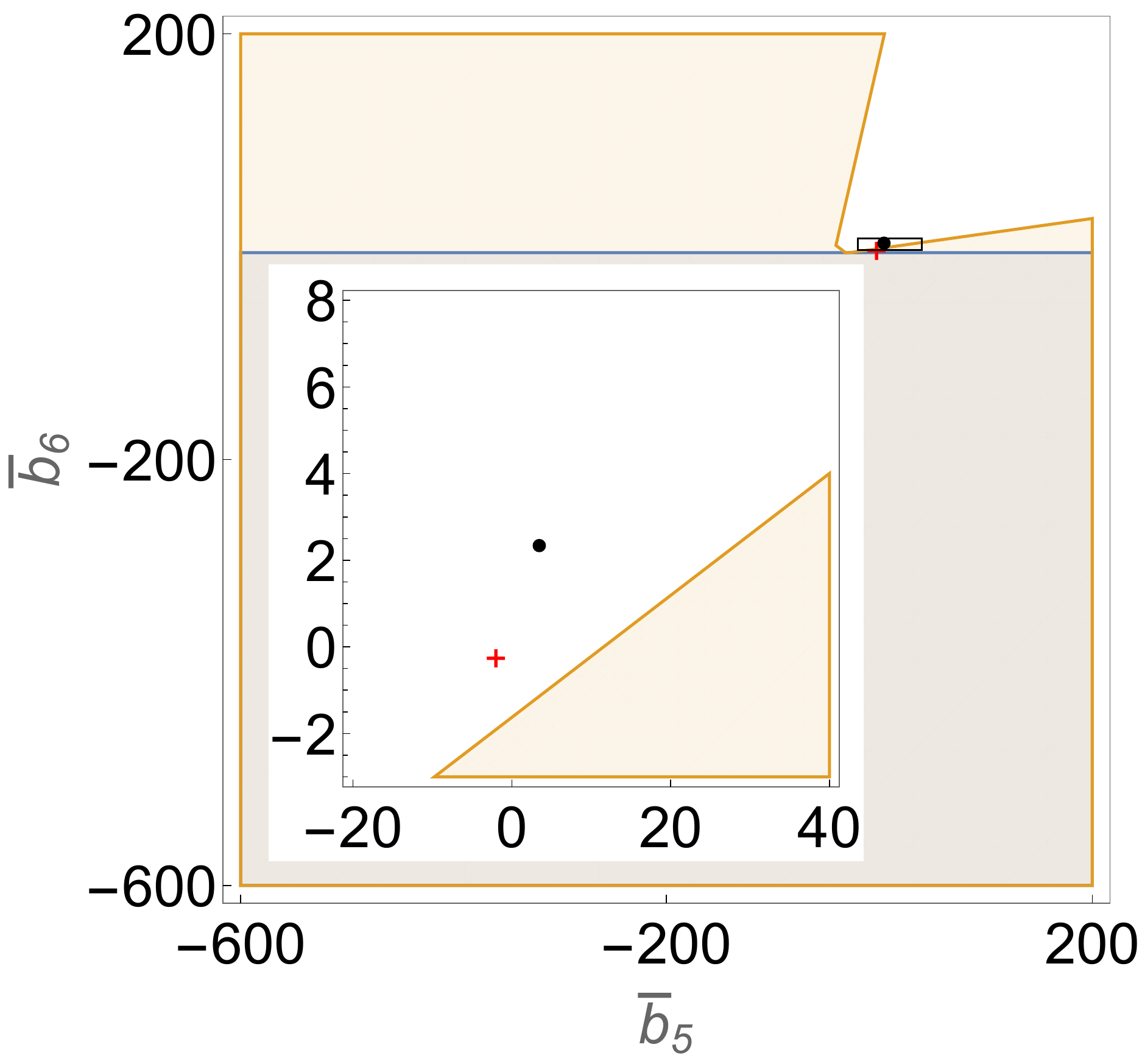}
    \end{subfigure}
     \begin{subfigure}{}
      \includegraphics[width=0.316\textwidth]{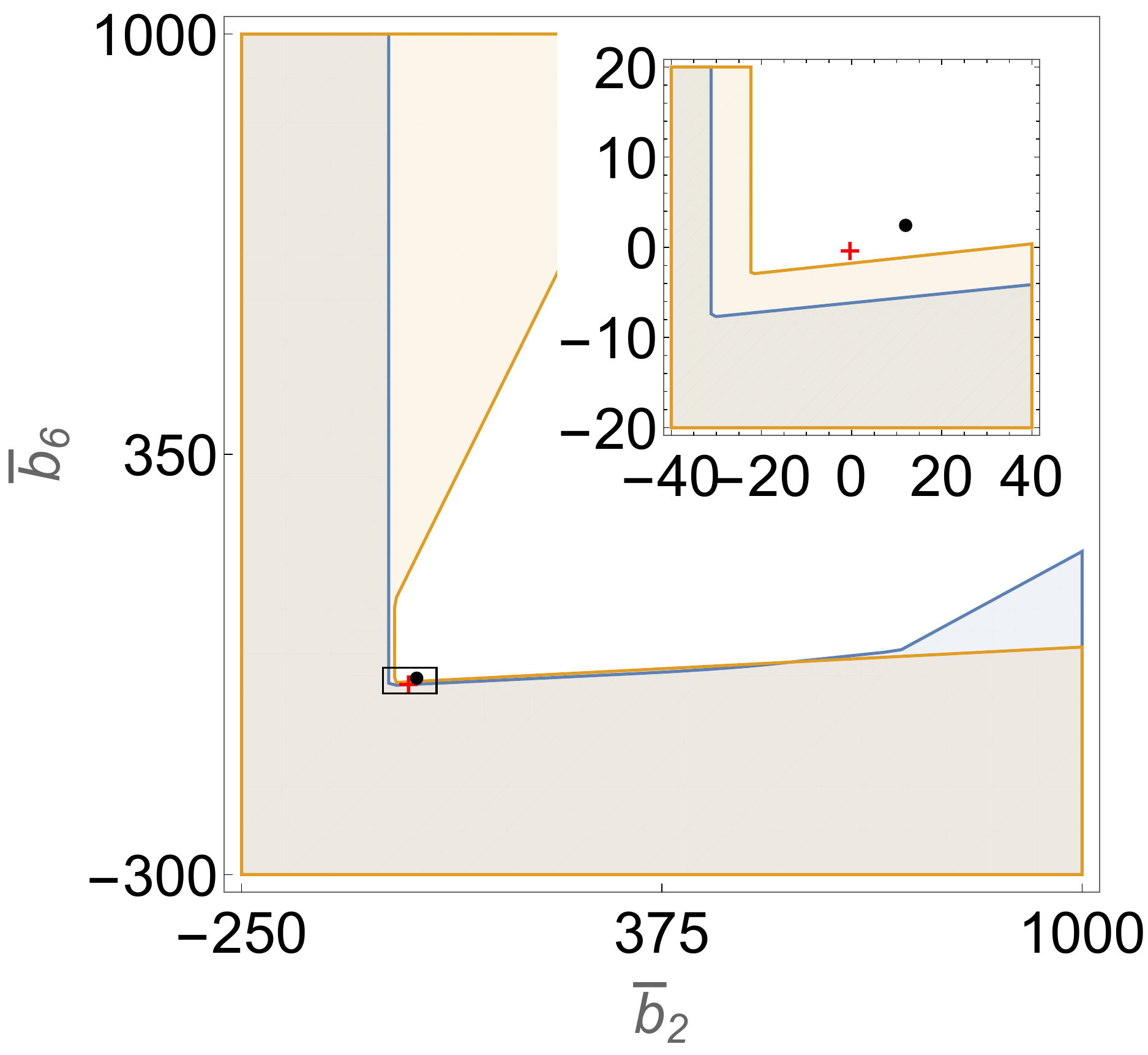}
    \end{subfigure}
    \begin{subfigure}{}
      \includegraphics[width=0.313\textwidth]{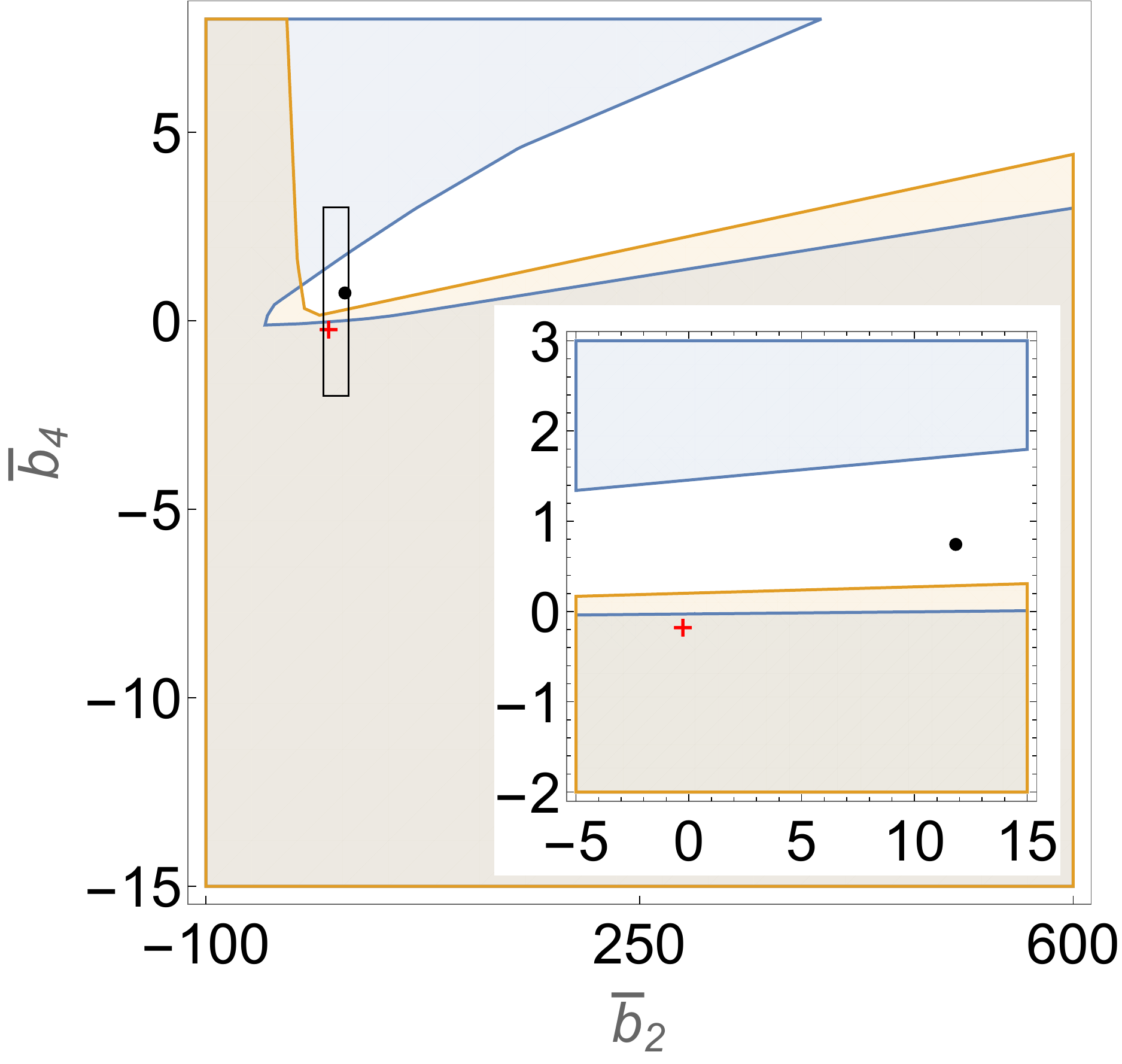}
    \end{subfigure}
    \caption{Continuation of Fig~\ref{fig:2D1}.} 
    \label{fig:2D2}
\end{figure}

There are six $b_i$ constants ($i=1,2,...,6$) which appear linearly in the amplitude (\ref{amplitude}) and are functions of the $\mc{O}(p^4)$ and $\mc{O}(p^6)$ LECs. In this section, we shall apply the $Y$ positivity bounds on the two-loop ChPT amplitude to get the strongest bounds on the $b_i$ constants for different choices of $\{\eta, t, N,M\}$. Specifically, we will apply $1430$ $Y$ bounds with $\eta=0,1$, $N<6$, $M<11$ and $13$ values of $t$.

The $b_i$ constants contain powers of the $4\pi$ factor and are not naturally order one, so instead we will present results in terms of 
\be
\bar{b}_i \equiv (16 \pi^{2})^{\zeta_i}  b_i, ~~~~ \zeta_i=1~{\rm for}~i=1,2,3,4, ~~ \zeta_i=2 ~ {\rm for}~i=5,6 .
\ee
The values of $\bar{b}_i$ from the $\mc{L}_0$ Weinberg Lagrangian with all the higher order LECs setting to zero up to two loops are given by
\begin{equation}
{\bar{b}^{0}_{1}=\frac{13}{18},} ~~ {\bar{b}^{0}_{2}=-\frac{2}{9},} ~~ {\bar{b}^{0}_{3}=-\frac{7}{12},} ~~ {\bar{b}^{0}_{4}=-\frac{5}{36},} ~~ {\bar{b}^{0}_{5}=\frac{-66029+2688\pi^2}{20736},} ~~ {\bar{b}^{0}_{6}=\frac{-11375+768\pi^2}{20736}}  .
\label{theo data}
\end{equation}
This is, however, not a good approximation of the amplitude to that order, even not considering the fact that the LECs in the higher order Lagrangian are needed to absorb the UV divergence from the loop integrals. A good fit of these constants is provided by Colangelo {\it et al.} \cite{Colangelo:2001df}
\begin{equation}
\begin{array}{lll}
{\bar{b}_{1}=-12.4 \pm 1.6,} & {\bar{b}_{2}=11.8 \pm 0.6,} & {\bar{b}_{3}=-0.33 \pm 0.07}, \\ {\bar{b}_{4}=0.74 \pm 0.01,} & {\bar{b}_{5}=3.58 \pm 0.37,} & {\bar{b}_{6}=2.35 \pm 0.02} .
\end{array}
\label{expri data1}
\end{equation}
where the uncertainties come from higher order corrections in the EFT and from the experimental data input when solving the Roy equations.

The positivity bounds on ChPT carve out a geometric shape in 6D space $(b_1,b_2,b_3,b_4,b_5,b_6)$. It is clear that the constrained $b_i$ space has to be convex. This is simply because if two points $b_i$ and $b'_i$ satisfy the positivity bounds $\sum_i a_i b_i>a_0$ and $\sum_i a_i b'_i>a_0$, then any point in between the two points $b''_i=\li b_i + (1-\li) b'_i$ also satisfies the positivity bounds. We cannot visualize a 6D constrained $b_i$ space, so we will look at the lower dimensional sections of the space with extra dimensions projected to the central values of the fit (\ref{expri data1}).

Let us look at the 2D projections of the constrained $b_i$ space (the 3D sections can be found in Appendix~\ref{sec:3D}). Setting the other $4$ parameters to the central values of the fit in Eq.~\eqref{expri data1}, there are $15$ pairs of $\{\bar{b}_{i}, \bar{b}_{j}\}$. We see from Figures~\ref{fig:2D1} and \ref{fig:2D2} that for most of these sections of the constrained $b_i$ space (except for $\{\bar{b}_{5}, \bar{b}_{6}\}$, $\{\bar{b}_{2}, \bar{b}_{6}\}$, $\{\bar{b}_{2}, \bar{b}_{4}\}$), the parameter space allowed by our positivity bounds are enclosed/compact regions. The boundary of the constrained $b_i$ space can be either straight lines or curly lines, the latter corresponding to choosing continuous values of $t$ in the positivity bounds. The black point represents the central value point of Eq.~\eqref{expri data1} and the red cross represents the parameters computed from the $\mc{L}_0$ Lagrangian with necessary counter terms. The $\mc{L}_0$ value is, not surprisingly, ruled out by our bounds in some sections, while the fit in Eq.~\eqref{expri data1} with its error bars are consistent with our positivity bounds. The constraints on the scale-independent coefficients $\bar{r}_i$ can be easily deduced from those on $b_i$ since they are linearly related to each other, see Appendix~\ref{appendix: loop fun}.

\section{Improved $Y$ bounds on ChPT}
\label{sec:impYb}

As discussed in Section \ref{sec:improbounds}, if the imaginary part of the amplitude can be accurately determined, one may subtract out the low energy contribution of the dispersion integral, and this will improve the positivity bounds.

\subsection{Structure of the bounds}

\begin{figure}[tb]
\begin{center}
\includegraphics[width=.45\linewidth]{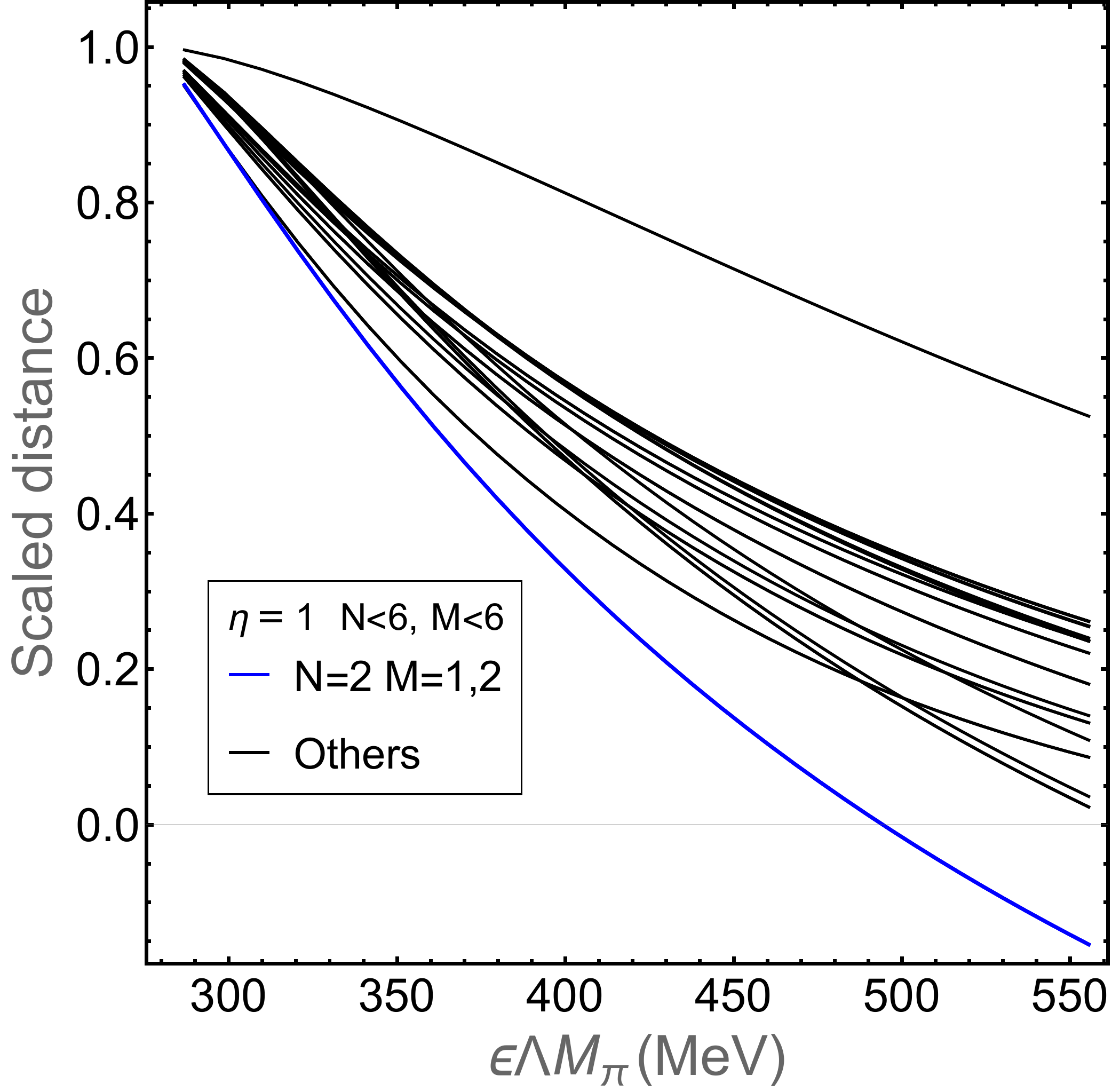}
\end{center}
\caption{Distances between the improved bound planes and a fiducial point of $b_{i}$ (the central value of the empirically fitted values in Eq.~\eqref{expri data1}) in the $(b_1,b_2,b_3,b_4,b_5,b_6)$ space for $\eta= 1$, $t=4$ at different energy scales. The distances are normalized to 1 at $\epi\Lambda=2$ ({\it i.e.}, the original $Y$ bounds without improved subtractions) to facilitate visualization in the plot, and a negative distance indicates that the positivity plane has excluded the fiducial point. The $\{N=2,M=1,2\}$ bounds are the first bounds to become negative at $\epi\Lambda M_\pi\simeq 490{\rm MeV}$, beyond which the improved bounds become invalid. }
\label{impro-struc}
\end{figure}

The dependence of the improved bounds on the parameters $\{\eta_1,\eta_2, t, N,M\}$ is very similar to that of the original $Y$ bounds. In particular, the improved dispersion relation is still $s\leftrightarrow u$ symmetric, so only $\eta=(\eta_1+\eta_2)/2$ appears in the bounds linearly, and we only need to consider the bounds with $\eta=0$ and $1$. We need to consider different $t$ and only need to consider the low orders of $N$ and $M$. However, for improved bounds, we also have the parameter $\epi\Lambda$ to choose. A small $\epi\Lambda$ does not improve the bounds very much, while, to achieve a sufficient accuracy, $\epi\Lambda$ cannot be too close to $\Lambda$ ($\epi \Lambda=4$ corresponding to the original positivity bounds for which there is no subtraction of the imaginary part of the amplitude). The possible choice of $\epi \Lambda$ is clearly limited by how well the EFT at a given order can approximate the imaginary part of the full amplitude. 
Indeed, we find that the improved $Y$ bounds will break down at energies far below $\Lambda$ in ChPT. Assuming the current experimental determination of the $b_i$ constants are more or less accurate, this can be used to set a rough scale when the EFT at a given order stops being an effective description of the underlying physics. In Figure \ref{impro-struc}, we plot the distance in the $b_i$ space between the positivity plane and the fiducial point of $b_i$ given by the experimentally fitted values in \eqref{expri data1}. A negative distance in the plot indicates that the positivity plane has excluded the fiducial point, which implies that the improved positivity bound breaks down around that scale, as a valid positivity bound should not exclude the relatively good experimental values. We see that the first bound to become invalid is that of $N=2,M=1,2$ when $\epi\Lambda M_\pi\simeq 490{\rm MeV}$, with the other bounds also becoming negative at around $600{\rm MeV}$. Thus, we should not use the improved positivity bounds beyond $\epi\Lambda M_\pi\simeq 490{\rm MeV}$ and preferably somewhat below that scale. Nevertheless, {\it a priori} the exact scale $\epi \Lambda$ at which the improved positivity bounds lose their accuracy is difficult to pin down, so we shall present the results for different $\epi \Lambda M_\pi$ below $490{\rm MeV}$.

Physically, the limit of the choices of $\epi\Lambda$  is due to the onset of the scalar isoscalar resonance $f_0(500)$ (also known as the $\sigma$ meson), which couples to the $\pi\pi$ $S$ wave with isospin 0 and has a pole at $449^{+22}_{-16} - i(275\pm12)$~MeV in the second Riemann sheet of the complex energy plane as determined from dispersive analyses~\cite{Pelaez:2015qba}.  It is not included explicitly in ChPT, and a pole cannot be obtained with a perturbative momentum expansion to any finite order.  Thus, perturbativity will break down at a scale where the $f_0(500)$ becomes important.

Another thing one needs to consider in applying the improved bounds, actually somewhat related to what was discussed above, is that we need to check whether the perturbative expansion of the bounds themselves is respected. Let us see how this is supposed to work. At low energies, the usual EFT power counting suggests that $\tilde B_{\epi\Lambda}(v,t)$ be expanded as
\be
B_{\epi\Lambda}(s,t)=\tilde B_{\epi\Lambda}(v,t) = B_0 \sum_{i=0}^\infty  \(\f{M_\pi^2}{\Lambda^2}\)^{i} f^{\epi\Lambda}_i(v,t) ,
\ee
where $B_0$ is a dimensionless constant and $f^{\epi\Lambda}_i(v,t)$ is a dimensionless function of dimensionless variables $v$ and $t$. This expansion is valid when $ f^{\epi\Lambda}_i(v,t)\sim v\sim t\sim 1$, as usually assumed. Plugging this into the improved $Y$ bounds, we get
\be
\sum_{i=0}^\infty   \(\f{M_\pi^2}{\Lambda^2}\)^{i} Y_{\epi\Lambda, i}^{(2 N, M)}(t) >0   ,
\ee
where $Y_{\epi\Lambda, i}^{(2 N, M)}(t)$ is similar to $Y_{\epi\Lambda}^{(2 N, M)}(t)$ with the replacement of $\tilde B_{\epi\Lambda}(v,t)$ with $f^{\epi\Lambda}_i(v,t)$. Assuming that higher order terms are smaller and truncating the expansion to a finite order, we get the positivity bounds for the EFT, and the truncation error may be estimated by the term after the truncation. For the original $Y$ bounds in ChPT, this perturbative structure is respected for a reasonably small ${M_\pi^2}/{\Lambda^2}$, where the two-loop contribution is smaller than the one-loop contribution. This, however, may not be so for the improved bounds with a large $\epi\Lambda$ subtraction. For an amplitude up to two loops, this can be verified, and we shall discard the bounds for which perturbativity is violated. Furthermore, even if perturbativity is respected for the expansion, at a practical level, it is also desirable that the improvements on the bounds gained by the $\epi\Lambda$ subtractions can outrun the extra uncertainties introduced by the subtractions {\it per se}.

\subsection{Bounds on $\bar l_1$ and $\bar l_2$}

\begin{figure}[tb]
\begin{center}
\includegraphics[width=.43\linewidth]{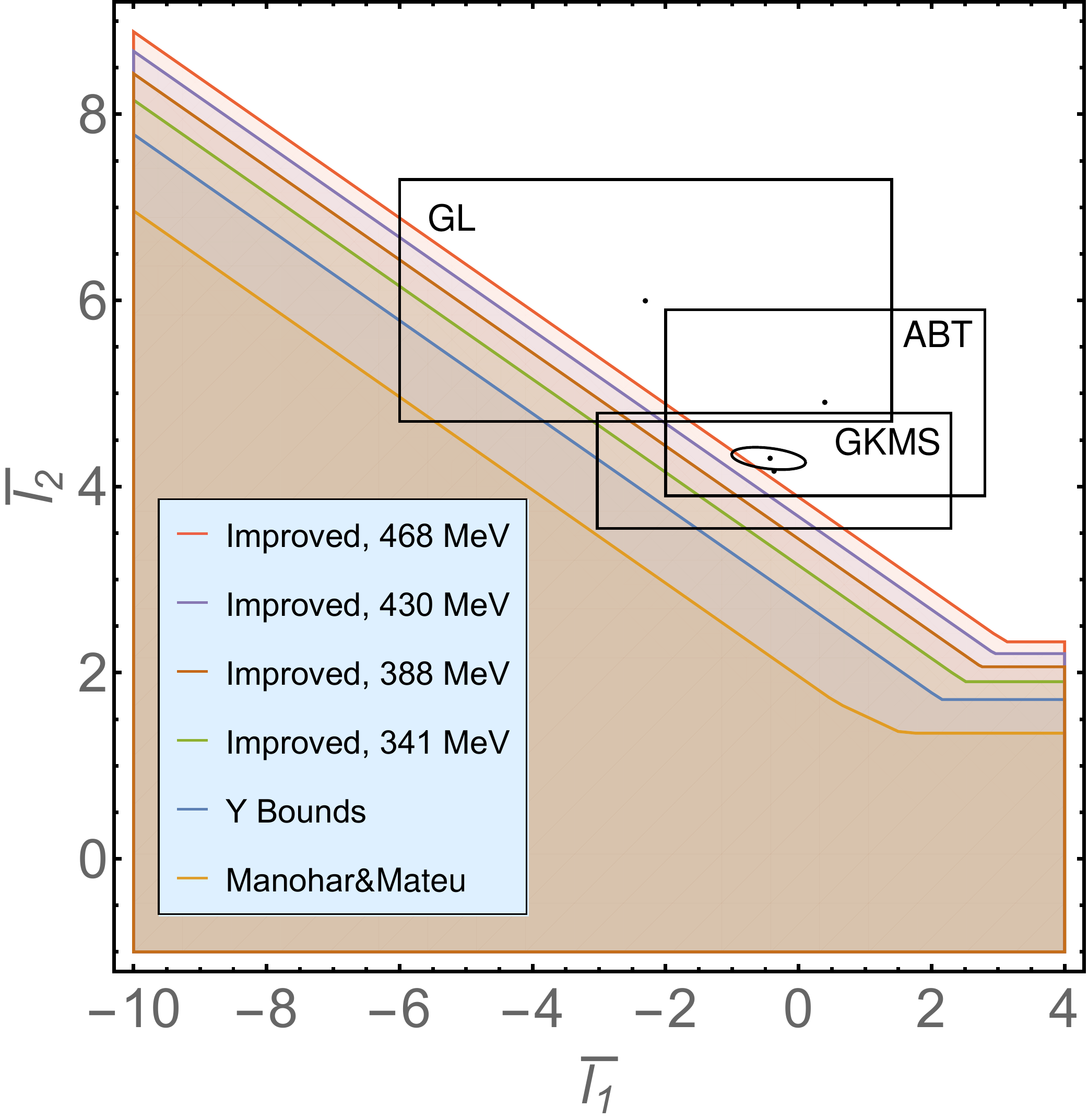}
\end{center}
\caption{Improved $Y$ bounds on $\bar l_1$ and $\bar l_2$ for the $\pi\pi$ scattering to one loop for different $\epi\Lambda$ subtractions. For example, the red line is for $\epi\Lambda M_\pi=468~{\rm MeV}$. ``Y bounds'' indicates no $\epi\Lambda$ subtraction and ``Manohar\&Mateu'' is the bounds from \cite{Manohar:2008tc}. The rectangles GL, ABT, GKMS and the small ellipse inside it are the ranges of the fitted values of $\bar l_1$ and $\bar l_2$ given in \cite{Gasser:1983yg}, \cite{Girlanda:1997ed}, \cite{Amoros:2000mc} and \cite{Colangelo:2001df} respectively. }
\label{fig: 1-loop-Imp}
\end{figure}

\begin{figure}[tb]
\begin{center}
\includegraphics[width=.8\linewidth]{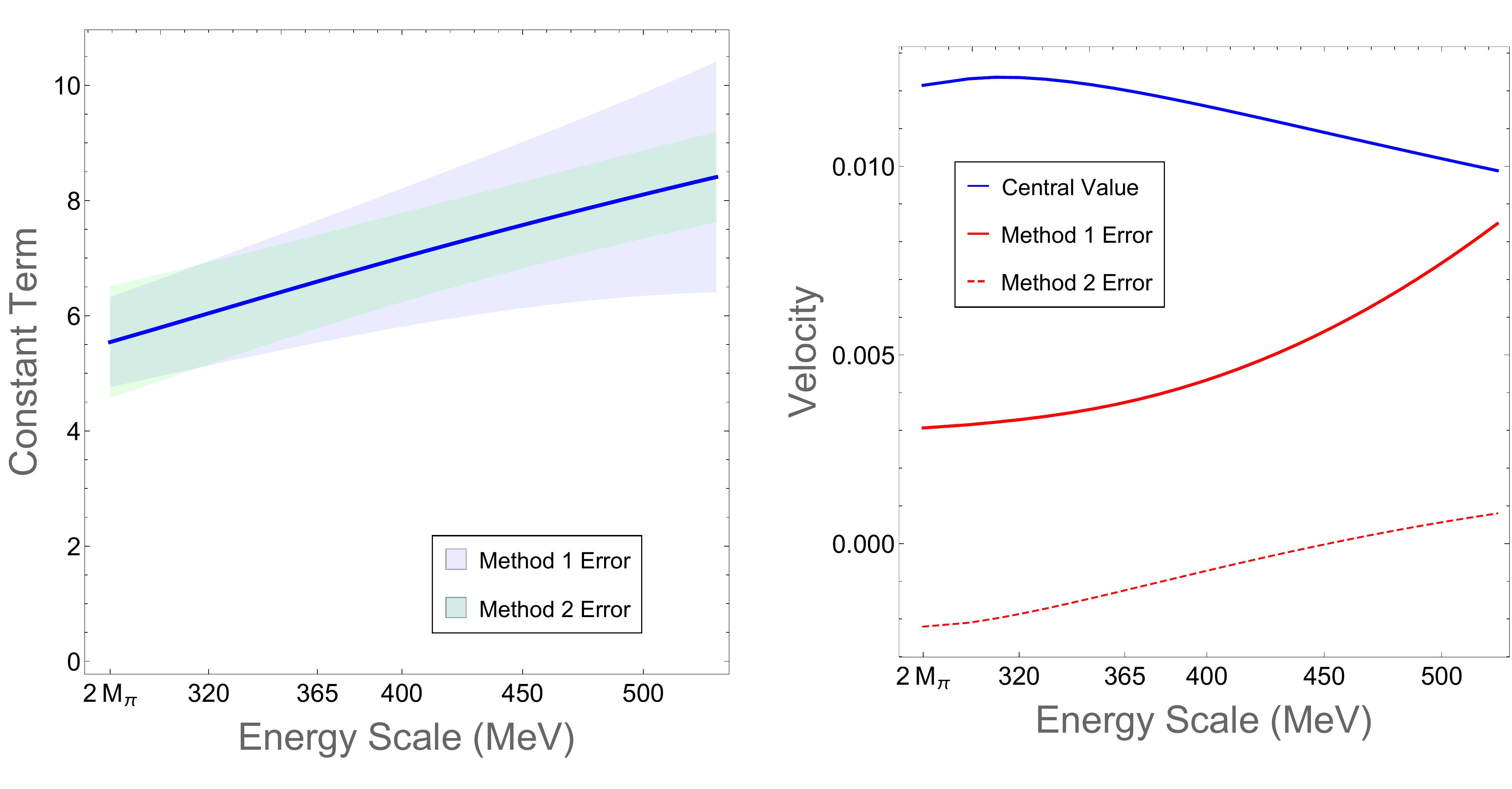}
\end{center}
\caption{Error estimates of the improved $\bar l_1$ and $\bar l_2$ bounds for different $\epi\Lambda$ subtractions (``Energy Scale" denotes the value of $\epi\Lambda M_{\pi}$). The left subfigure shows the error estimates of the $\bar l_{2}+2\bar l_2$ bounds (``Constant Term'' denotes the right hand side of $\bar l_{2}+2\bar l_2 > {\rm Constant~Term}$, which depends on the $\epi\Lambda$ subtraction). The blue solid line is the $\mc{O}(p^4)$ value of the Constant Term, and the error is estimated with the value of the Constant Term at $\mc{O}(p^6)$. Two error estimation methods are represented: Method 1 (light blue region) is to only compute the two loop contribution at $\mc{O}(p^6)$ and multiply it by a factor of 3, which is to roughly account for the badly known tree and one loop contributions at $\mc{O}(p^6)$; Method 2 (light green region) is to set all the $b_{i}$ constants to the central values of their estimates provided by Colangelo {\it et al.} \cite{Colangelo:2001df} at $\mc{O}(p^6)$ (see Eq.~\eqref{expri data1}). The right subfigure shows the growth rate of the $\mc{O}(p^4)$ value of the Constant Term (blue line, "Central Value") and the error estimates (red solid line for Method 1 and red dashed line for Method 2) for the left subfigure. We see that for $\epi\Lambda$ below $490$MeV the improvement of the bounds outruns the increase of the error estimates.}
\label{fig: error-eta1}
\end{figure}

We first use the improved $Y$ bounds to constrain the LECs $\bar l_1$ and $\bar l_2$ at NLO. As mentioned above, we shall discard the improved bounds where perturbativity breaks down, for which we need to compare the $\mc{O}(p^4)$ and $\mc{O}(p^6)$ contributions. As mentioned above, improved positivity bounds become invalid when $\epi\Lambda M_\pi> 490$~MeV. For improved subtractions below 490 MeV, we will further check whether the improvements on the constraints on $\bar l_1$ and $\bar l_2$ can outrun the errors introduced by the very subtraction procedure. While the constraints on $\bar l_1$ and $\bar l_2$ can be obtained with the $\mc{O}(p^4)$ amplitude, we can estimate errors from the higher orders. As mentioned before, the LECs needed to evaluate the $\mc{O}(p^6)$ amplitude are badly known. We will however use two different methods to estimate the errors: Method 1 is again to only compute the two loop contribution at $\mc{O}(p^6)$ and multiply it by a factor of 3, to roughly account for the badly known tree and one loop contributions at $\mc{O}(p^6)$; Method 2 is to set all the $b_{i}$ constants to the central values of their estimates provided by Colangelo {\it et al.} \cite{Colangelo:2001df} at $\mc{O}(p^6)$ (see Eq.~\eqref{expri data1}). As with Method 1, Method 2 is not a rigorous procedure either, but one can see in Figure \ref{fig: error-eta1} that the two methods are mostly consistent with each other. For the parameter space of $\bar l_1$ and $\bar l_2$,  it is the improved $\bar{l}_1  +2  \bar{l}_2$ bounds that provide essential improvements on the constraints as compared to the original bounds. See Figure \ref{fig: error-eta1} for the error estimates of the improved $\bar{l}_1  +2  \bar{l}_2$ bounds for different $\epi\Lambda$ subtractions, which shows that the errors increase relatively slower than the improvements on the bounds.  One can count a couple of reasons for this. First, in the construction of the improved $Y$ bounds, $\mathcal{M}^{2}$ is increased from $2+t / 2$ to $\epsilon^{2} \Lambda^{2}+t / 2-2$, which suppresses the errors as the improved $Y$ bounds contain various factors of $1/\mathcal{M}^{2}$. Also, the $\mc{O}(p^6)$ contribution from the subtraction integral of the improved bound is much smaller than the $\mc{O}(p^6)$ contribution from the original $B(s,t)$ amplitude, the former being less than $\sim2\%$ of the later for $\epi\Lambda M_\pi$ up to 550MeV. Therefore, the improvements on the bounds gained by the $\epi\Lambda$ subtractions in this case appear to outrun the extra errors introduced by the subtractions (See the right subfigure of Figure \ref{fig: error-eta1}).

To illustrate the results, we choose to look at 4 choices for $\epi\Lambda M_\pi$: 341~MeV, 388~MeV, 430~MeV and 468~MeV, and vary different $\{\eta, t, N,M\}$ to get the strongest bounds. Not surprisingly, the constraints are stronger for large $\epi\Lambda$; see Figure \ref{fig: 1-loop-Imp} for the results. For this particular case, the shape of the strongest bounds are unchanged after the $\epi\Lambda$ subtraction, and the improved bounds shift the bounds upwards.

Note that the improved bounds are also independent of the pion mass and decay constant at the one-loop level, and similar to the original $Y$ bounds, the bounds with $N>1$ give rise to trivial results. We have checked that the improved bounds $Y_{\epsilon\Lambda}^{(2 N, M)}(t)={\rm const}>0$ are satisfied for $2\leq\epsilon \Lambda\leq 10$, $1<N<6, M<11$ and $13$ values of $t$.

\subsection{Bounds on the $b_i$ constants}

We also want to use the improved bounds to enhance the bounds on the $b_i$ constants.  Again,  we shall discard the improved bounds where perturbativity breaks down, and we choose to look at 4 choices for $\epi\Lambda M_\pi$: 341~MeV, 388~MeV, 430~MeV and 468~MeV, and vary different $\{\eta, t, N,M\}$ to get the strongest bounds. Similarly, we see that greater $\epi\Lambda$ leads to better constraints on $b_i$; see Figure \ref{fig:2D-Imp1} and \ref{fig:2D-Imp2}.

\begin{figure}[H]
    \centering 
    \begin{subfigure}{}
      \includegraphics[width=0.308\textwidth]{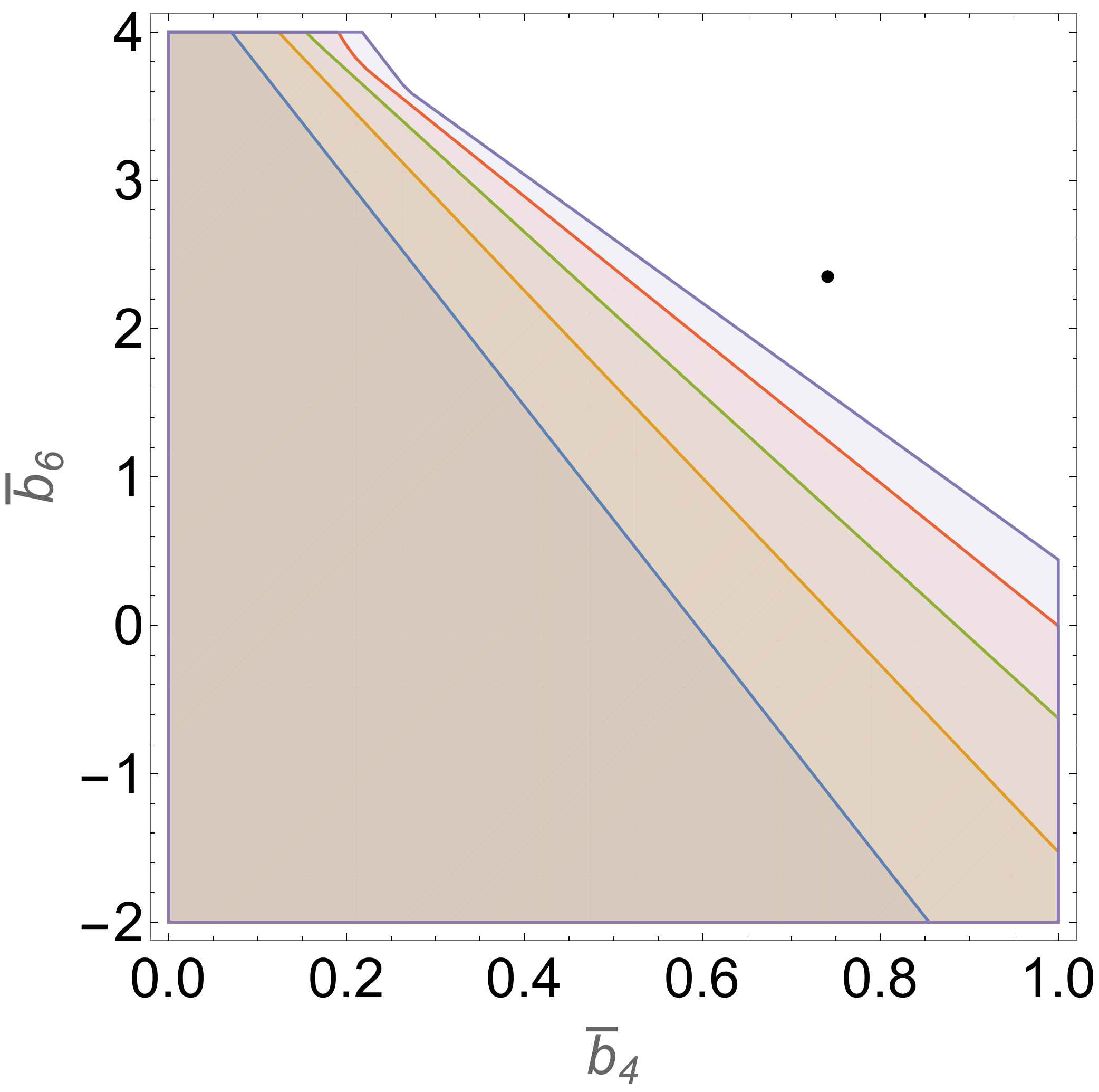}
    \end{subfigure}
     \begin{subfigure}{}
      \includegraphics[width=0.295\textwidth]{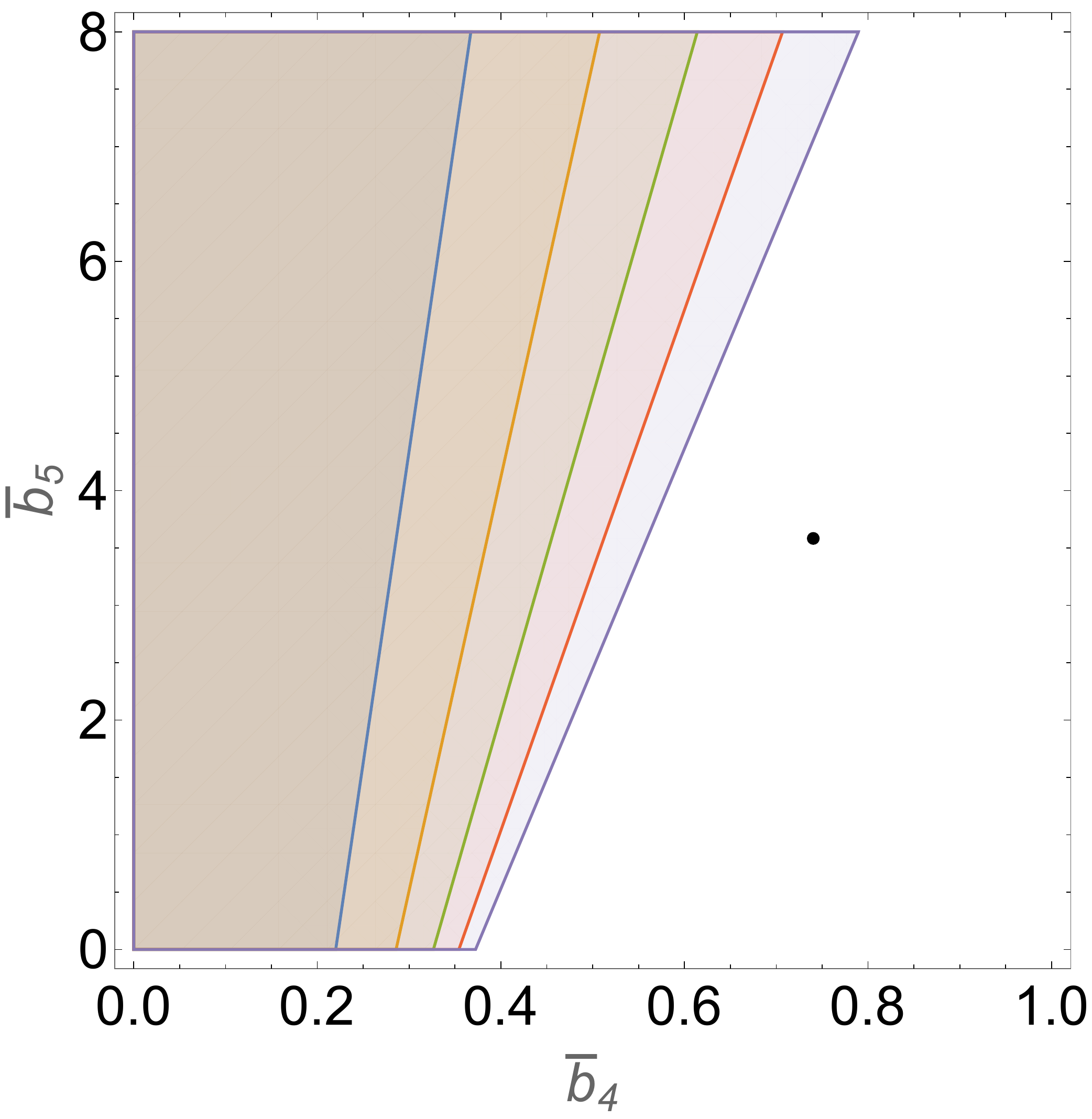}
    \end{subfigure}
    \begin{subfigure}{}
      \includegraphics[width=0.308\textwidth]{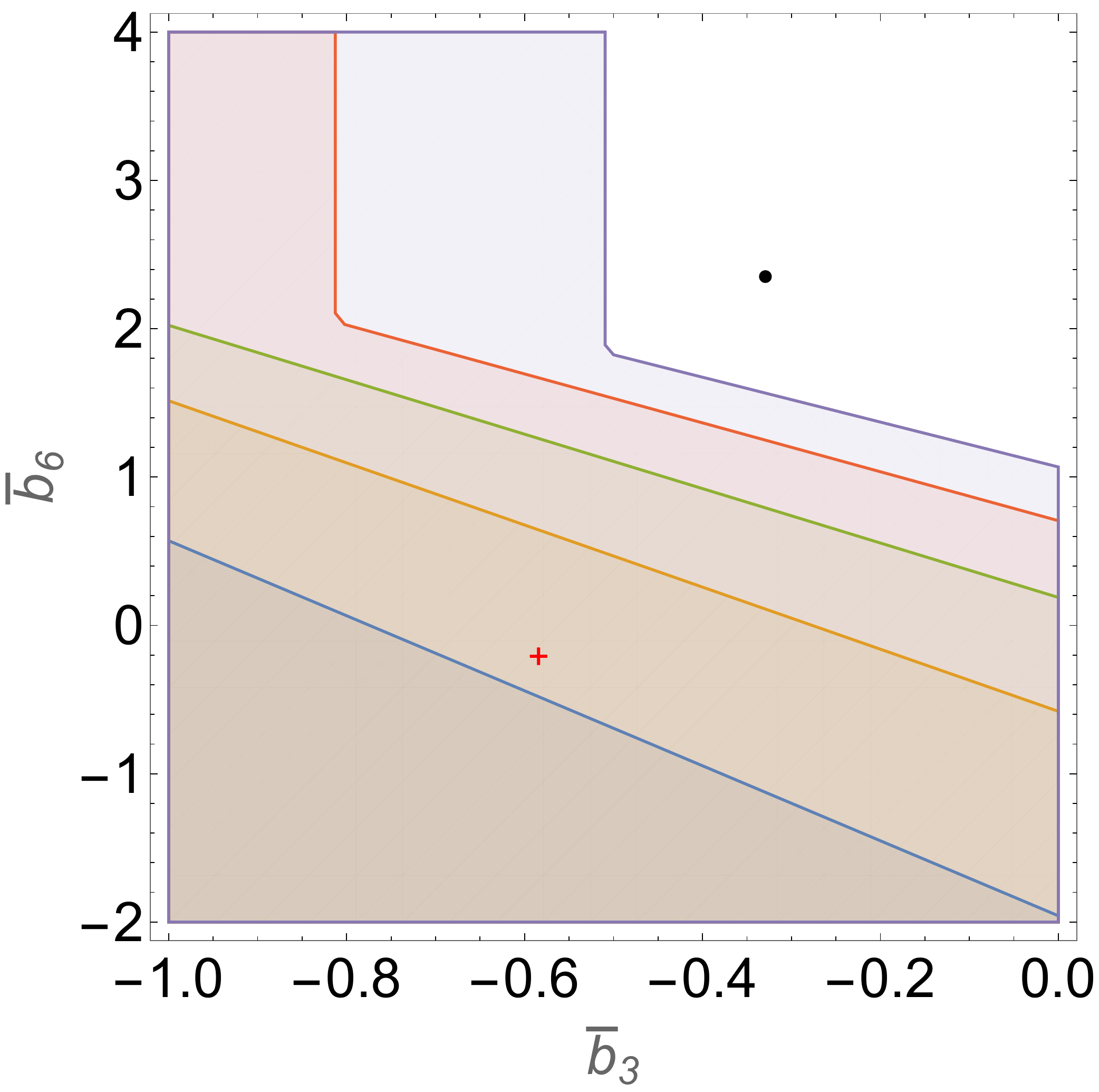}
    \end{subfigure}
    \\
    \centering 
    \begin{subfigure}{}
      \includegraphics[width=0.316\textwidth]{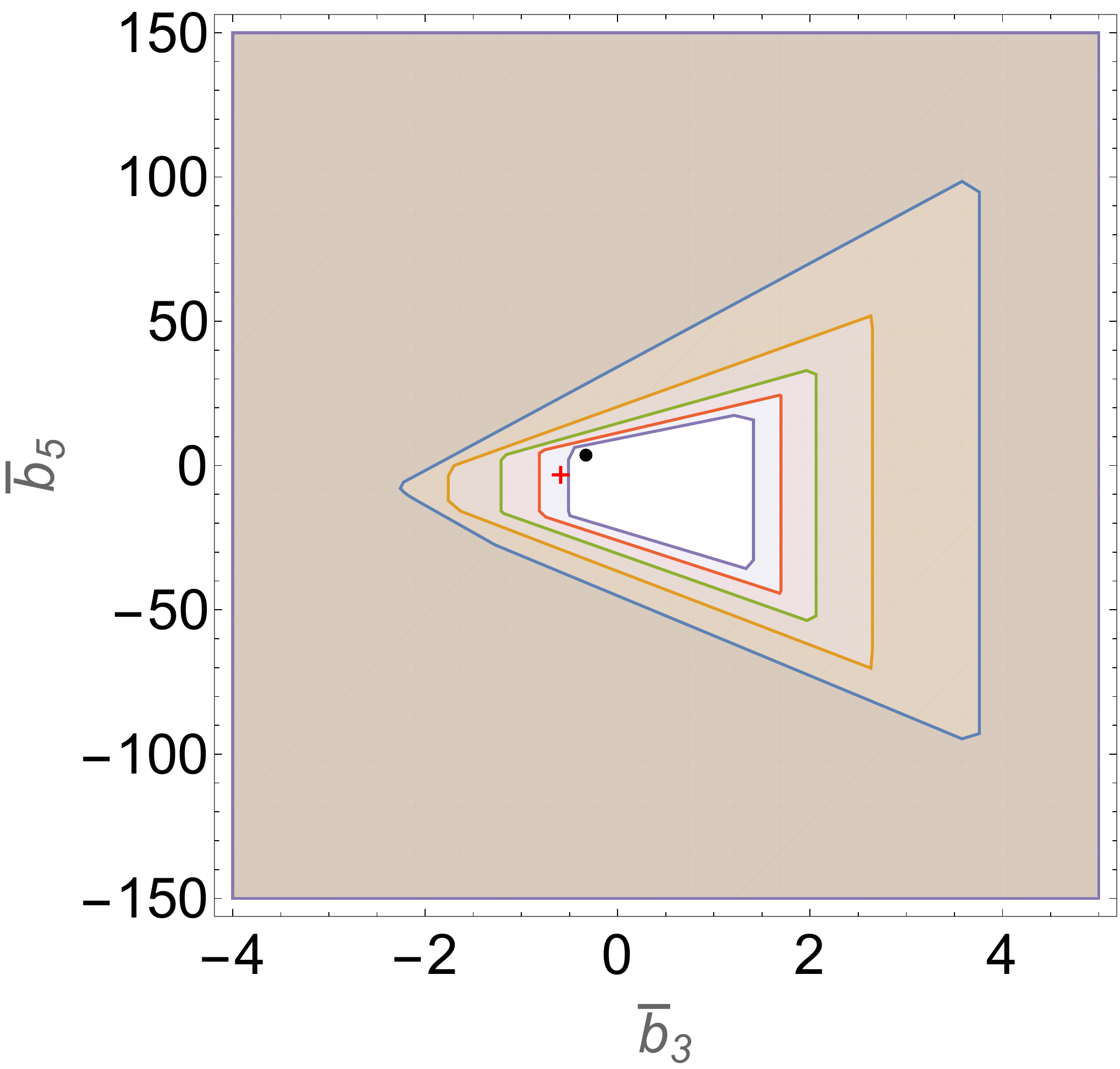}
    \end{subfigure}
     \begin{subfigure}{}
      \includegraphics[width=0.319\textwidth]{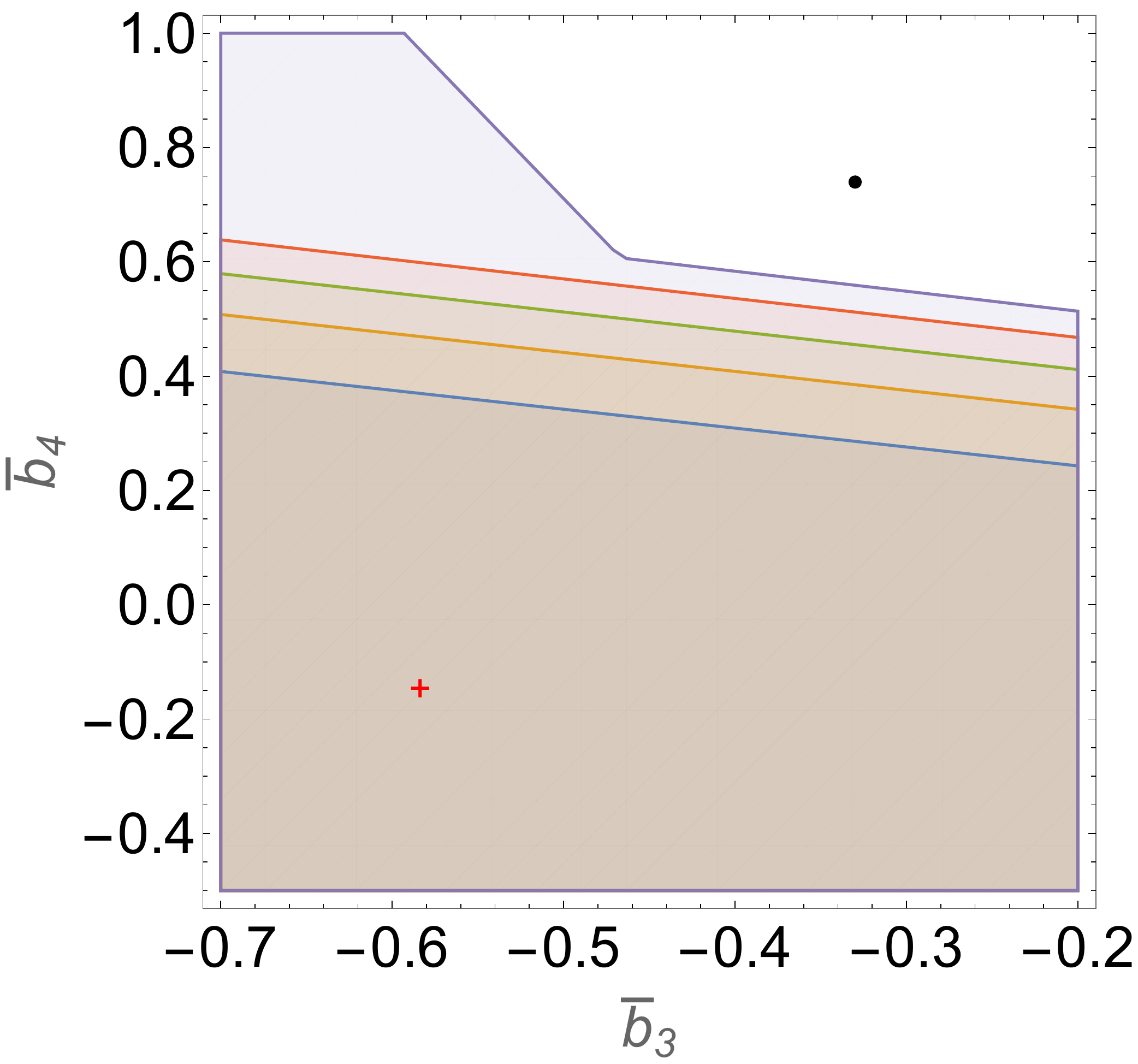}
    \end{subfigure}
    \begin{subfigure}{}
      \includegraphics[width=0.31\textwidth]{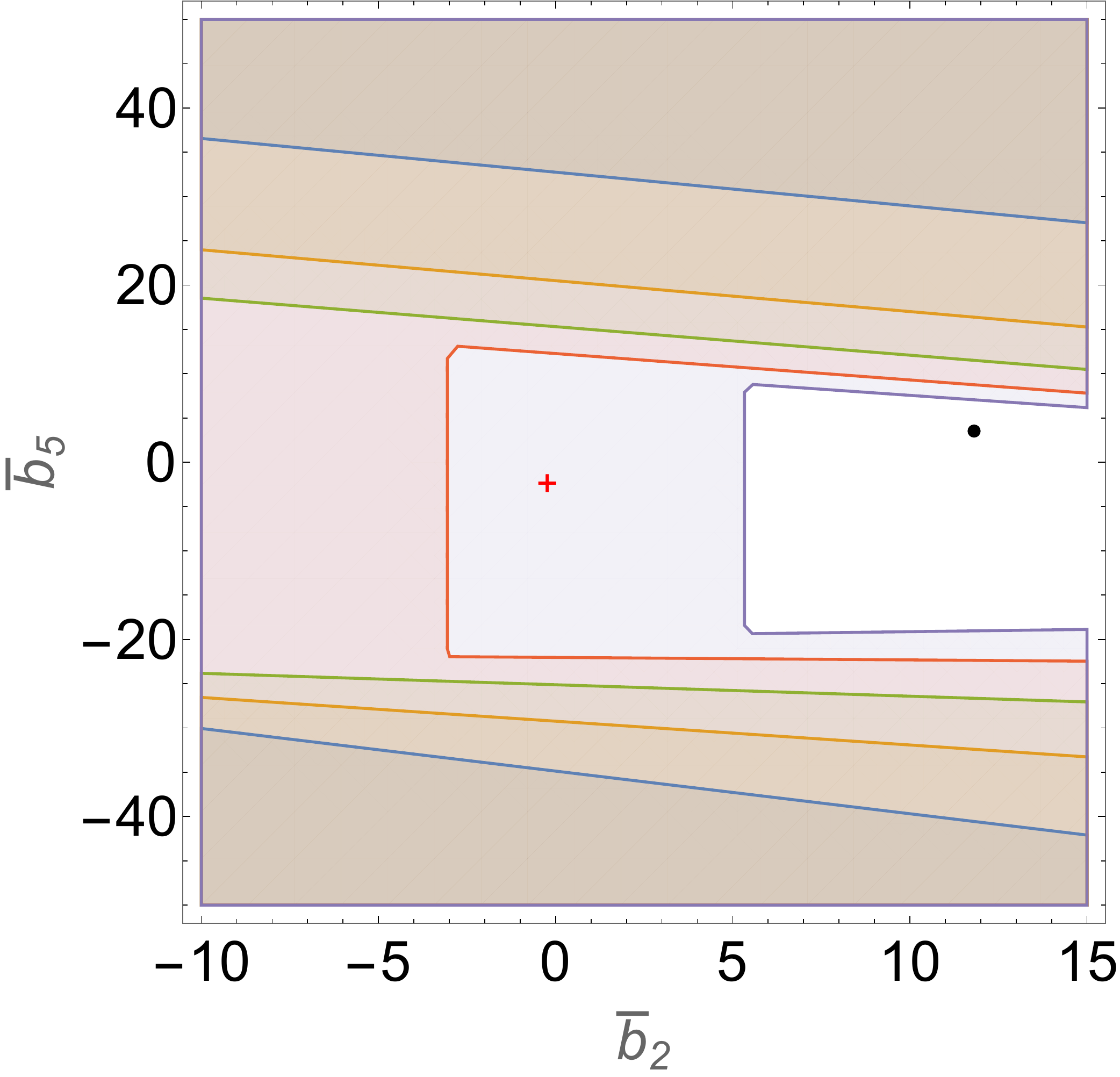}
    \end{subfigure}
    \\
    \centering 
    \begin{subfigure}{}
      \includegraphics[width=0.31\textwidth]{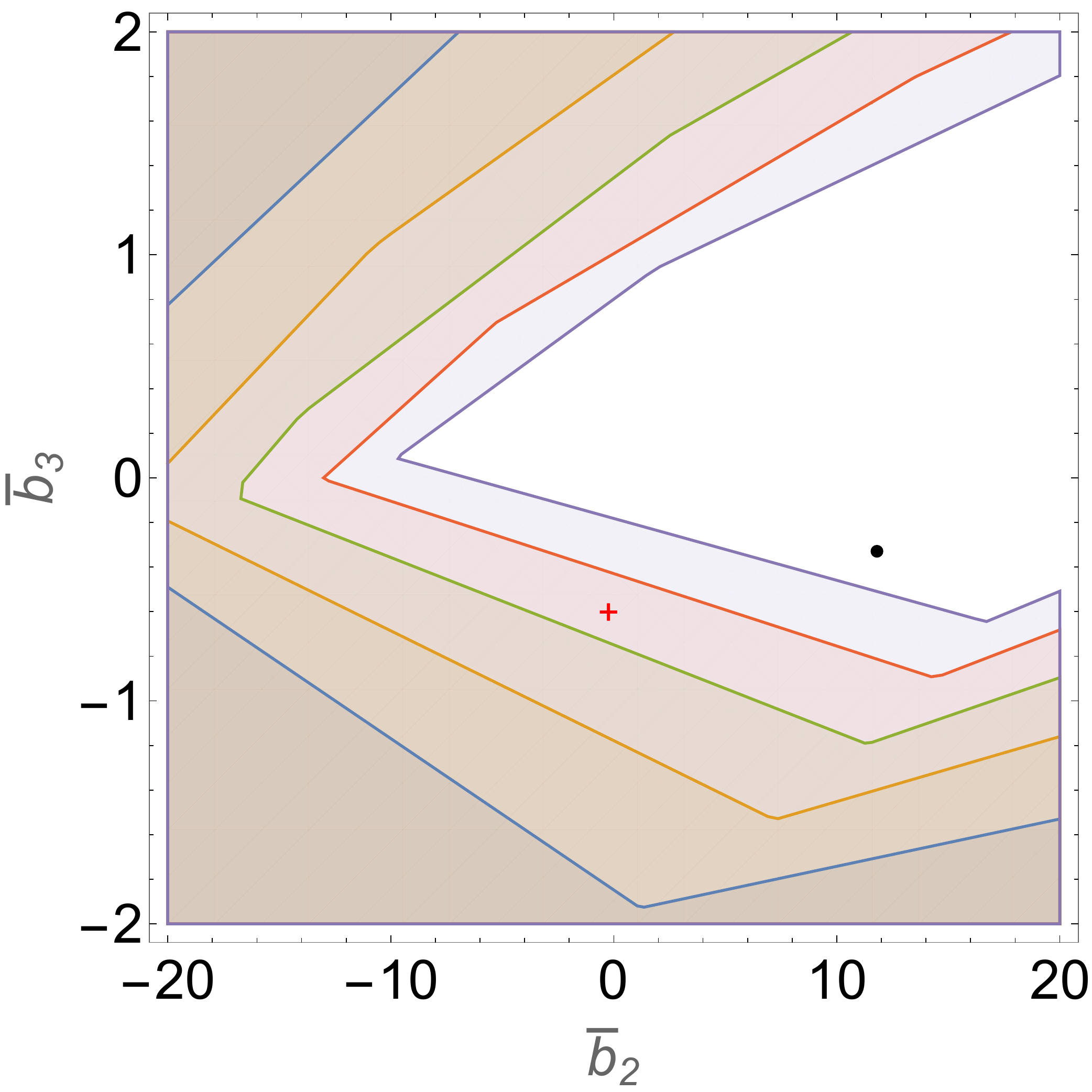}
    \end{subfigure}
     \begin{subfigure}{}
      \includegraphics[width=0.31\textwidth]{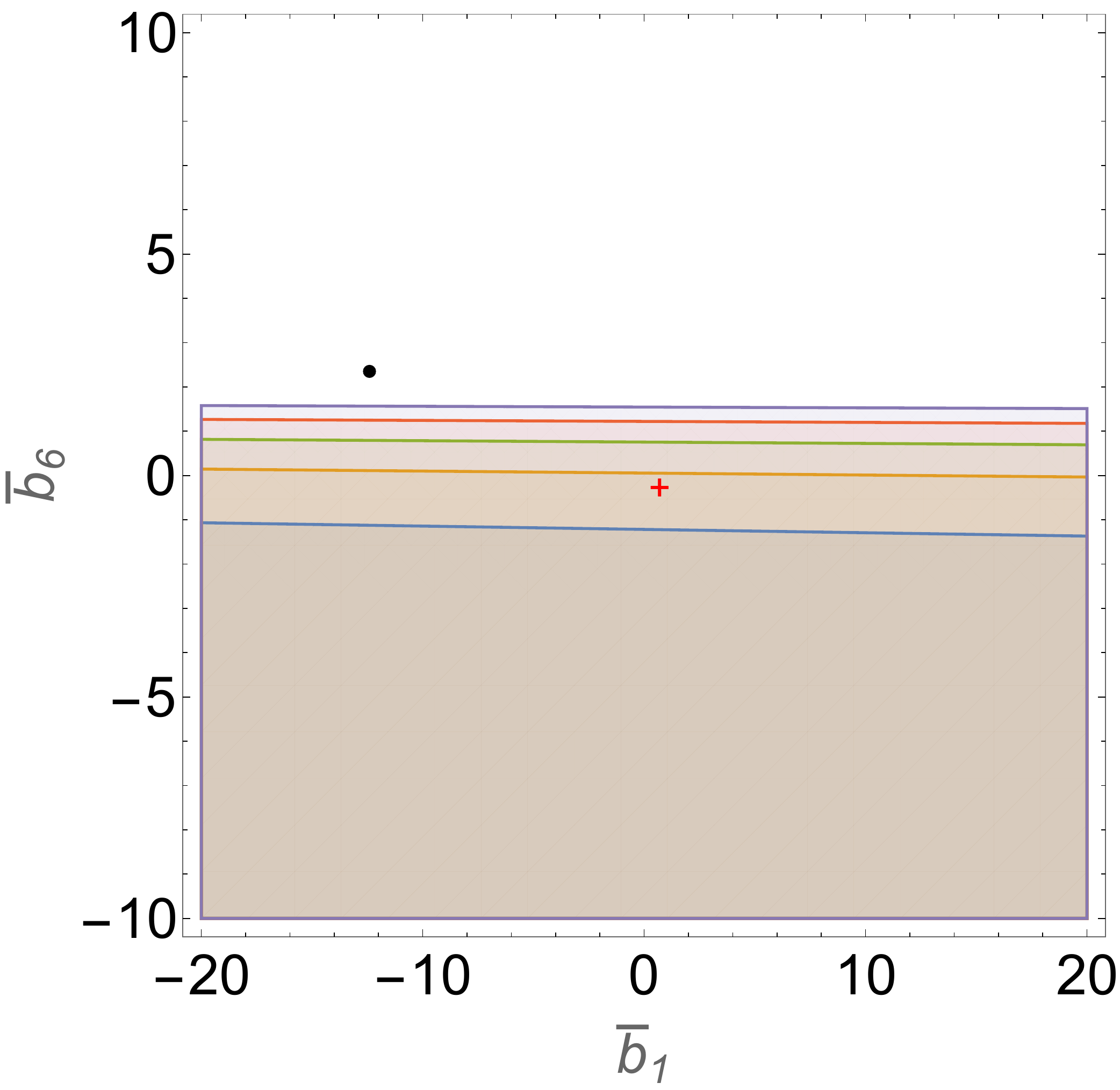}
    \end{subfigure}
    \begin{subfigure}{}
      \includegraphics[width=0.323\textwidth]{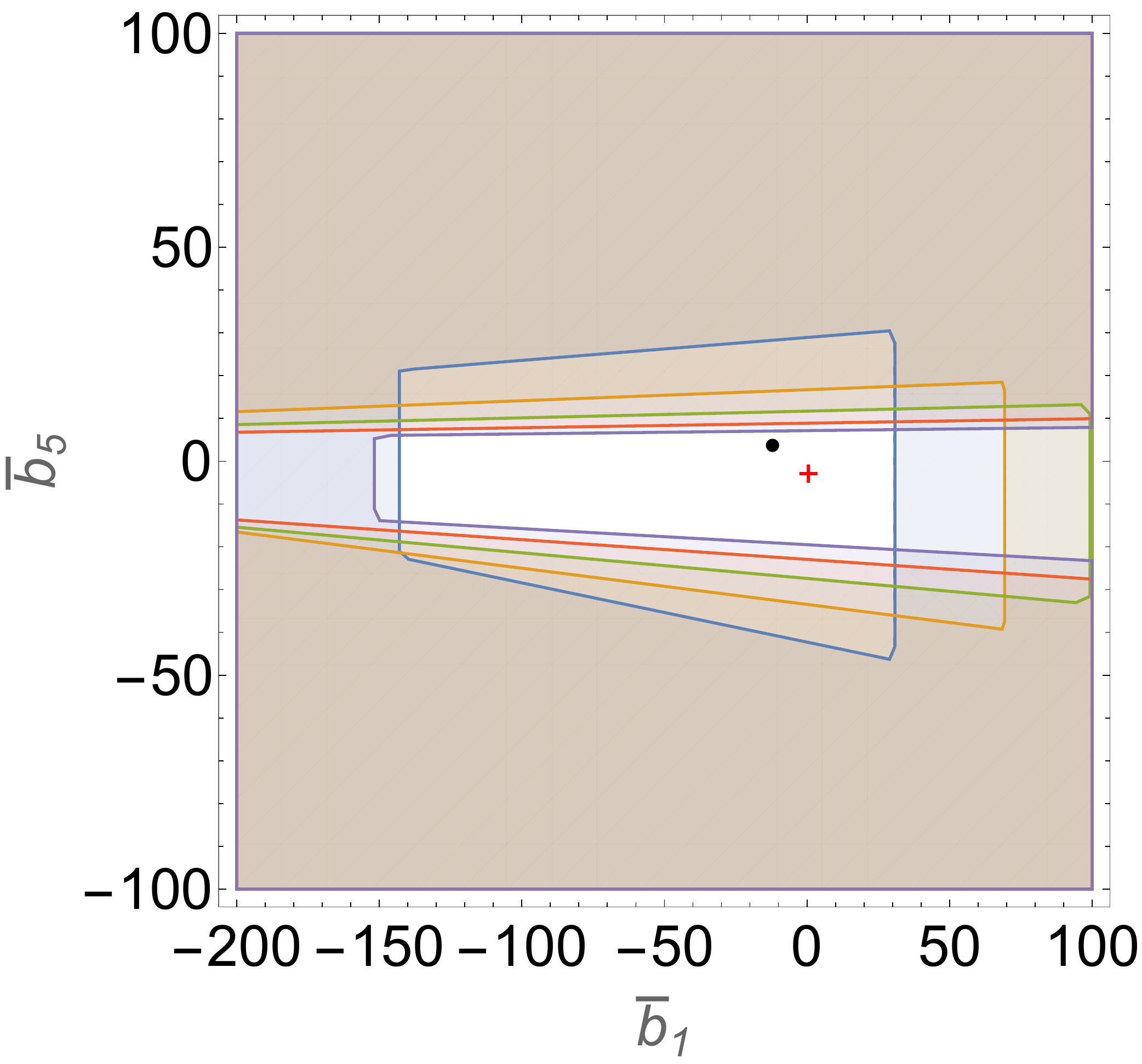}
    \end{subfigure}
    \caption{2D sections of the improved constrained $b_i$ space. The 2D sections are obtained by setting the other $4$ parameters to the central values of the fit \eqref{expri data1}. The black point represents the central values of the fit \eqref{expri data1} with inputs from the experimental data and theoretical estimates, and the red cross represents the theoretical point computed from the Weinberg Lagrangian. Different lines corresponds to different choices of $\epi\Lambda  M_\pi$: $341$~MeV (Orange), $388$~MeV (green), $430$~MeV (Red), $468$~MeV (Purple), original $Y$ bounds (blue). To be continued in Figure~\ref{fig:2D-Imp2}.} 
    \label{fig:2D-Imp1}
\end{figure}

\begin{figure}[H]
    \centering 
    \begin{subfigure}{}
      \includegraphics[width=0.31\textwidth]{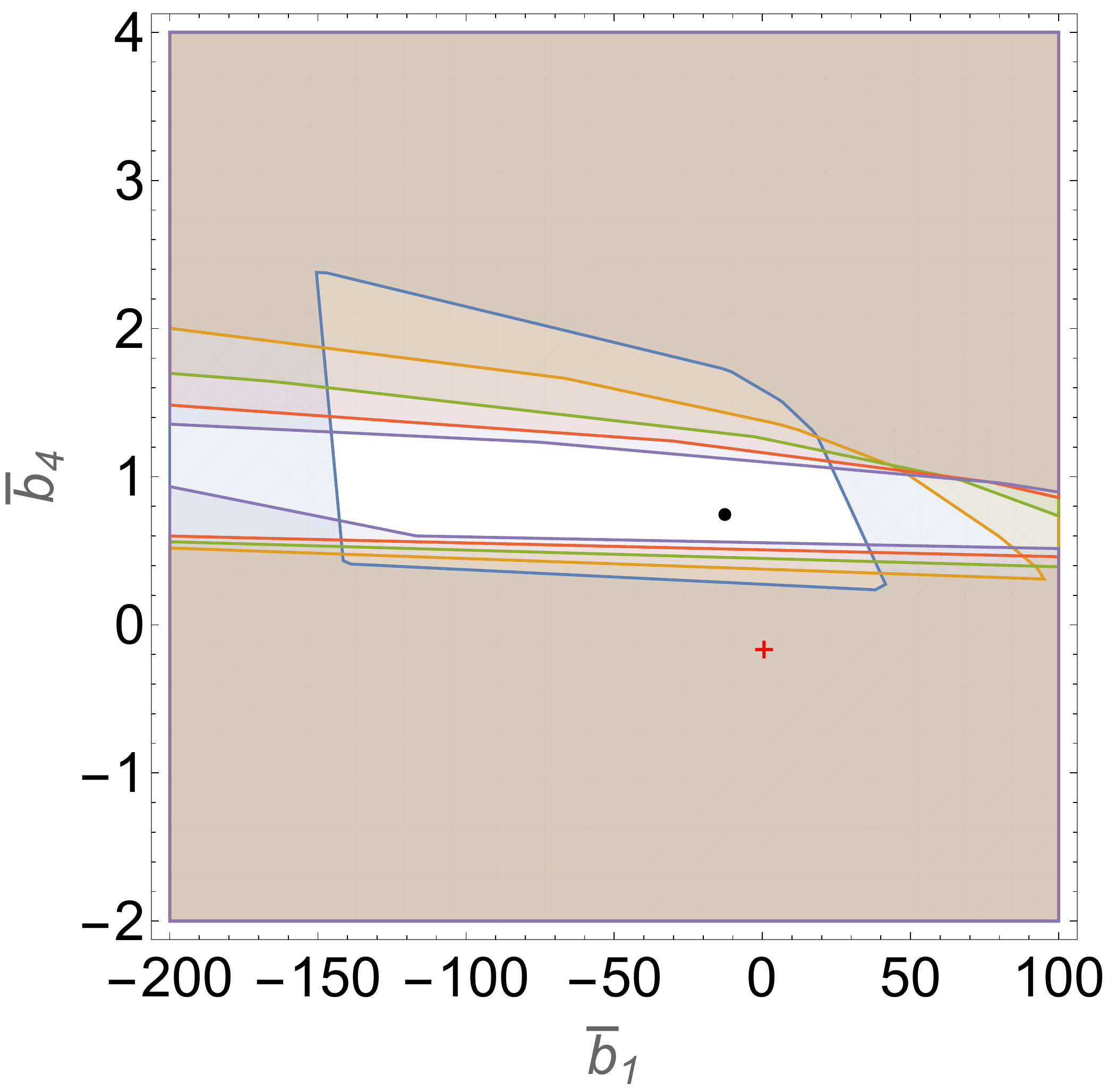}
    \end{subfigure}
     \begin{subfigure}{}
      \includegraphics[width=0.308\textwidth]{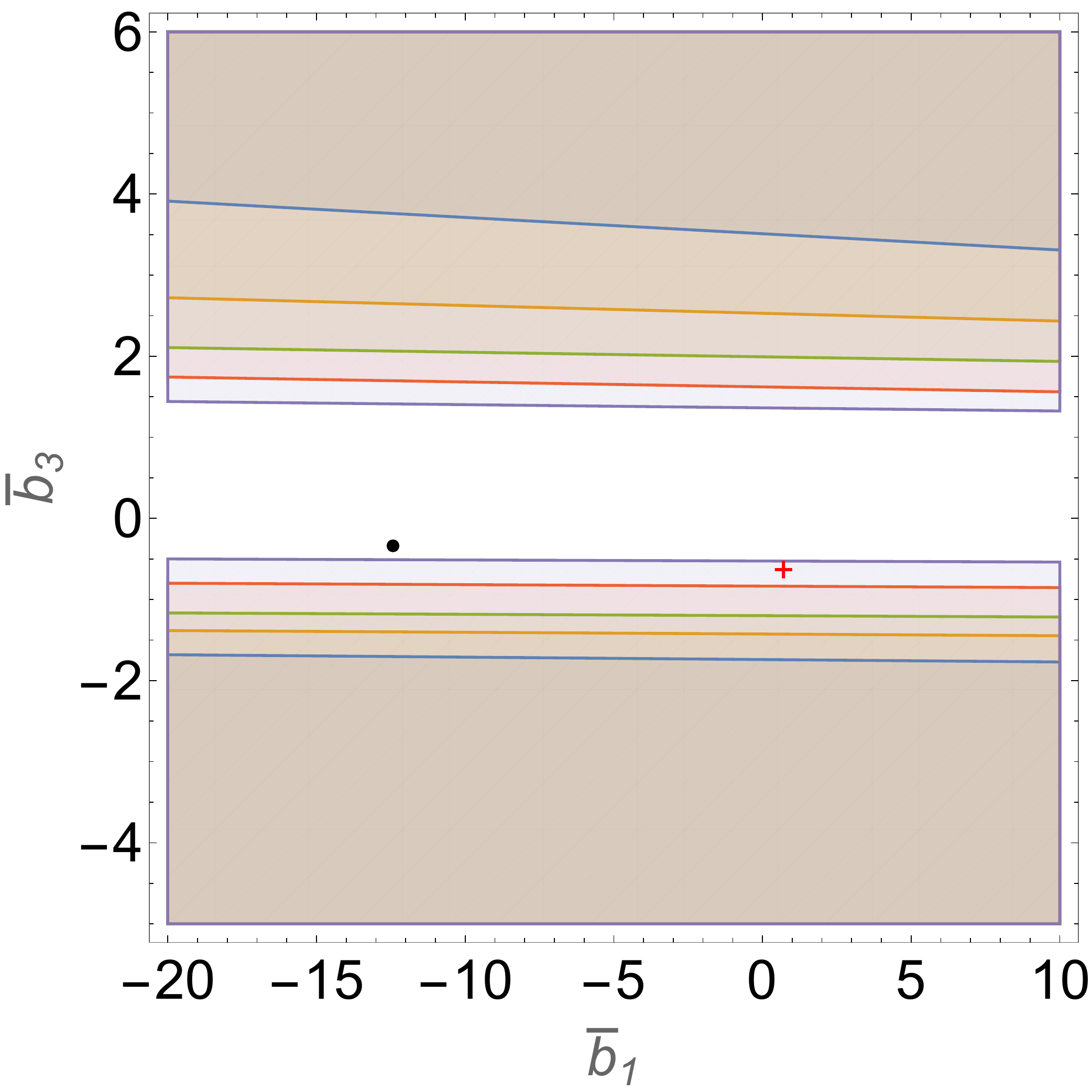}
    \end{subfigure}
    \begin{subfigure}{}
      \includegraphics[width=0.31\textwidth]{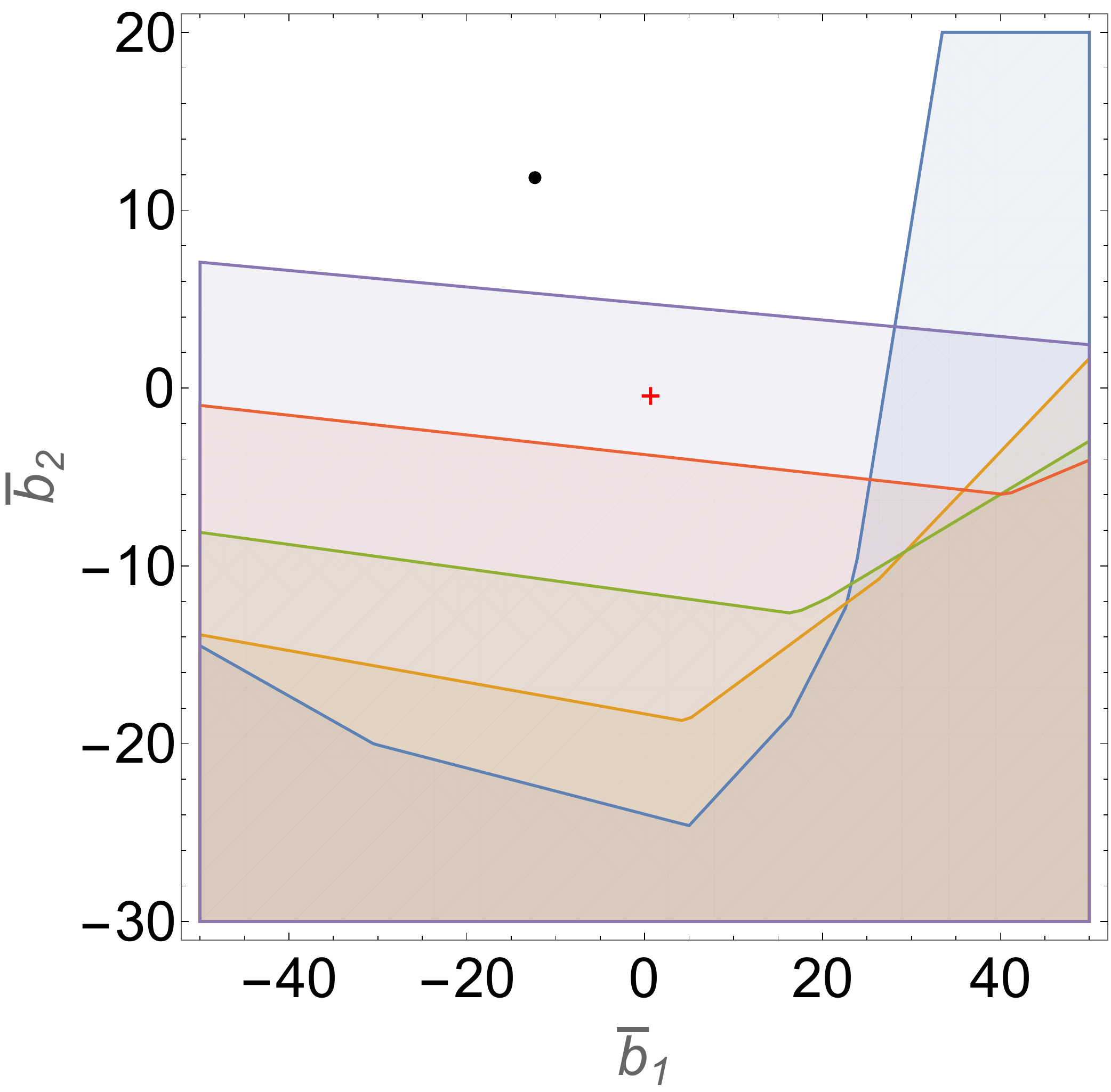}
    \end{subfigure}
    \\
      \centering 
    \begin{subfigure}{}
      \includegraphics[width=0.31\textwidth]{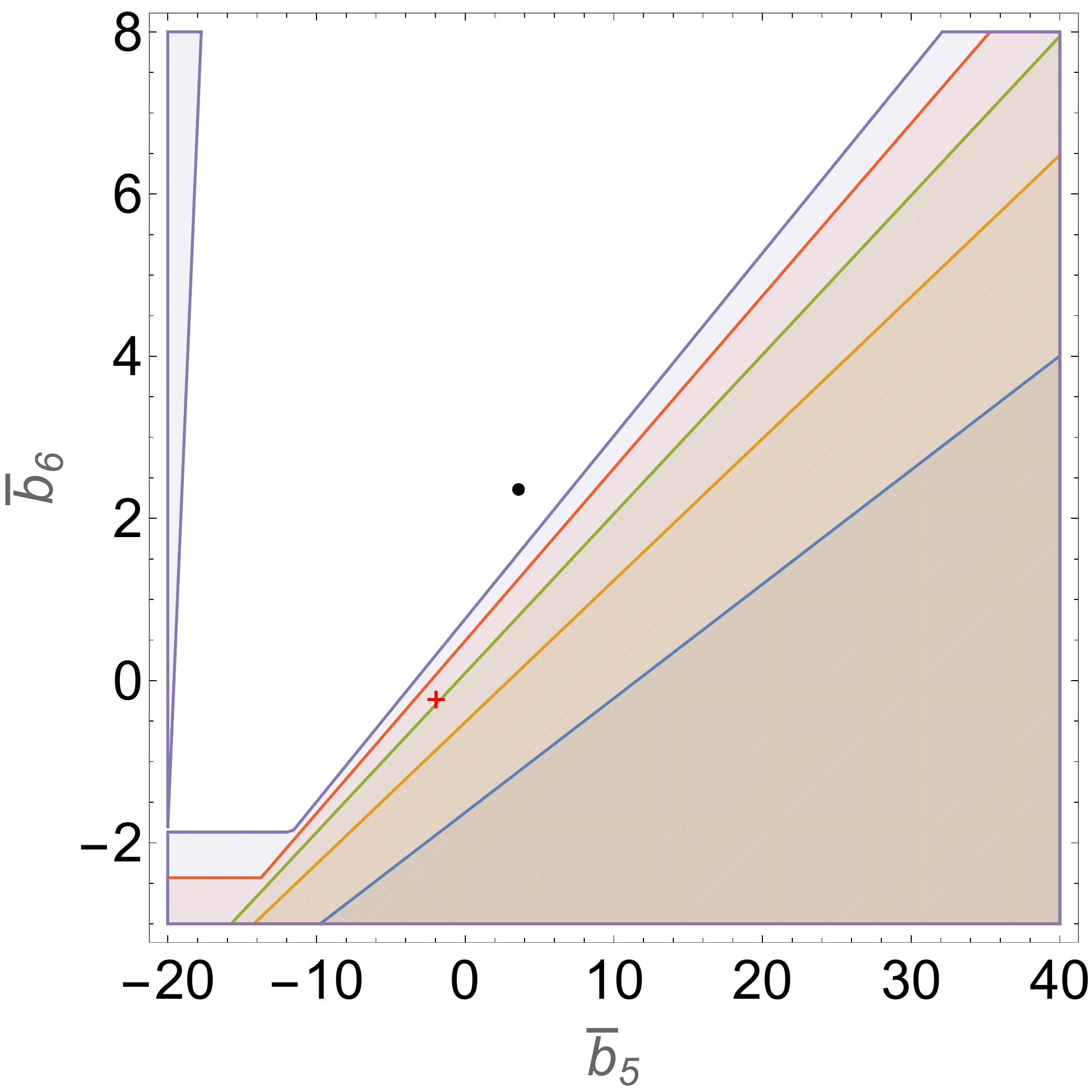}
    \end{subfigure}
     \begin{subfigure}{}
      \includegraphics[width=0.314\textwidth]{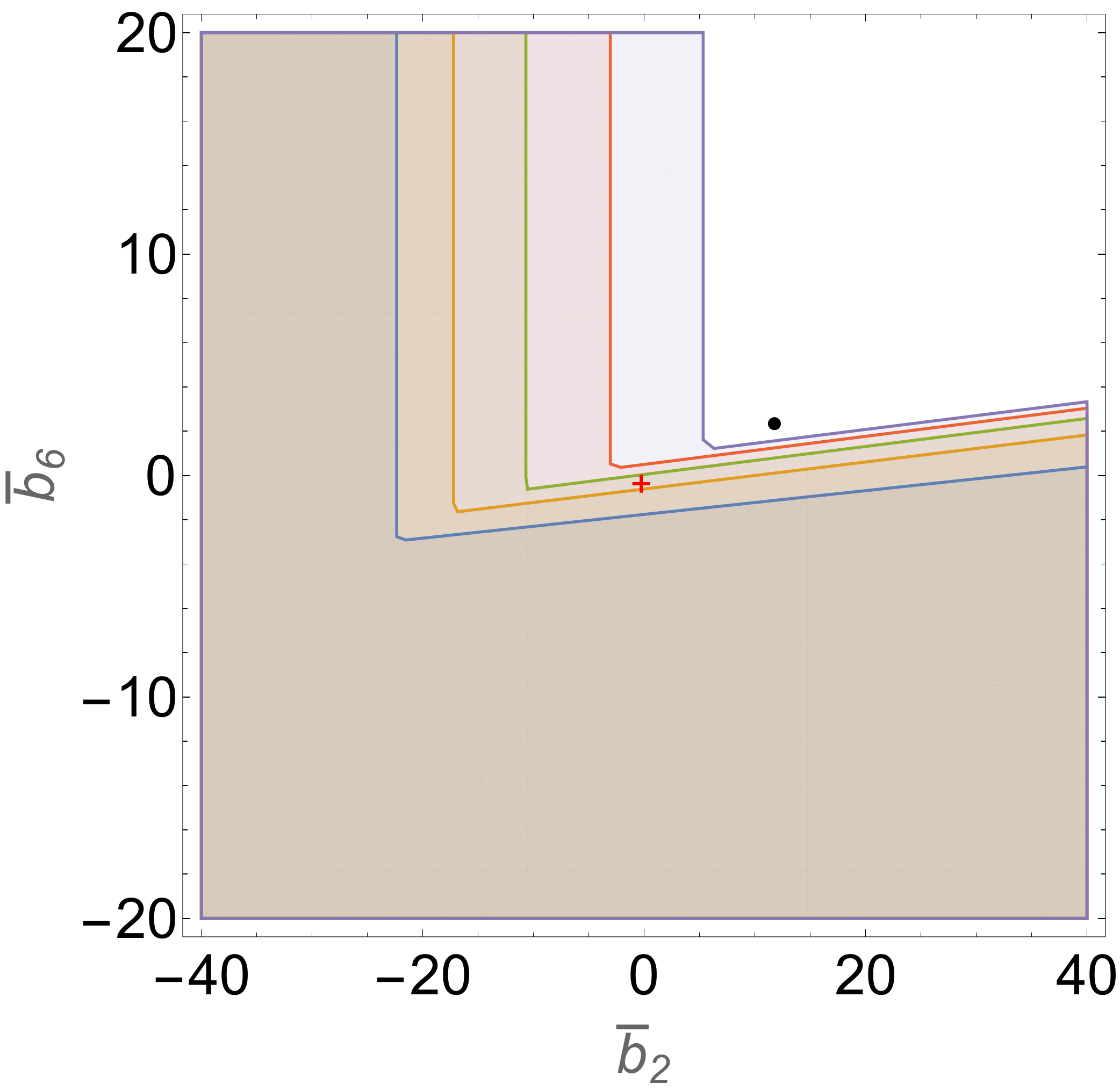}
    \end{subfigure}
    \begin{subfigure}{}
      \includegraphics[width=0.31\textwidth]{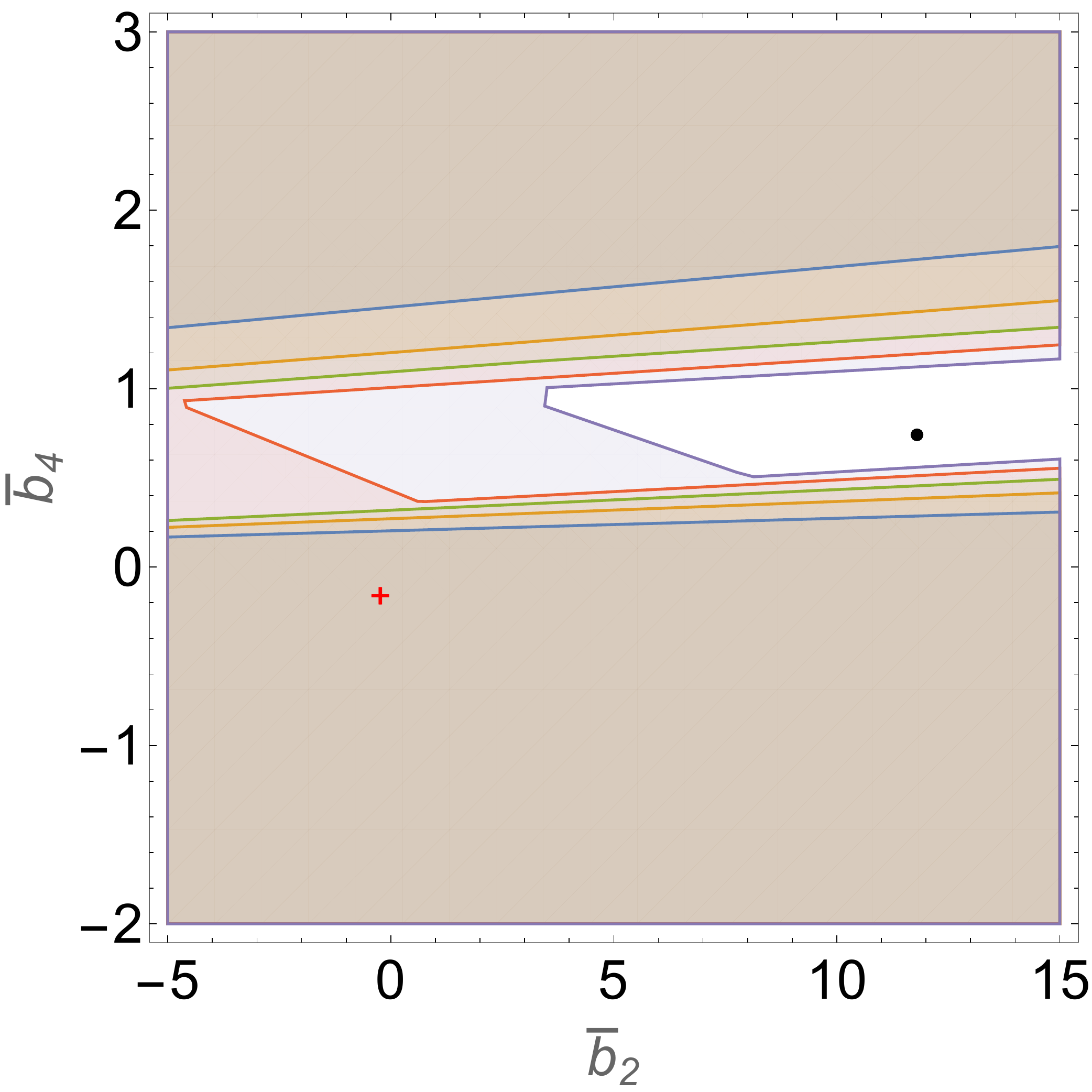}
    \end{subfigure}
    \caption{Continuation of Figure~\ref{fig:2D-Imp1}.} 
    \label{fig:2D-Imp2}
\end{figure}

\subsection{Pad{\'e} approximation}
\label{sec:pade}

Unitarity is only perturbatively respected in ChPT. It is well-known that the ChPT amplitude at leading orders violates unitarity at relatively low energy scales because of the existence of the $f_0(500)$ resonance. There are various methods to resum the ChPT scattering amplitudes in order to restore the exact unitarity~\cite{Truong:1988zp,Dobado:1989qm,Truong:1991gv,Dobado:1996ps,Oller:1997ti,Oller:1997ng,Oller:1998hw,Oller:1998zr,Hannah:1999ev,Nieves:1999bx,Oller:2000fj,Oller:2000ma,GomezNicola:2001as,Pelaez:2015qba} (for works on unitarized ChPT at two loops, see Refs.~\cite{Nieves:1999bx,Pelaez:2006nj}); the meson-meson scattering data can be described in such nonperturbative approaches up to around 1.2~GeV, far higher than that of the perturbative ChPT. A very convenient and extensively used method to restore unitarity up to close to the cutoff scale is to make use of a mathematical tool, called Pad{\'e} approximation~\cite{Truong:1988zp,Dobado:1989qm,Dobado:1996ps,Oller:1997ng,Oller:1998hw,Hannah:1999ev}. We want to check whether the same trick can be applied to the improved positivity bounds.

In the Pad{\'e} unitarization, the Pad{\'e} approximation is applied to the partial waves of the isospin amplitude 
\begin{equation}
T^{I}(s, t, u)=16 \pi \sqrt{\f{s}{s-4}}\sum_{\ell}(2 \ell+1) P_{\ell}(\cos \theta) T_{I}^{\ell}(s)  .
\end{equation}
In our case, the ChPT amplitude is calculated to two loops, $T_{{\ell}}^{I}(s)=T_{{\ell}, 1}^{I}(s)+T_{{\ell}, 2}^{I}(s)+T_{{\ell}, 3}^{I}(s)$ with subscripts 1,2,3 indicating the order of $x_2$, so we can take the [1,2] Pad{\'e} approximation, which is to replace $T_{I}^{\ell}(s)$ with
\begin{equation}
T_{{\ell}}^{I[1,2]}(s)=\frac{T_{{\ell}, 1}^{I}(s)}{1-\frac{T_{{\ell}, 2}^{I}(s)}{T_{{\ell}, 1}^{I}(s)}-\frac{T_{{\ell}, 3}^{I}(s)}{T_{{\ell}, 1}^{I}(s)}+\left(\frac{T_{{\ell}, 2}^{I}(s)}{T_{{\ell}, 1}^{I}(s)}\right)^{2}}  .
\end{equation}
Since perturbative unitarity is satisfied order by order, we can show that the unitarized partial wave amplitude $T_{{\ell}}^{I[1,2]}(s)$ satisfies the unitarity relation $\operatorname{Im} T_{{\ell}}^{I[1,2]}(s)= \left|T_{{\ell}}^{I[1,2]}(s)\right|^{2}$, which is very useful in many circumstances (for reviews, see Refs.~\cite{Oller:2000ma,Pelaez:2015qba,Oller:2019opk}).

\begin{figure}[tb]
\begin{center}
\includegraphics[width=.45\linewidth]{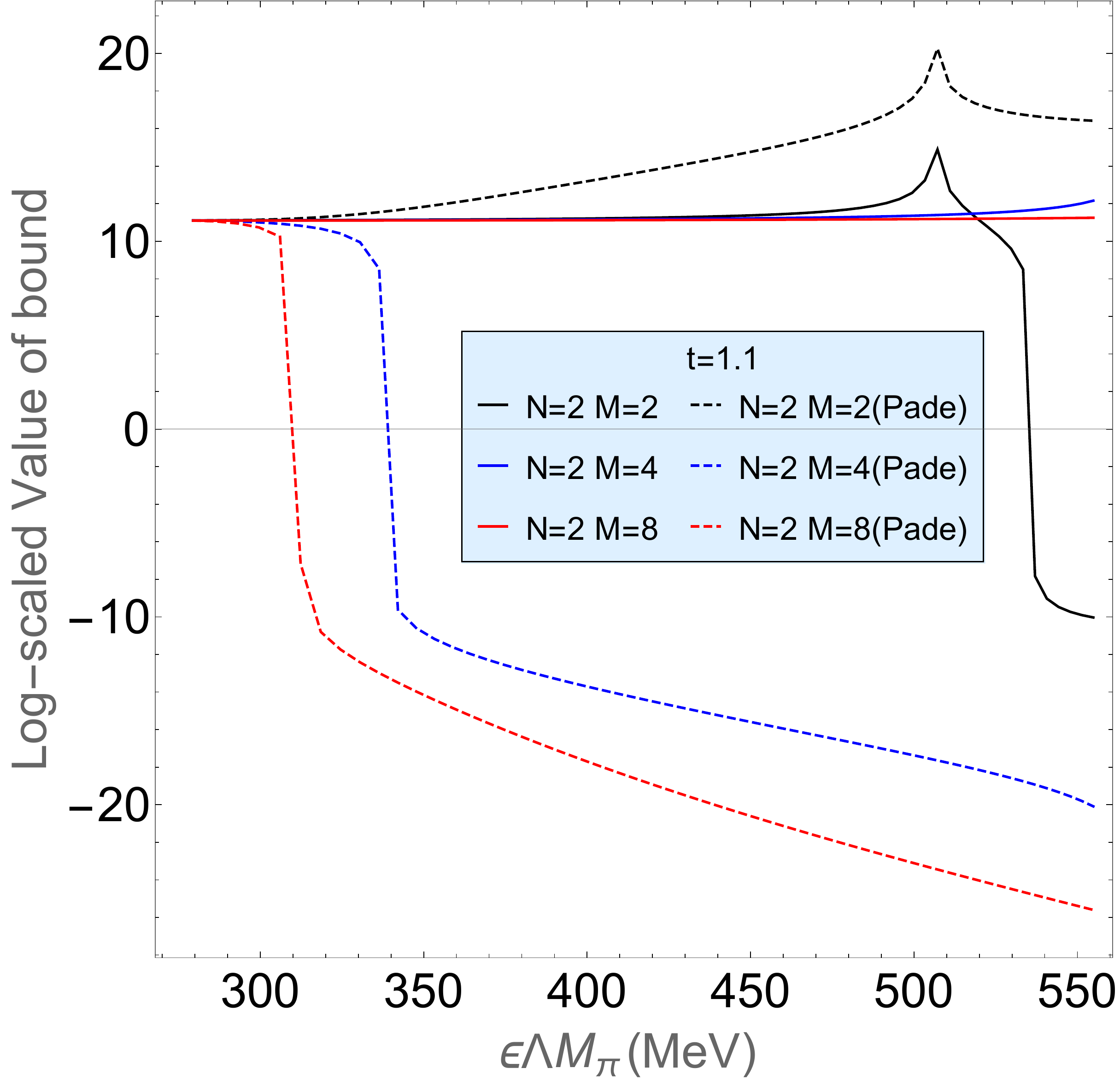}
\end{center}
\caption{Comparison of the improved $Y$ bounds with (solid lines) and without (dashed lines)  the Pad{\'e} approximation. The example is for the $\pi^{0} \pi^{0}\rightarrow \pi^{0} \pi^{0}$ scattering amplitude up to two loops. The longitudinal axis indicates the logarithmically scaled values of the bounds.}
\label{Pade}
\end{figure}

However, we find that the unitarized Pad{\'e} amplitude actually significantly lower the value of $\epi\Lambda$ that can be used to subtract the dispersion integral in the improved positivity bounds. In other words, in a sense, the Pad{\'e} amplitude has worse dispersive properties than the original amplitude.  For example, if we Pad{\'e} unitarize the $\pi^{0} \pi^{0}\rightarrow \pi^{0} \pi^{0}$ amplitude, using it for the improved $Y$ bounds, and employ the same $b_i$ constants in Eq.~\eqref{expri data1}, the energy scale $\epi\Lambda M_\pi$ at which the improved $t=1.1,N=2, M=8$ bound becomes negative is at $305$~MeV; see Figure~\ref{Pade}. In comparison, using the original amplitude, the $t=1.1,N=2,M=2$ bound only breaks down after $500$~MeV. One need, however, to bear in mind that in principle the LECs in the unitarized amplitudes should take different values than those determined from the ChPT amplitude. For example, it was found previously for the scattering between the pseudo-Goldstone bosons and charmed mesons: the LECs determined from lattice QCD data using the unitarized ChPT in that case do not fulfill the positivity bounds derived for the perturbative amplitudes~\cite{Du:2016tgp}. In any case, the simple Pad{\'e} unitarization procedure does not improve the analytic properties of the amplitude in terms of dispersion relations. Other undesirable properties of the Pad{\'e} unitarization have been noticed previously, such as predicting spurious physical sheet resonances~\cite{Ang:2001bd,Qin:2002hk} and incorrect coefficients for the leading chiral logarithms in higher loop diagrams~\cite{Gasser:1990bv}.

\section{Summary}
\label{sec:concl}

We have applied the generalized positivity bounds (the $Y$ bounds and the improved $Y$ bounds) to ChPT to NNLO. This allows us to constrain the $\bar l_i$ LECs and the $b_i$ constants, which are combinations of LECs, to be within convex regions respectively. The constrained regions are convex because the $Y$ bounds produce inequalities that are linear in $\bar l_i$ and $b_i$. We see that constraints from the new bounds are stronger than the constraints obtained by the previous positivity bounds. For the $\bar l_i$ constants, although the values fitted from experimental data combined with other theoretical estimates are widely believed to be relatively accurate by now, the constraints from the positivity bounds are still interesting because of the cleanness in its assumptions, which are merely the fundamental principles of quantum field theory such as unitarity and analyticity. Also, the bounds we obtained for the $\bar l_i$ constants are independent of the pion mass and the pion decay constant, so these bounds apply to any ChPT with the same underlying symmetry, not just for the ChPT derived from QCD. For the bounds on the six $b_i$ constants, we see that most of its 2D sections near the empirically fitted central values are enclosed, with the constraints in some directions stronger than the others. Moreover, we have applied the improved positivity bounds to constrain the $\bar l_i$ and $b_i$ constants, which can further enhance the bounds.

Using the improved positivity, we can detect an energy scale at which ChPT as an EFT must break down. This is because when the $\epi\Lambda$ subtraction is set sufficiently high, the empirically fitted values of the LECs will be in conflict with the positivity bounds. For ChPT from QCD, this scale is around 490~MeV, consistent with the existence of the $f_0(500)$ resonance. A well-known method to ``magically'' restore unitarity is to apply the Pad{\'e} approximant for the partial waves of the isospin amplitude. However, we find that the Pad{\'e} method is rather unsatisfactory to restore the dispersion relation, as the improved bounds with the Pad{\'e} unitarized amplitude actually break down at much lower energy scales.

\acknowledgments

We would like to thank De-Liang Yao, Zhi-Hui Guo, Zhi-Guang Xiao and Han-Qing Zheng for helpful discussions. SYZ acknowledges support from the starting grant from University of Science and Technology of China under grant No.~KY2030000089 and GG2030040375 and is also supported by National Natural Science Foundation of China (NSFC) under grant No.~11947301. The work of FKG is supported in part by NSFC under grants No.~11835015, No.~11947302 and No.~11961141012, by NSFC and  Deutsche Forschungsgemeinschaft through the funds provided to the Sino-German Collaborative Research Center  CRC110 ``Symmetries and the Emergence of Structure in QCD"  (NSFC Grant No.~11621131001), by the Chinese Academy of Sciences (CAS) under Grants No.~QYZDB-SSW-SYS013 and No.~XDPB09, and by the CAS Center for Excellence in Particle Physics (CCEPP). CZ is supported by IHEP under Contract No.~Y7515540U1.

\appendix

\section{Loop functions and $b_i$ constants}
\label{appendix: loop fun}

Here we list the explicit expressions for the loop functions and the $b_i$ constants. The loop functions $F^{(i)}(s)$ and $G^{(i)}(s, t)$ are defined as follows \cite{Bijnens:1995yn}
\begin{align} 
F^{(1)}(s) &=\frac{1}{2} \overline{J}(s)\left(s^{2}-1\right), 
\\ 
G^{(1)}(s, t) &=\frac{1}{6} \overline{J}(t)\left(14-4 s-10 t+s t+2 t^{2}\right) ,
\\ 
F^{(2)}(s) &=\overline{J}(s)\left\{\frac{1}{16 \pi^{2}}\left(\frac{503}{108} s^{3}-\frac{929}{54} s^{2}+\frac{887}{27} s-\frac{140}{9}\right) +b_{1}(4 s-3)+b_{2}\left(s^{2}+4 s-4\right)  \right.
\nn
&~~~\left. +\frac{b_{3}}{3}\left(8 s^{3}-21 s^{2}+48 s-32\right)+\frac{b_{4}}{3}\left(16 s^{3}-71 s^{2}+112 s-48\right)\right\} 
\nn 
&~~~+\frac{1}{18} K_{1}(s)\left\{20 s^{3}-119 s^{2}+210 s-135-\frac{9}{16} \pi^{2}(s-4)\right\} 
\nn 
&~~~+\frac{1}{32} K_{2}(s)\left\{s \pi^{2}-24\right\}+\frac{1}{9} K_{3}(s)\left\{3 s^{2}-17 s+9\right\} ,
\\
 G^{(2)}(s, t) &=\overline{J}(t)\left\{\frac{1}{16 \pi^{2}}\left[\frac{412}{27}-\frac{s}{54}\left(t^{2}+5 t+159\right)-t\left(\frac{267}{216} t^{2}-\frac{727}{108} t+\frac{1571}{108}\right)\right]\right.
 \nn 
 &~~~+b_{1}(2-t)+\frac{b_{2}}{3}(t-4)(2 t+s-5)-\frac{b_{3}}{6}(t-4)^{2}(3 t+2 s-8) 
 \nn 
 &~~~\left. +\frac{b_{4}}{6}\left(2 s(3 t-4)(t-4)-32 t+40 t^{2}-11 t^{3}\right)\right\} 
 \nn 
 & ~~~+\frac{1}{36} K_{1}(t)\left\{174+8 s-10 t^{3}+72 t^{2}-185 t-\frac{\pi^{2}}{16}(t-4)(3 s-8)\right\}
 \nn
 &~~~+\frac{1}{9} K_{2}(t)\left\{1+4 s+\frac{\pi^{2}}{64} t(3 s-8)\right\}
 \nn
 &~~~+\frac{1}{9} K_{3}(t)\left\{1+3 st-s+3t^2-9t\right\}+\frac{5}{3} K_{4}(t)\left\{4-2s-t\right\},
\end{align}
where the functions $\overline{J}$ and $K_{i}$ are given by
\bal
\left(\begin{array}{c}{\overline{J}} \\ {K_{1}} \\ {K_{2}} \\ {K_{3}}\end{array}\right) &=\left(\begin{array}{cccc}{0} & {0} & {z} & {-4 \bar N} \\ {0} & {z} & {0} & {0} \\ {0} & {z^{2}} & {0} & {8} \\ {\bar N z s^{-1}} & {0} & {\pi^{2}(\bar N s)^{-1}} & {\pi^{2}}\end{array}\right)\left(\begin{array}{c}{h^{3}} \\ {h^{2}} \\ {h} \\ {-\left(2 \bar N^{2}\right)^{-1}}\end{array}\right),
\\
K_{4} &=\frac{1}{s z}\left(\frac{1}{2} K_{1}+\frac{1}{3} K_{3}+\frac{1}{\bar N} \overline{J}+\frac{\left(\pi^{2}-6\right) s}{12 \bar N^{2}}\right),
\eal
with
\begin{equation}
h(s)=\frac{1}{
\bar N \sqrt{z}} \ln \frac{\sqrt{z}-1}{\sqrt{z}+1} \quad, \quad z=1-\frac{4}{s}, \quad \bar N=16 \pi^{2}  .
\end{equation}
The constants $b_{1},b_2, ..., b_{6}$ are given by 
\begin{align}
b_{1}&= 8 l_{1}^{r}+2 l_{3}^{r}-2 l_{4}^{r}+\frac{7}{6} L+\frac{1}{16 \pi^{2}} \frac{13}{18} 
\nn
& ~~~ + x_{2}\left\{\frac{1}{16 \pi^{2}}\left[\frac{56}{9} l_{1}^{r}+\frac{80}{9} l_{2}^{r}+15 l_{3}^{r}+\frac{26}{9} l_{4}^{r}+\frac{47}{108} L-\frac{17}{216}+\frac{1}{16 \pi^{2}} \frac{3509}{1296}\right]\right.
\nn
&~~~\left.+\frac{1}{6}\left[4 k_{1}+28 k_{2}-6 k_{3}+13 k_{4}\right]+\left[32 l_{1}^{r}+12 l_{3}^{r}-5 l_{4}^{r}\right] l_{4}^{r}-8 l_{3}^{r 2}+r_{1}^{r}\right\},\\
b_{2}&=-8 l_{1}^{r}+2 l_{4}^{r}-\frac{2}{3} L-\frac{1}{16 \pi^{2}} \frac{2}{9}
\nn
&~~~+x_{2}\left\{\frac{1}{16 \pi^{2}}\left[-24 l_{1}^{r}-\frac{166}{9} l_{2}^{r}-18 l_{3}^{r}-\frac{8}{9} l_{4}^{r}-\frac{203}{54} L+\frac{317}{3456}-\frac{1}{16 \pi^{2}} \frac{1789}{432}\right]\right. 
\nn
&~~~\left.-\frac{1}{6}\left[54 k_{1}+62 k_{2}+15 k_{3}+10 k_{4}\right]-\left[32 l_{1}^{r}+4 l_{3}^{r}-5 l_{4}^{r}\right] l_{4}^{r}+r_{2}^{r}\right\},
\\
b_{3}&=2 l_{1}^{r}+\frac{1}{2} l_{2}^{r}-\frac{1}{2} L-\frac{1}{16 \pi^{2}} \frac{7}{12}
\nn
&~~~+x_{2}\left\{\frac{1}{16 \pi^{2}}\left[\frac{178}{9} l_{1}^{r}+\frac{38}{3} l_{2}^{r}-\frac{7}{3} l_{4}^{r}-\frac{365}{216} L-\frac{311}{6912}+\frac{1}{16 \pi^{2}} \frac{7063}{864}\right]\right. 
\nn
&~~~\left.+2\left[4 l_{1}^{r}+l_{2}^{r}\right] l_{4}^{r}+\frac{1}{6}\left[38 k_{1}+30 k_{2}-3 k_{4}\right]+r_{3}^{r}\right\},
\\
b_{4}&=\frac{1}{2} l_{2}^{r}-\frac{1}{6} L-\frac{1}{16 \pi^{2}} \frac{5}{36}
\\
&~~~+x_{2}\left\{\frac{1}{16 \pi^{2}}\left[\frac{10}{9} l_{1}^{r}+\frac{4}{9} l_{2}^{r}-\frac{5}{9} l_{4}^{r}+\frac{47}{216} L+\frac{17}{3456}+\frac{1}{16 \pi^{2}} \frac{1655}{2592}\right]\right.
\nn
&~~~\left.+2 l_{2}^{r} l_{4}^{r}-\frac{1}{6}\left[k_{1}+4 k_{2}+k_{4}\right]+r_{4}^{r}\right\},
\\
b_{5}&=\frac{1}{16 \pi^{2}}\left[-\frac{31}{6} l_{1}^{r}-\frac{145}{36} l_{2}^{r}+\frac{625}{288} L+\frac{7}{864}-\frac{1}{16 \pi^{2}} \frac{66029}{20736}\right]-\frac{21}{16} k_{1}-\frac{107}{96} k_{2}+r_{5}^{r},
\\
b_{6}&=\frac{1}{16 \pi^{2}}\left[-\frac{7}{18} l_{1}^{r}-\frac{35}{36} l_{2}^{r}+\frac{257}{864} L+\frac{1}{432}-\frac{1}{16 \pi^{2}} \frac{11375}{20736}\right]-\frac{5}{48} k_{1}-\frac{25}{96} k_{2}+r_{6}^{r},
\end{align}
where $L=\frac{1}{16 \pi^{2}} \ln \frac{M_\pi^{2}}{\mu^{2}}$ and $k_{i}=\left(4 r_{i}^{r}-\gamma_{i} L\right) L$  with $\gamma_{1}={1}/{3},~ \gamma_{2}={2}/{3}, ~\gamma_{3}=-{1}/{2},~ \gamma_{4}=2$. $r_i^r$ are linear combinations of $c_i^r$, the renormalized LECs of $\mc{L}_6$; see Ref.~\cite{Bijnens:1999hw} for the explicit relations. The scale dependence of $l^r_i(\mu)$ and $r_{i}^{r}(\mu)$ can be separated out as follows
\begin{align}
l_{i}^{r}(\mu)&=\frac{\gamma_i}{32 \pi^{2}}\left(\bar{l}_{i}+\ln \frac{M_{\pi}^{2}}{\mu^{2}}\right), \\
r_{i}^{r}(\mu)&=d^{(2)}_i \left(\ln \frac{M_{\pi}^{2}}{\mu^{2}}\right)^2+d^{(1)}_i \ln \frac{M_{\pi}^{2}}{\mu^{2}}+\bar{r}_i,
\end{align}
where $\bar{l}_{i}$ and $\bar{r}_{i}$ are scale independent LECs and $d^{(1)}_i$ and  $d^{(2)}_i$ are fixed by $\mu {\d b_{i}}/{\d \mu}=0$. With these, we can write $\bar{r}_{i}$ in the following form
\begin{equation}
\bar{r}_i=q_i \cdot b_i+h_i ,
\end{equation}
where $q_1=q_2=q_3=q_4={1}/{x_2},~ q_5=q_6=1$, and \\
\begin{align}
h_1&=\frac{\bar{l}_3{}^2}{512 \pi ^4}+\frac{5 \bar{l}_4{}^2}{256 \pi
   ^4}+\frac{3 \bar{l}_4 \bar{l}_3}{256 \pi ^4}+\frac{15 \bar{l}_3}{1024 \pi ^4}-\frac{7 \bar{l}_1}{1728 \pi ^4}-\frac{5 \bar{l}_2}{432 \pi ^4}-\frac{\bar{l}_1 \bar{l}_4}{48 \pi ^4}-\frac{13 \bar{l}_4}{1152 \pi ^4}+\frac{17}{3456 \pi
   ^2}-\frac{3509}{331776 \pi ^4}
\nn
& ~~~+\frac{1}{x_2}\left(-\frac{\bar{l}_1}{12 \pi ^2}+\frac{\bar{l}_3}{32 \pi ^2}+\frac{\bar{l}_4}{8 \pi ^2}-\frac{13}{288 \pi ^2}\right),\\
   h_2&=-\frac{5 \bar{l}_4{}^2}{256 \pi ^4}+\frac{\bar{l}_1 \bar{l}_4}{48 \pi ^4}-\frac{\bar{l}_3 \bar{l}_4}{256 \pi ^4}+\frac{\bar{l}_4}{288 \pi
   ^4}+\frac{\bar{l}_1}{64 \pi ^4}+\frac{83 \bar{l}_2}{3456 \pi ^4}-\frac{9 \bar{l}_3}{512 \pi ^4}-\frac{317}{55296 \pi ^2}+\frac{1789}{110592 \pi ^4}
   \nn
   &~~~+\frac{1}{x_2}\left(\frac{\bar{l}_1}{12 \pi ^2}-\frac{\bar{l}_4}{8 \pi ^2}+\frac{1}{72 \pi ^2}\right),\\
   h_3&=-\frac{\bar{l}_4 \bar{l}_1}{192 \pi ^4}-\frac{89 \bar{l}_1}{6912 \pi ^4}-\frac{19 \bar{l}_2}{1152 \pi ^4}-\frac{\bar{l}_2 \bar{l}_4}{384 \pi ^4}+\frac{7 \bar{l}_4}{768
   \pi ^4}+\frac{311}{110592 \pi ^2}-\frac{7063}{221184 \pi ^4}
   \nn
& ~~~+\frac{1}{x_2}\left(-\frac{\bar{l}_1}{48 \pi ^2}-\frac{\bar{l}_2}{96 \pi ^2}+\frac{7}{192 \pi ^2}\right),\\
   h_4&=-\frac{5 \bar{l}_1}{6912 \pi ^4}-\frac{\bar{l}_2}{1728 \pi ^4}-\frac{\bar{l}_2 \bar{l}_4}{384 \pi ^4}+\frac{5 \bar{l}_4}{2304 \pi ^4}-\frac{17}{55296 \pi
   ^2}-\frac{1655}{663552 \pi ^4}
   \nn
& ~~~+\frac{1}{x_2}\left(\frac{5}{576 \pi ^2}-\frac{\bar{l}_2}{96 \pi ^2}\right),\\
   h_5&=\frac{31 \bar{l}_1}{9216 \pi ^4}+\frac{145 \bar{l}_2}{27648 \pi ^4}-\frac{7}{13824 \pi ^2}+\frac{66029}{5308416 \pi ^4},\\
   h_6&=\frac{7 \bar{l}_1}{27648 \pi ^4}+\frac{35 \bar{l}_2}{27648 \pi ^4}-\frac{1}{6912 \pi ^2}+\frac{11375}{5308416 \pi ^4}.
\end{align}

\section{3D sections of the constrained $b_i$ space}
\label{sec:3D}

Here we list the plots of the 3D sections of the constrained $b_i$ space for the original $Y$ bounds. The 3D sections are obtained by setting the three of the six $b_i$ parameters to the central values of the fit \eqref{expri data1}. See Figures~\ref{fig:3D1} and \ref{fig:3D2}.

\begin{figure}[H]
    \centering 
    \begin{subfigure}{}
      \includegraphics[width=0.29\textwidth]{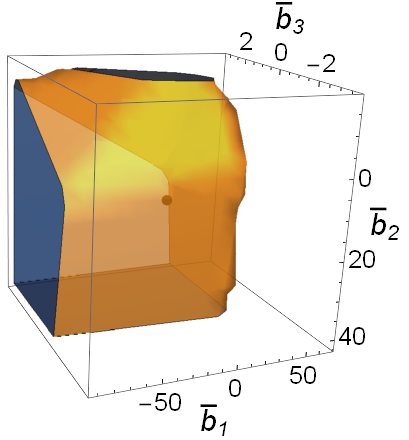}
    \end{subfigure}
     \begin{subfigure}{}
      \includegraphics[width=0.29\textwidth]{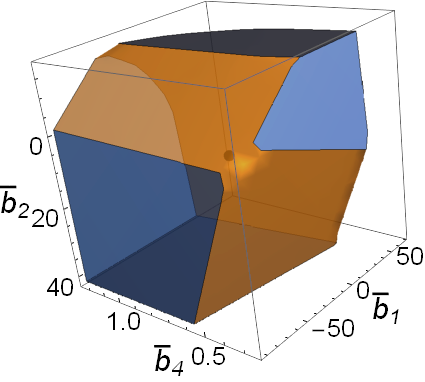}
    \end{subfigure}
    \begin{subfigure}{}
      \includegraphics[width=0.29\textwidth]{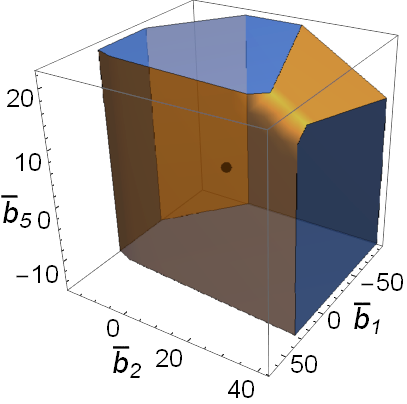}
    \end{subfigure}
    \\
    \centering 
    \begin{subfigure}{}
      \includegraphics[width=0.3\textwidth]{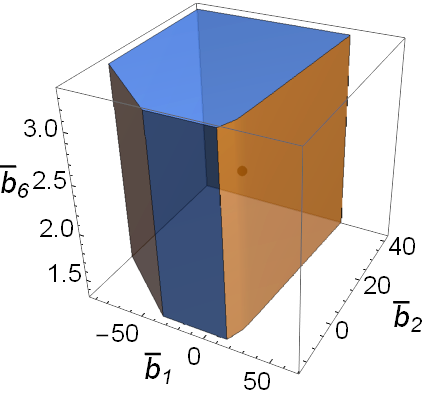}
    \end{subfigure}
     \begin{subfigure}{}
      \includegraphics[width=0.3\textwidth]{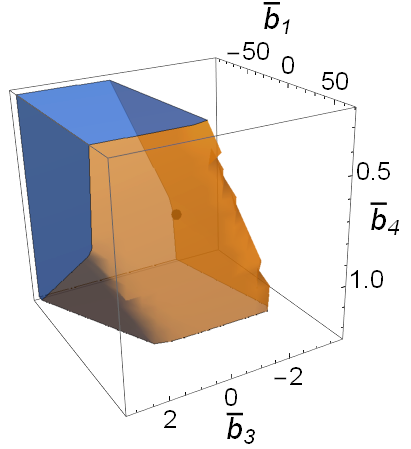}
    \end{subfigure}
    \begin{subfigure}{}
      \includegraphics[width=0.295\textwidth]{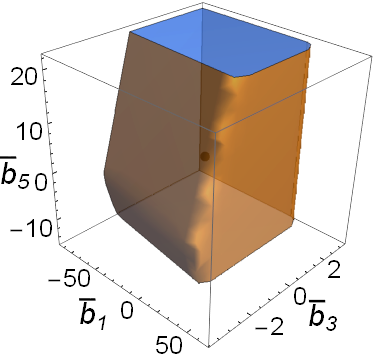}
    \end{subfigure}  
     \\
     \centering     
    \begin{subfigure}{}
      \includegraphics[width=0.3\textwidth]{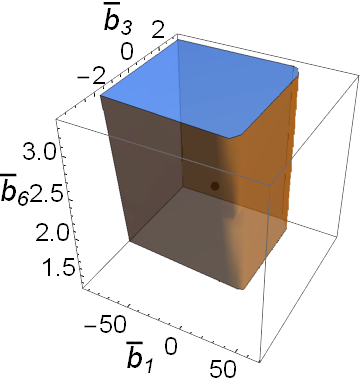}
    \end{subfigure}
     \begin{subfigure}{}
      \includegraphics[width=0.3\textwidth]{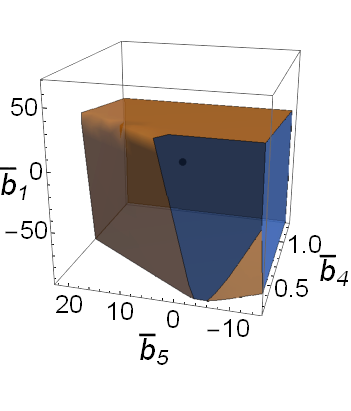}
    \end{subfigure}
    \begin{subfigure}{}
      \includegraphics[width=0.295\textwidth]{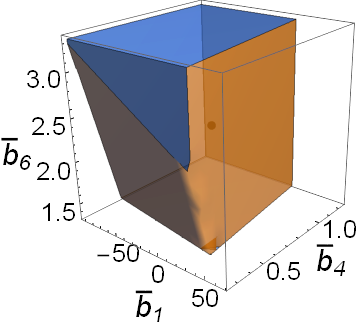}
    \end{subfigure}
    \caption{3D sections of the constrained $b_i$ space. The 3D sections are obtained by setting the other $3$ parameters to the central values of the fit \eqref{expri data1}. The black point represents the central values of the fit \eqref{expri data1} with inputs from the experimental data and theoretical estimates. To be continued in Figure~\ref{fig:3D2}.} 
    \label{fig:3D1}
\end{figure}

\begin{figure}[H]
  \centering 
    \begin{subfigure}{}
      \includegraphics[width=0.316\textwidth]{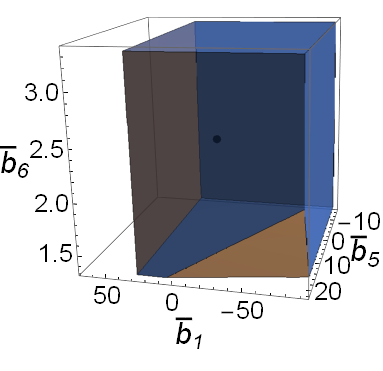}
    \end{subfigure}
     \begin{subfigure}{}
      \includegraphics[width=0.319\textwidth]{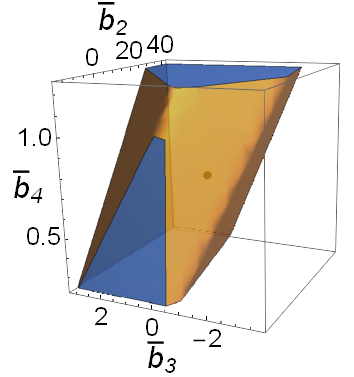}
    \end{subfigure}
    \begin{subfigure}{}
      \includegraphics[width=0.31\textwidth]{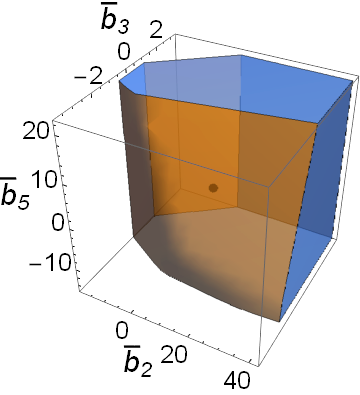}
    \end{subfigure}  
    \\
    \centering 
    \begin{subfigure}{}
      \includegraphics[width=0.319\textwidth]{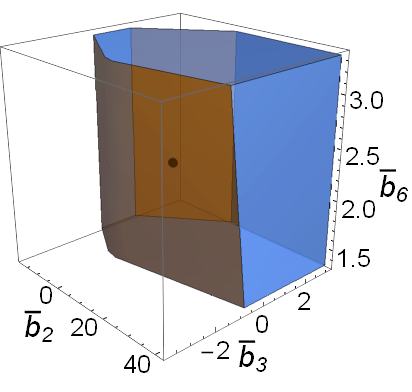}
    \end{subfigure}
     \begin{subfigure}{}
      \includegraphics[width=0.31\textwidth]{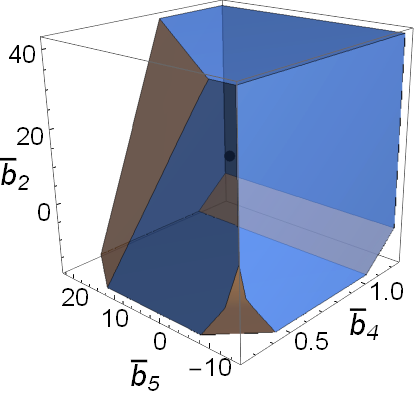}
    \end{subfigure}
    \begin{subfigure}{}
      \includegraphics[width=0.31\textwidth]{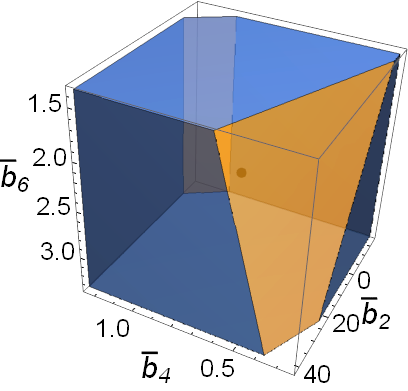}
    \end{subfigure}
    \\
    \centering 
    \begin{subfigure}{}
      \includegraphics[width=0.312\textwidth]{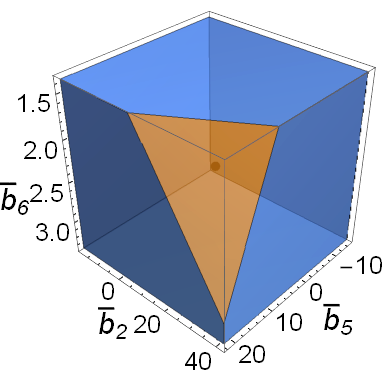}
    \end{subfigure}
     \begin{subfigure}{}
      \includegraphics[width=0.322\textwidth]{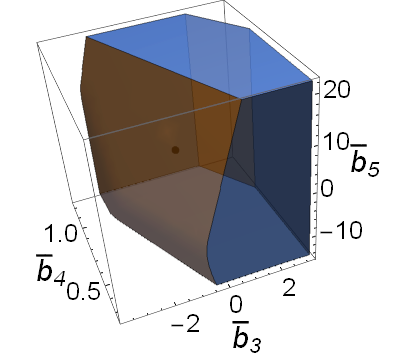}
    \end{subfigure}
    \begin{subfigure}{}
      \includegraphics[width=0.31\textwidth]{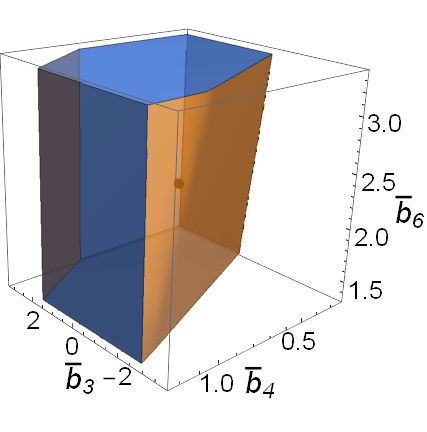}
    \end{subfigure}
   \\
    \centering 
    \begin{subfigure}{}
      \includegraphics[width=0.345\textwidth]{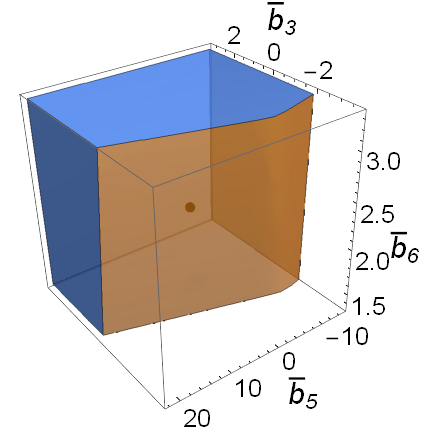}
    \end{subfigure}
     \begin{subfigure}{}
      \includegraphics[width=0.31\textwidth]{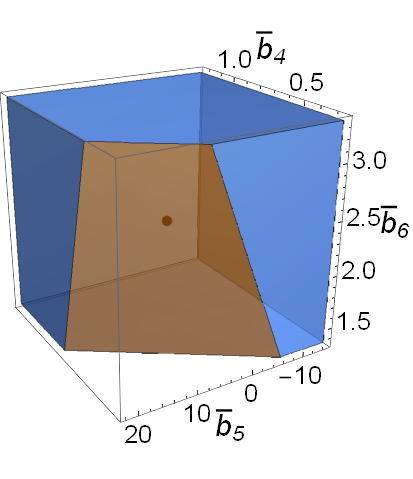}
    \end{subfigure}
    \caption{Continuation of Figure~\ref{fig:3D1}.} 
    \label{fig:3D2}
\end{figure}

\bibliographystyle{JHEP}
\bibliography{refs1}

\end{document}